\documentclass[a4paper,11pt]{article}
\pdfoutput=1 
\usepackage{jheppub}
\usepackage[T1]{fontenc} 
\usepackage{color}
\usepackage{fancybox}
\usepackage{amsmath}
\usepackage{amsfonts}
\usepackage{slashed}
\usepackage{amssymb}
\usepackage{indentfirst}
\usepackage{stmaryrd}
\usepackage{subcaption}
\usepackage{graphicx}
\usepackage{cleveref}
\usepackage{textcomp}
\usepackage[toc]{appendix}
\usepackage[font={small,it}]{caption}
\usepackage[justification=centering]{caption}

\allowdisplaybreaks

\newcommand{\be}{\begin{equation}}
\newcommand{\bea}{\begin{eqnarray}}

\newcommand{\ee}{\end{equation}}
\newcommand{\eea}{\end{eqnarray}}

\newcommand{\half}{\frac{1}{2}}

\title{\boldmath Simplified $D=11$ Pure Spinor $b$ Ghost}

\author{Nathan Berkovits$^{*}$, Max Guillen$^{*}$}

\affiliation{$^{*}$ICTP South American Institute for Fundamental Reserch\\
Instituto de F\'{i}sica Te\'{o}rica, UNESP-Universidade Estadual Paulista\\ R. Dr. Bento T. Ferraz 271, Bl. II, S\~{a}o Paulo 01140-070, SP, Brazil}

\emailAdd{nberkovi@ift.unesp.br, luismax@ift.unesp.br}

\abstract{A $b$-ghost was constructed for the $D=11$ non-minimal pure spinor superparticle by requiring that $\{Q , b\} = T$ where $Q = \Lambda^{\alpha}D_{\alpha} + R^{\alpha}\bar{W}_{\alpha}$ is the usual non-minimal pure spinor BRST operator. As was done for the $D=10$ $b$-ghost, we will show that the $D=11$ $b$-ghost can be simplified by introducing an $SO(10,1)$ fermionic vector $\bar{\Sigma}^{i}$ constructed out of the fermionic spinor $D_{\alpha}$ and pure spinor variables. This simplified version will be shown to satisfy $\{Q, b\} = T$ and $\{b , b\} =$ BRST - trivial.}


\begin{document} 
\hfill{}
\maketitle

\section{Introduction}

The $D=11$ pure spinor superparticle is a useful tool to describe $D=11$ linearized supergravity in a manifestly covariant way \cite{Berkovits:2002uc}. This formalism describes physical states as elements of the cohomology of a BRST operator defined by $Q_{min} = \Lambda^{\alpha}D_{\alpha}$, where $\Lambda^{\alpha}$ is a $D=11$ pure spinor\footnote{In this paper, a d=11 pure spinor $\Lambda^{\alpha}$
will be defined to satisfy $\Lambda\Gamma^{a}\Lambda = 0$. A d=11 pure spinor is sometimes \cite{Howe:1991bx} defined to satisfy both $\Lambda\Gamma^{a}\Lambda = 0$ and $\Lambda\Gamma^{ab}\Lambda = 0$.} satisfying the constraint $\Lambda\Gamma^{a}\Lambda = 0$, $a$ is an $SO(10,1)$ vector index, and $D_{\alpha}$ are the first-class constraints of the $D=11$ Brink-Schwarz-like superparticle \cite{Green:1999by}. The spectrum found by using this formalism coincides with that obtained via the BV quantization of $D=11$ linearized supergravity and includes the graviton, gravitino, and 3-form at ghost-number 3, as well as their ghosts and antifields at other ghost number \cite{Cederwall:2001dx,Berkovits:2002uc}, each one of them satisfying certain equations of motion and gauge invariances as dictated by the BV prescription.

\vspace{2mm}
Motivated by the non-minimal version of the pure spinor superstring \cite{Berkovits:2005bt}, Cederwall formulated the $D=11$ non-minimal pure spinor superparticle by introducing a new set of variables $\bar{\Lambda}_{\alpha}$, $R_{\beta}$ and their respective momenta $\bar{W}^{\alpha}$, $S^{\beta}$, where $\bar{\Lambda}_{\alpha}$ is a $D=11$ bosonic spinor and $R_{\beta}$ is a $D=11$ fermionic spinor satisfying the constraints $\bar{\Lambda}\Gamma^{a}\bar{\Lambda} = 0$ and $\bar{\Lambda}\Gamma^{a}R = 0$ \cite{Cederwall:2009ez,Cederwall:2010tn}. In order for the new variables to not affect the physical spectrum, the BRST operator should be modified to $Q = \Lambda^{\alpha}D_{\alpha} + R_{\alpha}\bar{W}^{\alpha}$, as in the quartet argument of \cite{Kugo:1979gm}. In the non-minimal pure spinor formalism of superstring, one can formulate a consistent prescription to compute scattering amplitudes by constructing a non-fundamental $b$ ghost satisfying $\{Q, b\} = T$. Therefore, it is important to know if a similar $b$ ghost can be constructed
 in the $D=11$ superparticle case.


\vspace{2mm}
The $D=11$ $b$-ghost was first constructed in \cite{Cederwall:2012es} in terms of quantities which are not manifestly invariant under the gauge symmetries of $w_\alpha$ generated by $\Lambda \Gamma^a \Lambda=0$. This $b$-ghost was later shown in \citep{Karlsson:2014xva} to be $Q$-equivalent to one written in terms of the gauge-invariant quantities $N_{ab}$ and $J$, and we   will focus on this manifestly gauge-invariant version of the $b$-ghost. 

\vspace{2mm}
The complicated form of the $b$-ghost in
\citep{Karlsson:2014xva} makes it difficult to treat, so for instance its nilpotency property $\{b,b\}$ has not yet been analyzed. A similar complication exists in $D=10$ dimensions, however, it was shown in \cite{Berkovits:2013pla} that the $D=10$ $b$-ghost could be simplified by defining new fermionic vector variables. In this paper, a similar simplification involving fermionic vector variables will be found for the $D=11$ $b$-ghost which will simplify the computations of $\{Q,b\}=T$ and $\{b,b\}$.

\vspace{2mm}
The paper is organized as follows: In section 2 we review the $D=10$ non-minimal pure spinor superparticle, constructing the corresponding pure spinor $b$-ghost and its simplification. In section 3 we review the $D=11$ pure spinor superparticle, constructing the manifestly gauge-invariant $b$-ghost and explaining how to translate the simplification of the $D=10$ $b$-ghost to the $D=11$ $b$-ghost by defining the $SO(10,1)$ composite fermionic vector $\bar{\Sigma}^{j}$. Finally we construct the simplified $D=11$ $b$-ghost and show that it satisfies the relations $\{Q , b\} = T$ and $\{b , b\} =$ BRST-trivial. Some comments are given at the end of the paper concerning the relation between the $b$-ghost found in \cite{Karlsson:2014xva} and this simplified $b$-ghost.

\section{$D=10$ non-minimal pure spinor superparticle}
The $D=10$ (\emph{minimal}) pure spinor superparticle action is given by \cite{Berkovits:2001rb}:
\begin{equation}
S = \int d\tau (\dot{X}^{m}P_{m} + \dot{\theta}^{\mu}p_{\mu} - \frac{1}{2}P^{m}P_{m} + \dot{\lambda}^{\mu}w_{\mu})
\end{equation}
where $m$, $\mu$ are $SO(9,1)$ vector/spinor indices, $\theta^{\mu}$ is an $SO(9,1)$ Majorana-Weyl spinor, $p_{\mu}$ is its corresponding conjugate momentum and $P^{m}$ is the momentum. The variable $\lambda^{\mu}$ is a $D=10$ pure spinor satisfying the constraint $\lambda\gamma^{m}\lambda = 0$ where $m$ is an $SO(9,1)$ vector index, and $w_{\mu}$ is its corresponding conjugate momentum. Because of the pure spinor constraint this $SO(9,1)$ antichiral spinor is defined up to the gauge transformation $\delta w_{\mu} = (\gamma^{m}\lambda)_{\mu}f_{m}$, where $f_{m}$ is an arbitrary vector. The $SO(9,1)$ gamma matrices denoted by $\gamma^{m}$ satisfy the Clifford algebra $(\gamma^{m})_{\mu\nu}(\gamma^{n})^{\nu\rho} + (\gamma^{n})_{\mu\nu}(\gamma^{m})^{\nu\rho} = 2\eta^{mn}\delta^{\rho}_{\mu}$. The physical states are defined as elements of the cohomology of the BRST operator $Q = \lambda^{\mu}d_{\mu}$, where $d_{\mu}=p_\mu - P_m (\gamma^m \theta)_\mu$ are the first-class constraints of the $D=10$ Brink-Schwarz superparticle \cite{Brink:1981nb}. The spectrum turns out to describe the BV version of $D=10$ (abelian) Super Yang-Mills \cite{Berkovits:2001rb,Berkovits:2002zk,Bedoya:2009np}. 

\vspace{2mm}
In the \emph{non-minimal} version of the pure spinor superparticle \cite{Berkovits:2005bt}\cite{Bjornsson:2010wm}, one  introduces a new pure anti-Weyl spinor $\bar{\lambda}_{\mu}$, and a fermionic field $r_{\mu}$ satisfying the constraint $\bar{\lambda}\gamma^{m}r = 0$, together with their respective conjugate momenta $\bar{w}^{\mu}$, $s^{\mu}$. In order to not affect the cohomology corresponding to $Q_{min}$, the \emph{non-minimal} BRST operator is defined as $Q_{non-min} = \lambda^{\mu}d_{\mu} + \bar{w}^{\mu}r_{\mu}$. Thus the $D=10$ non-minimal pure spinor superparticle is described by the action:
\begin{equation}
S = \int d\tau (\dot{X}^{m}P_{m} + \dot{\theta}^{\mu}p_{\mu} - \frac{1}{2}P^{m}P_{m} + \dot{\lambda}^{\mu}w_{\mu} + \bar{w}^{\mu}\dot{\bar{\lambda}}_{\mu} + \dot{r}_{\mu} s^{\mu}) 
\end{equation}
and the BRST operator $Q = \lambda^{\mu}d_{\mu} + \bar{w}^{\mu}r_{\mu}$. By construction, the physical spectrum also describes BV $D=10$ (abelian) Super Yang-Mills.


\subsection{$D=10$ b-ghost}
As discussed in \cite{Bjornsson:2010wm,Bjornsson:2010wu} a consistent scattering amplitude prescription can be defined using a composite $b$-ghost satisfying $\{Q, b\} = T$, where $Q$ is the non-minimal BRST operator and $T = -\frac{1}{2}P^{a}P_{a}$ is the stress-energy tensor. This superparticle $b$-ghost is obtained by dropping the worldsheet non-zero modes in the superstring $b$ ghost and is
\begin{eqnarray}\label{eeq5}
b &=& \frac{1}{2}\frac{(\bar{\lambda}\gamma_{m}d)}{\bar{\lambda}\lambda}P^{m} - \frac{1}{192}\frac{(\bar{\lambda}\gamma^{mnp}r)[(d\gamma_{mnp}d) + 24N_{mn}P_{p}]}{(\bar{\lambda}\lambda)^{2}} + \frac{1}{16}\frac{(r\gamma_{mnp}r)(\bar{\lambda}\gamma^{m}d)N^{np}}{(\bar{\lambda}\lambda)^{3}} \nonumber \\ 
&& - \frac{1}{128}\frac{(r\gamma_{mnp}r)(\bar{\lambda}\gamma^{pqr}r)N^{mn}N_{qr}}{(\bar{\lambda}\lambda)^{4}}
\end{eqnarray}
where $N_{mn} = \half \lambda \gamma_{mn} w$.

The complicated nature of this expression makes it difficult to prove nilpotence \cite{Jusinskas:2013yca}, however it was shown in 
\cite{Berkovits:2013pla} that the $b$-ghost can be simplified by introducing an $SO(9,1)$ composite fermionic vector $\bar{\Gamma}^{m}$ satisfying the constraint $ (\gamma_{m}\bar{\lambda})^{\mu}\bar{\Gamma}^{m} = 0$. In the expression (\ref{eeq5}), the terms involving $d_{\mu}$ always appear in the combination
\begin{equation}
\bar{\Gamma}^{m} = \frac{1}{2}\frac{(\bar{\lambda}\gamma^{m}d)}{(\bar{\lambda}\lambda)} - \frac{1}{8}\frac{(\bar{\lambda}\gamma^{mnp}r)N_{np}}{(\bar{\lambda}\lambda)^{2}, }
\end{equation}
and using this $\bar{\Gamma}^{m}$, the $b$-ghost can be written in the simpler form:
\begin{equation}\label{eeq6}
b = P^{m}\bar{\Gamma}_{m} - \frac{1}{4}\frac{(\lambda\gamma^{mn}r)}{(\bar{\lambda}\lambda)}\bar{\Gamma}_{m}\bar{\Gamma}_{n}
\end{equation} 
This simplified $D=10$ $b$-ghost was shown to satisfy the property $\{Q , b\} = T$ in \cite{Berkovits:2014ama}, and as shown in Appendix I, the nilpotence property $\{b,b\} =0$ easily follows from
$\{\bar\Gamma_m, \bar\Gamma_n\}=0$ and $[\bar\Gamma_m, \bar\lambda\lambda]=0$.

\section{$D=11$ non-minimal pure spinor superparticle}
The $D=11$ non-minimal pure spinor superparticle action is given by \cite{Berkovits:2002uc}
\begin{equation}\label{eeq1}
S = \int d\tau (\dot{X}^{a}P_{a} + \dot{\Theta}^{\alpha}P_{\alpha} - \frac{1}{2}P^{a}P_{a} + \dot{\Lambda}^{\alpha}W_{\alpha} + \dot{\bar{\Lambda}}_{\alpha}\bar{W}^{\alpha} + \dot{R}_{\alpha}S^{\alpha})
\end{equation}
We use letters of the beginning of the Greek alphabet ($\alpha, \beta, \ldots$) to denote $SO(10,1)$ spinor indices and henceforth we will use Latin letters ($a, b, \ldots, l, m, \ldots$) to denote $SO(10,1)$ vector indices, unless otherwise stated. In \eqref{eeq1} $\Theta^{\alpha}$ is an $SO(10,1)$ Majorana spinor and $P_{\alpha}$ is its corresponding conjugate momentum, and $P_{a}$ is the momentum for $X^a$. The variables $\Lambda^{\alpha}$, $\bar{\Lambda}_{\alpha}$ are $D=11$ pure spinors and $W_{\alpha}$, $\bar{W}^{\alpha}$ are their respective conjugate momenta, $R_{\alpha}$ is an $SO(10,1)$ fermionic spinor satisfying $\bar{\Lambda}\Gamma^{a}R = 0$ and $S^{\alpha}$ is its corresponding conjugate momentum. The $SO(10,1)$ gamma matrices denoted by $\Gamma^{a}$ satisfy the Clifford algebra $(\Gamma^{a})_{\alpha\beta}(\Gamma^{b})^{\beta\sigma} + (\Gamma^{b})_{\alpha\beta}(\Gamma^{a})^{\beta\sigma} = 2\eta^{ab}\delta^{\sigma}_{\alpha}$. In $D=11$ dimensions there exist an antisymmetric spinor metric $C_{\alpha\beta}$ (and its inverse $(C^{-1})^{\alpha\beta}$) which allows us to lower (and raise) spinor indices (e.g. $(\Gamma^{a})^{\alpha\beta} = C^{\alpha\sigma}C^{\beta\delta}(\Gamma^{a})_{\sigma\delta}$, $(\Gamma^{a})^{\alpha}_{\hspace{2mm}\beta} = C^{\alpha\sigma}(\Gamma^{a})_{\sigma\beta}$, etc).

\vspace{2mm}
The physical states described by this theory are defined as elements of the cohomology of the BRST operator $Q = \Lambda^{\alpha}D_{\alpha} + R_{\alpha}\bar{W}^{\alpha}$ where $D_\alpha = P_\alpha - P_a (\Gamma^a \Theta)_\alpha$ and describe $D=11$ linearized supergravity.

\subsection{$D=11$ $b$-ghost and its simplification}
As in the $D=10$ case, a composite $D=11$ $b$-ghost can be constructed satisfying the properties $\{Q , b\} = T$ where $T = -P^a P_a$, and was found in \cite{Cederwall:2012es,Karlsson:2014xva,Karlsson:2015qda} to be:
\begin{align}\label{eq16}
b &= \frac{1}{2}\eta^{-1}(\bar{\Lambda}\Gamma_{ab}\bar{\Lambda})(\Lambda\Gamma^{ab}\Gamma^{i}D)P_{i} + \eta^{-2}L^{(1)}_{ab,cd}[(\Lambda\Gamma^{a}D)(\Lambda\Gamma^{bcd}D) + 2(\Lambda\Gamma^{abc}_{\hspace{4mm}ij}\Lambda)N^{di}P^{j} \notag \\
& + \frac{2}{3}(\eta^{b}_{\hspace{2mm}p}\eta^{d}_{\hspace{2mm}q} - \eta^{bd}\eta_{pq})(\Lambda\Gamma^{apcij}\Lambda)N_{ij}P^{q}] - \frac{1}{3}\eta^{-3}L^{(2)}_{ab,cd,ef}\{(\Lambda\Gamma^{abcij}\Lambda)(\Lambda\Gamma^{def}D)N_{ij} \notag \\
 & - 12[ (\Lambda\Gamma^{abcei}\Lambda)\eta^{fj} - \frac{2}{3}\eta^{f[a}(\Lambda\Gamma^{bce]ij}\Lambda](\Lambda\Gamma^{d}D)N_{ij}\} \notag \\
& + \frac{4}{3}\eta^{-4}L^{(3)}_{ab,cd,ef,gh}(\Lambda\Gamma^{abcij}\Lambda)[(\Lambda\Gamma^{defgk}\Lambda)\eta^{hl} - \frac{2}{3}\eta^{h[d}(\Lambda\Gamma^{efg]kl}\Lambda)]\{N_{ij},N_{kl}\}
\end{align}
where
\begin{eqnarray}
\eta &=& (\Lambda\Gamma^{ab}\Lambda)(\bar{\Lambda}\Gamma_{ab}\bar{\Lambda})\\
L^{(n)}_{a_{0}b_{0}, a_{1}b_{1}, \ldots, a_{n}b_{n}} &=& (\bar{\Lambda}\Gamma_{\llbracket a_{0}b_{0}}\bar{\Lambda})(\bar{\Lambda}\Gamma_{a_{1}b_{1}}R)\ldots (\bar{\Lambda}\Gamma_{a_{n}b_{n}\rrbracket}R)
\end{eqnarray}
and $\llbracket \rrbracket$ means antisymmetrization between each pair of indices. The $D=11$ ghost current is defined by $N_{ij} = \Lambda\Gamma_{ij}W$.


\vspace{2mm}
To simplify this complicated expression for the D=11 $b$-ghost, we shall
mimic the procedure explained above for the D=10 $b$-ghost and look for a similar object to $\bar{\Gamma}^{m}$. A hint comes from looking at the quantity multiplying the momentum $P^{i}$ in the expression for the D=11 $b$-ghost:
\begin{eqnarray}
b &=& P^{i}[\frac{1}{2}\eta^{-1}(\bar{\Lambda}\Gamma_{ab}\bar{\Lambda})(\Lambda\Gamma^{ab}\Gamma_{i}D) + \eta^{-2}L^{(1)}_{ab,cd}[2(\Lambda\Gamma^{abc}_{\hspace{4mm}ki}\Lambda)N^{dk}\nonumber \\
&& 
+ \frac{2}{3}(\eta^{b}_{\hspace{2mm}p}\eta^{d}_{\hspace{2mm}i} - \eta^{bd}\eta_{pi})(\Lambda\Gamma^{apcqj}\Lambda)N_{qj}]] + \ldots
\end{eqnarray}
Therefore our candidate to play the analog role to $\bar{\Gamma}^{m}$ is:
\begin{equation}\label{eq406}
\bar{\Sigma}^{i} = \bar{\Sigma}^{i}_{0} + \frac{2}{\eta^{2}}L^{(1)}_{ab,cd}(\Lambda\Gamma^{abcki}\Lambda)N^{d}_{\hspace{1mm}k} + \frac{2}{3\eta^{2}}L^{(1)\hspace{2mm}i}_{ab,c}(\Lambda\Gamma^{abcqj}\Lambda)N_{qj} - \frac{2}{3\eta^{2}}L^{(1)\hspace{2mm}d}_{ad,c}(\Lambda\Gamma^{aicqj}\Lambda)N_{qj}
\end{equation}
where $\bar{\Sigma}^{i}_{0} = \frac{1}{2}\eta^{-1}(\bar{\Lambda}\Gamma_{ab}\bar{\Lambda})(\Lambda\Gamma^{ab}\Gamma^{i}D)$ is the only term containing $D_{\alpha}$'s. 
Using the identities \eqref{app1}, \eqref{app2} in Appendix B, one finds that
$\bar{\Sigma}^{j}$ satisfies the constraint:
\begin{equation}
(\bar{\Lambda}\Gamma_{ab}\bar{\Lambda})\bar{\Sigma}^{a} = 0.
\end{equation}

\vspace{2mm}
Furthermore, it will be shown in Appendix \ref{apC} that the $D_{\alpha}$'s appearing in $\bar{\Sigma}_{0}^{i}$ are the same as those appearing in the b-ghost. Therefore a plausible assumption for the simplification of the b-ghost would be
$b = P^{i}\bar{\Sigma}_{i} + O(\bar{\Sigma}^{2})$. As will now be shown, the simplified form of 
the $b$-ghost satisfying $\{Q, b\} = T$ is indeed
\begin{align}
b & = P^{i}\bar{\Sigma}_{i} - \frac{2}{\eta}(\bar{\Lambda}\Gamma_{ac}R)(\Lambda\Gamma^{aj}\Lambda)\bar{\Sigma}_{j}\bar{\Sigma}^{c} - \frac{1}{\eta}(\bar{\Lambda}R)(\Lambda\Gamma^{jk}\Lambda)\bar{\Sigma}_{j}\bar{\Sigma}_{k} \label{eq28}
\end{align}

\subsection{Computation of $\{Q, \bar\Sigma^j\}$ }

To show that the $b$-ghost of \eqref{eq28} satisfies $\{Q, b\}=T$, it will be convenient to first compute
$\{Q,\bar\Sigma^i\}$ where, 
using the identities \eqref{app5}, \eqref{app10},
\begin{align}\label{eeq22}
\bar{\Sigma}^{i} & = \bar{\Sigma}_{0}^{i} + \frac{2}{\eta^{2}}(\bar{\Lambda}\Gamma_{ab}\bar{\Lambda})(\bar{\Lambda}\Gamma_{cd}R)(\Lambda\Gamma^{abcki}\Lambda)N^{d}_{\hspace{2mm}k} + \frac{2}{3\eta^{2}}(\bar{\Lambda}\Gamma_{ab}\bar{\Lambda})(\bar{\Lambda}\Gamma_{c}^{\hspace{2mm}i}R)(\Lambda\Gamma^{abcqj}\Lambda)N_{qj} \notag \\
&  - \frac{2}{3\eta^{2}}(\bar{\Lambda}\Gamma_{ac}\bar{\Lambda})(\bar{\Lambda}R)(\Lambda\Gamma^{aicqj}\Lambda)N_{qj}
\end{align}
Using equation \eqref{eeq22} and the identities \eqref{app19}, \eqref{app20}, \eqref{app21}, \eqref{app22}:
\begin{align}
\{Q, \bar{\Sigma}^{i}\} & = -P^{i} - \frac{2}{\eta}[(\bar{\Lambda}\Gamma^{mb}\bar{\Lambda})(\Lambda\Gamma_{b}^{\hspace{1mm}i}\Lambda) - (\bar{\Lambda}\Gamma^{ib}\bar{\Lambda})(\Lambda\Gamma_{b}^{\hspace{1mm}m}\Lambda)]P_{m} + \frac{2}{\eta}(\bar{\Lambda}\Gamma_{mn}R)(\Lambda\Gamma^{mn}\Lambda)\bar{\Sigma}^{i}_{0} \notag \\
& + \frac{4}{\eta}(\bar{\Lambda}\Gamma^{mn}R)(\Lambda\Gamma_{mn}\Lambda)(\bar{\Sigma}^{i} - \bar{\Sigma}^{i}_{0})-\frac{1}{\eta}(\bar{\Lambda}\Gamma_{ab}R)(\Lambda\Gamma^{ab}\Gamma^{i}D) \notag \\
& - \frac{2}{\eta^{2}}(\bar{\Lambda}\Gamma_{ab}\bar{\Lambda})(\bar{\Lambda}\Gamma_{cd}R)(\Lambda\Gamma^{abcki}\Lambda)(\Lambda\Gamma^{d}_{\hspace{1mm}k}D) - \frac{2}{3\eta^{2}}(\bar{\Lambda}\Gamma_{ab}\bar{\Lambda})(\bar{\Lambda}\Gamma_{c}^{\hspace{2mm}i}R)(\Lambda\Gamma^{abcdk}\Lambda)(\Lambda\Gamma_{dk}D)\notag \\
& - \frac{2}{3\eta^{2}}(\bar{\Lambda}\Gamma_{ab}\bar{\Lambda})(\bar{\Lambda}R)(\Lambda\Gamma^{iabdk}\Lambda)(\Lambda\Gamma_{dk}D) - \frac{4}{\eta^{2}}(\bar{\Lambda}\Gamma_{ab}R)(\bar{\Lambda}\Gamma_{cd}R)(\Lambda\Gamma^{abcki}\Lambda)N^{d}_{\hspace{1mm}k}\notag \\
& - \frac{4}{3\eta^{2}}(\bar{\Lambda}\Gamma_{ab}R)(\bar{\Lambda}\Gamma_{c}^{\hspace{2mm}i}R)(\Lambda\Gamma^{abcdk}\Lambda)N_{dk} - \frac{4}{3\eta^{2}}(\bar{\Lambda}\Gamma_{ab}R)(\bar{\Lambda}R)(\Lambda\Gamma^{iabdk}\Lambda)N_{dk} \notag \\
& - \frac{2}{3\eta^{2}}(\bar{\Lambda}\Gamma_{ab}\bar{\Lambda})(RR)(\Lambda\Gamma^{iabdk}\Lambda)N_{dk} \label{eq26}
\end{align}

As shown in Appendix \ref{apD1}, this expression in invariant under the same gauge transformations under which $\bar{\Sigma}_{0}^{i}$ is invariant:
\begin{equation}\label{eq400}
 \delta D_{\alpha} = (\Gamma^{ij}\Lambda\ensuremath{'})_{\alpha}f_{ij}
\end{equation} 
where $(\Lambda\ensuremath{'})^{\alpha} = \frac{1}{2\eta}(\bar{\Lambda}\Gamma_{mn}\bar{\Lambda})(\Lambda\Gamma^{mn})^{\alpha}$ is a pure spinor, and $f_{ij}$ is an antisymmetric gauge parameter. Therefore we can write all $D_\alpha$'s in this object in terms of $\bar{\Sigma}^{i}_{0}$, and the result is (see Appendix \ref{apD1}):
\begin{align}
\{Q, \bar{\Sigma}^{i}\} & = -P^{i} - \frac{2}{\eta}[(\bar{\Lambda}\Gamma^{mb}\bar{\Lambda})(\Lambda\Gamma_{b}^{\hspace{1mm}i}\Lambda) - (\bar{\Lambda}\Gamma^{ib}\bar{\Lambda})(\Lambda\Gamma_{b}^{\hspace{1mm}m}\Lambda)]P_{m} + \frac{4}{\eta}(\bar{\Lambda}\Gamma_{mn}R)(\Lambda\Gamma^{mn}\Lambda)(\bar{\Sigma}^{i} - \bar{\Sigma}^{i}_{0}) \notag \\
& - \frac{2}{\eta}(\bar{\Lambda}\Gamma^{ci}R)(\Lambda\Gamma_{ck}\Lambda)\bar{\Sigma}_{0}^{k} + \frac{4}{\eta}(\bar{\Lambda}\Gamma_{cd}R)(\Lambda\Gamma^{ci}\Lambda)\bar{\Sigma}_{0}^{d} + \frac{2}{\eta}(\bar{\Lambda}R)(\Lambda\Gamma^{ik}\Lambda)\bar{\Sigma}_{0\,k} \notag \\
& - \frac{2}{\eta^{2}}(\bar{\Lambda}\Gamma_{cd}R)(\Lambda\Gamma^{cd}\Lambda)(\bar{\Lambda}\Gamma^{in}\bar{\Lambda})(\Lambda\Gamma^{nk}\Lambda)\bar{\Sigma}_{0\,k} - \frac{4}{\eta^{2}}(\bar{\Lambda}\Gamma_{ab}R)(\bar{\Lambda}\Gamma_{cd}R)(\Lambda\Gamma^{abcki}\Lambda)N^{d}_{\hspace{1mm}k}\notag\\
& - \frac{4}{3\eta^{2}}(\bar{\Lambda}\Gamma_{ab}R)(\bar{\Lambda}\Gamma_{c}^{\hspace{2mm}i}R)(\Lambda\Gamma^{abcdk}\Lambda)N_{dk} - \frac{4}{3\eta^{2}}(\bar{\Lambda}\Gamma_{ab}R)(\bar{\Lambda}R)(\Lambda\Gamma^{iabdk}\Lambda)N_{dk}\notag \\
& - \frac{2}{3\eta^{2}}(\bar{\Lambda}\Gamma_{ab}\bar{\Lambda})(RR)(\Lambda\Gamma^{iabdk}\Lambda)N_{dk}\label{eq401}
\end{align}

After plugging \eqref{eeq22} into \eqref{eq401}, all of the terms explicitly depending on $N_{ab}$ are cancelled and we get (see appendix \ref{apD2}): 
\begin{align}\label{eq500}
\{Q, \bar{\Sigma}^{i}\} & = -P^{i} - \frac{2}{\eta}[(\bar{\Lambda}\Gamma^{mb}\bar{\Lambda})(\Lambda\Gamma_{b}^{\hspace{1mm}i}\Lambda) - (\bar{\Lambda}\Gamma^{ib}\bar{\Lambda})(\Lambda\Gamma_{b}^{\hspace{1mm}m}\Lambda)]P_{m} -\frac{2}{\eta}(\bar{\Lambda}\Gamma^{ci}R)(\Lambda\Gamma_{ck}\Lambda)\bar{\Sigma}^{k} \notag \\
& + \frac{4}{\eta}(\bar{\Lambda}\Gamma_{cd}R)(\Lambda\Gamma^{ci}\Lambda)\bar{\Sigma}^{d} \notag  + \frac{2}{\eta}(\bar{\Lambda}R)(\Lambda\Gamma^{ik}\Lambda)\bar{\Sigma}_{k} - \frac{2}{\eta^{2}}(\bar{\Lambda}\Gamma_{cd}R)(\Lambda\Gamma^{cd}\Lambda)(\bar{\Lambda}\Gamma^{in}\bar{\Lambda})(\Lambda\Gamma^{nk}\Lambda)\bar{\Sigma}_{k}\\
\end{align}

\subsection{$\{Q, b\} = T$}

Using \eqref{eq500}  it is now straightforward to compute $\{Q , b\}$:
\begin{align}
\{Q , b\} & = P^{i}\{Q , \bar{\Sigma}_{i}\} - \frac{4}{\eta^{2}}(\Lambda\Gamma^{mn}\Lambda)(\bar{\Lambda}\Gamma_{mn}R)(\bar{\Lambda}\Gamma^{aj}R)(\Lambda\Gamma_{ak}\Lambda)\bar{\Sigma}^{k}\bar{\Sigma}_{j} \notag \\
& + \frac{2}{\eta}(\bar{\Lambda}\Gamma^{aj}R)(\Lambda\Gamma_{ak}\Lambda)(\{Q , \bar{\Sigma}^{k}\})\bar{\Sigma}_{j} - \frac{2}{\eta}(\bar{\Lambda}\Gamma^{aj}R)(\Lambda\Gamma_{ak}\Lambda)\bar{\Sigma}^{k}(\{Q , \bar{\Sigma}_{j}\}) \notag \\
& - \frac{2}{\eta^{2}}(\Lambda\Gamma^{mn}\Lambda)(\bar{\Lambda}\Gamma_{mn}R)(\bar{\Lambda}R)(\Lambda\Gamma^{jk}\Lambda)\bar{\Sigma}_{j}\bar{\Sigma}_{k} + \frac{1}{\eta}(RR)(\Lambda\Gamma^{jk}\Lambda)\bar{\Sigma}_{j}\bar{\Sigma}_{k} \notag \\
& + \frac{1}{\eta}(\bar{\Lambda}R)(\Lambda\Gamma^{jk}\Lambda)(\{Q , \bar{\Sigma}_{j}\})\bar{\Sigma}_{k} - \frac{1}{\eta}(\bar{\Lambda}R)(\Lambda\Gamma^{jk}\Lambda)\bar{\Sigma}_{j}(\{Q , \bar{\Sigma}_{k}\}) \label{eq27}
\end{align}
To make the computations transparent, each term in \eqref{eq27} involving $\{Q , \bar{\Sigma}_{i}\} $ will be simplified separately:
\begin{align}
M_{1} & = P^{i}\{Q , \bar{\Sigma}_{i}\} \notag \\
& = P_{i}\{-P^{i} - \frac{2}{\eta}[(\Lambda\Gamma^{ib}\Lambda)(\bar{\Lambda}\Gamma_{bm}\bar{\Lambda}) - (\Lambda\Gamma^{mb}\Lambda)(\bar{\Lambda}\Gamma^{bi}\bar{\Lambda})]P_{m} \notag \\
& - \frac{2}{\eta}(\bar{\Lambda}\Gamma^{ci}R)(\Lambda\Gamma_{ck}\Lambda)\bar{\Sigma}^{k} + \frac{4}{\eta}(\bar{\Lambda}\Gamma_{cd}R)(\Lambda\Gamma^{ci}\Lambda)\bar{\Sigma}^{d} \notag \\
& + \frac{2}{\eta}(\bar{\Lambda}R)(\Lambda\Gamma^{ik}\Lambda)\bar{\Sigma}_{k} - \frac{2}{\eta^{2}}(\bar{\Lambda}\Gamma^{cd}R)(\Lambda\Gamma_{cd}\Lambda)(\bar{\Lambda}\Gamma^{in}\bar{\Lambda})(\Lambda\Gamma_{nk}\Lambda)\bar{\Sigma}^{k}\}\notag \\
& = -P^{2} - \frac{2}{\eta}(\bar{\Lambda}\Gamma^{ci}R)(\Lambda\Gamma_{ck}\Lambda)P_{i}\bar{\Sigma}^{k} + \frac{4}{\eta}(\bar{\Lambda}\Gamma_{cd}R)(\Lambda\Gamma^{ci}\Lambda)P_{i}\bar{\Sigma}^{d} \notag \\
& + \frac{2}{\eta}(\bar{\Lambda}R)(\Lambda\Gamma^{ik}\Lambda)P_{i}\bar{\Sigma}_{k} - \frac{2}{\eta^{2}}(\bar{\Lambda}\Gamma^{cd}R)(\Lambda\Gamma_{cd}\Lambda)(\bar{\Lambda}\Gamma^{in}\bar{\Lambda})(\Lambda\Gamma_{nk}\Lambda)P_{i}\bar{\Sigma}^{k}
\end{align}
\begin{align}
M_{2} & = \frac{2}{\eta}(\bar{\Lambda}\Gamma^{aj}R)(\Lambda\Gamma_{ak}\Lambda)(\{Q , \bar{\Sigma}^{k}\})\bar{\Sigma}_{j}\notag \\
& = \frac{2}{\eta}(\bar{\Lambda}\Gamma^{aj}R)(\Lambda\Gamma_{ak}\Lambda)[-P^{k} + \frac{2}{\eta}(\Lambda\Gamma_{mb}\Lambda)(\bar{\Lambda}\Gamma^{bk}\bar{\Lambda})P^{m} \notag - \frac{2}{\eta}(\bar{\Lambda}\Gamma^{ck}R)(\Lambda\Gamma_{cp}\Lambda)\bar{\Sigma}^{p}\\
& - \frac{2}{\eta^{2}}(\bar{\Lambda}\Gamma^{cd}R)(\Lambda\Gamma_{cd}\Lambda)(\bar{\Lambda}\Gamma^{kn}\bar{\Lambda})(\Lambda\Gamma_{np}\Lambda)\bar{\Sigma}^{p}]\bar{\Sigma}_{j}\notag \\
& = -\frac{2}{\eta}{\bar{\Lambda}\Gamma^{aj}R}(\Lambda\Gamma_{ak}\Lambda)P^{k}\bar{\Sigma}_{j} + \frac{4}{\eta^{2}}(\bar{\Lambda}\Gamma^{aj}R)(\Lambda\Gamma_{ak}\Lambda)(\Lambda\Gamma_{mb}\Lambda)(\bar{\Lambda}\Gamma^{bk}\bar{\Lambda})P^{m}\bar{\Sigma}_{j}\notag \\
& - \frac{4}{\eta^{2}}(\bar{\Lambda}\Gamma^{aj}R)(\Lambda\Gamma_{ak}\Lambda)(\bar{\Lambda}\Gamma^{ck}R)(\Lambda\Gamma_{cp}\Lambda)\bar{\Sigma}^{p}\bar{\Sigma}_{j} \notag \\
& - \frac{4}{\eta^{3}}(\bar{\Lambda}\Gamma^{aj}R)(\Lambda\Gamma_{ak}\Lambda)(\bar{\Lambda}\Gamma^{cd}R)(\Lambda\Gamma_{cd}\Lambda)(\bar{\Lambda}\Gamma^{kn}\bar{\Lambda})(\Lambda\Gamma_{np}\Lambda)\bar{\Sigma}^{p}\bar{\Sigma}_{j} \notag
\end{align}
Using \eqref{app2}, we get
\begin{align}
M_{2} & = -\frac{2}{\eta}({\bar{\Lambda}\Gamma^{aj}R})(\Lambda\Gamma_{ak}\Lambda)P^{k}\bar{\Sigma}_{j} + \frac{2}{\eta}(\bar{\Lambda}\Gamma^{aj}R)(\Lambda\Gamma_{ma}\Lambda)P^{m}\bar{\Sigma}_{j} \notag \\ 
& - \frac{2}{\eta^{2}}(\bar{\Lambda}\Gamma^{aj}R)(\Lambda\Gamma_{ap}\Lambda)(\bar{\Lambda}\Gamma^{ck}R)(\Lambda\Gamma_{ck}\Lambda)\bar{\Sigma}^{p}\bar{\Sigma}_{j} \notag \\
& - \frac{2}{\eta^{2}}(\bar{\Lambda}\Gamma^{aj}R)(\Lambda\Gamma_{pa}\Lambda)(\bar{\Lambda}\Gamma^{cd}R)(\Lambda\Gamma_{cd}\Lambda)\bar{\Sigma}^{p}\bar{\Sigma}_{j} \notag \\
& = -\frac{4}{\eta}({\bar{\Lambda}\Gamma^{aj}R})(\Lambda\Gamma_{ak}\Lambda)P^{k}\bar{\Sigma}_{j}
\end{align}

\begin{align}
M_{3} & = - \frac{2}{\eta}(\bar{\Lambda}\Gamma^{aj}R)(\Lambda\Gamma_{ak}\Lambda)\bar{\Sigma}^{k}(\{Q , \bar{\Sigma}_{j}\}) \notag \\
& = - \frac{2}{\eta}(\bar{\Lambda}\Gamma^{aj}R)(\Lambda\Gamma_{ak}\Lambda)\bar{\Sigma}^{k}\{-P_{j} - \frac{2}{\eta}[(\Lambda\Gamma_{jb}\Lambda)(\bar{\Lambda}\Gamma^{bm}\bar{\Lambda}) - (\Lambda\Gamma^{mb}\Lambda)(\bar{\Lambda}\Gamma_{bj}\bar{\Lambda})]P_{m} \notag \\
& - \frac{2}{\eta}(\bar{\Lambda}\Gamma_{cj}R))(\Lambda\Gamma^{cp}\Lambda)\bar{\Sigma}_{p} + \frac{4}{\eta}(\bar{\Lambda}\Gamma^{cd}R)(\Lambda\Gamma_{cj}\Lambda)\bar{\Sigma}_{d} + \frac{2}{\eta}(\bar{\Lambda}R)(\Lambda\Gamma_{jp}\Lambda)\bar{\Sigma}^{p} \notag \\ & 
- \frac{2}{\eta^{2}}(\bar{\Lambda}\Gamma^{cd}R)(\Lambda\Gamma_{cd}\Lambda)(\bar{\Lambda}\Gamma_{jn}\bar{\Lambda})(\Lambda\Gamma^{np}\Lambda)\bar{\Sigma}_{p}\} \notag \\
& = \frac{2}{\eta}(\bar{\Lambda}\Gamma^{aj}R)(\Lambda\Gamma_{ak}\Lambda)\bar{\Sigma}^{k}P_{j} + \frac{4}{\eta^{2}}(\bar{\Lambda}\Gamma^{aj}R)(\Lambda\Gamma_{ak}\Lambda)\bar{\Sigma}^{k}(\Lambda\Gamma_{jb}\Lambda)(\bar{\Lambda}\Gamma^{bm}\bar{\Lambda})P_{m} \notag \\
& - \frac{4}{\eta^{2}}(\bar{\Lambda}\Gamma^{aj}R)(\Lambda\Gamma_{ak}\Lambda)\bar{\Sigma}^{k}(\Lambda\Gamma^{mb}\Lambda)(\bar{\Lambda}\Gamma_{bj}\bar{\Lambda})P_{m} + \frac{4}{\eta^{2}}(\bar{\Lambda}\Gamma^{aj}R)(\Lambda\Gamma_{ak}\Lambda)\bar{\Sigma}^{k}(\bar{\Lambda}\Gamma^{cj}R)(\Lambda\Gamma_{cp}\Lambda)\bar{\Sigma}^{p} \notag \\
& - \frac{8}{\eta^{2}}(\bar{\Lambda}\Gamma^{aj}R)(\Lambda\Gamma_{ak}\Lambda)\bar{\Sigma}^{k}(\bar{\Lambda}\Gamma^{cd}R)(\Lambda\Gamma_{cj}\Lambda)\bar{\Sigma}_{d} - \frac{4}{\eta^{2}}(\bar{\Lambda}\Gamma^{aj}R)(\Lambda\Gamma_{ak}R)\bar{\Sigma}^{k}(\bar{\Lambda}R)(\Lambda\Gamma_{jp}\Lambda)\bar{\Sigma}^{p} \notag \\
& + \frac{4}{\eta^{3}}(\bar{\Lambda}\Gamma^{aj}R)(\Lambda\Gamma_{ak}\Lambda)\bar{\Sigma}^{k}(\bar{\Lambda}\Gamma^{cd}R)(\Lambda\Gamma_{cd}\Lambda)(\bar{\Lambda}\Gamma_{jn}\bar{\Lambda})(\Lambda\Gamma^{np}\Lambda)\bar{\Sigma}_{p}\notag
\end{align}
Using \eqref{app2}, \eqref{app12}, \eqref{app25}:
\begin{align}
M_{3} & = \frac{2}{\eta}(\bar{\Lambda}\Gamma^{aj}R)(\Lambda\Gamma_{ak}\Lambda)\bar{\Sigma}^{k}P_{j} + \frac{2}{\eta^{2}}(\bar{\Lambda}\Gamma^{aj}R)(\Lambda\Gamma_{aj}\Lambda)(\Lambda\Gamma^{kb}\Lambda)(\bar{\Lambda}\Gamma_{bm}\bar{\Lambda})\bar{\Sigma}_{k}P^{m} \notag \\
& + \frac{2}{\eta}(\bar{\Lambda}R)(\Lambda\Gamma^{mk}\Lambda)P_{m}\bar{\Sigma}_{k} - \frac{2}{\eta^{2}}(\bar{\Lambda}\Gamma_{ac}R)(\Lambda\Gamma^{ac}\Lambda)(\bar{\Lambda}R)(\Lambda\Gamma^{kp}\Lambda)\bar{\Sigma}_{k}\bar{\Sigma}_{p} \notag \\
& - \frac{1}{\eta}(RR)(\Lambda\Gamma^{kp}\Lambda)\bar{\Sigma}_{k}\bar{\Sigma}_{p} + \frac{4}{\eta^{2}}(\bar{\Lambda}\Gamma^{aj}R)(\Lambda\Gamma_{aj}\Lambda)(\bar{\Lambda}\Gamma^{cd}R)(\Lambda\Gamma_{ck}\Lambda)\bar{\Sigma}^{k}\bar{\Sigma}_{d} \notag \\
& + \frac{2}{\eta^{2}}(\bar{\Lambda}\Gamma^{aj}R)(\Lambda\Gamma_{aj}\Lambda)(\bar{\Lambda}R)(\Lambda\Gamma^{kp}\Lambda)\bar{\Sigma}_{k}\bar{\Sigma}_{p} - \frac{2}{\eta^{2}}(\bar{\Lambda}R)(\Lambda\Gamma^{kp}\Lambda)(\bar{\Lambda}\Gamma_{cd}R)(\Lambda\Gamma^{cd}\Lambda)\bar{\Sigma}_{k}\bar{\Sigma}_{p}\notag \\
& = \frac{2}{\eta}(\bar{\Lambda}\Gamma^{aj}R)(\Lambda\Gamma_{ak}\Lambda)\bar{\Sigma}^{k}P_{j} + \frac{2}{\eta^{2}}(\bar{\Lambda}\Gamma^{aj}R)(\Lambda\Gamma_{aj}\Lambda)(\Lambda\Gamma^{kb}\Lambda)(\bar{\Lambda}\Gamma_{bm}\bar{\Lambda})\bar{\Sigma}_{k}P^{m} \notag \\
& + \frac{2}{\eta}(\bar{\Lambda}R)(\Lambda\Gamma^{mk}\Lambda)P_{m}\bar{\Sigma}_{k} - \frac{1}{\eta}(RR)(\Lambda\Gamma^{kp}\Lambda)\bar{\Sigma}_{k}\bar{\Sigma}_{p} + \frac{4}{\eta^{2}}(\bar{\Lambda}\Gamma^{aj}R)(\Lambda\Gamma_{aj}\Lambda)(\bar{\Lambda}\Gamma^{cd}R)(\Lambda\Gamma_{ck}\Lambda)\bar{\Sigma}^{k}\bar{\Sigma}_{d} \notag \\
& - \frac{2}{\eta^{2}}(\bar{\Lambda}R)(\Lambda\Gamma^{kp}\Lambda)(\bar{\Lambda}\Gamma_{cd}R)(\Lambda\Gamma^{cd}\Lambda)\bar{\Sigma}_{k}\bar{\Sigma}_{p}
\end{align}
\begin{align}
M_{4} & = \frac{1}{\eta}(\bar{\Lambda}R)(\Lambda\Gamma^{jk}\Lambda)(\{Q , \bar{\Sigma}_{j}\})\bar{\Sigma}_{k} \notag \\
& = -  \frac{1}{\eta}(\bar{\Lambda}R)(\Lambda\Gamma^{jk}\Lambda)P_{j}\bar{\Sigma}_{k} + \frac{2}{\eta^{2}}(\bar{\Lambda}R)(\Lambda\Gamma^{jk}\Lambda)(\Lambda\Gamma^{mb}\Lambda)(\bar{\Lambda}\Gamma_{bj}\bar{\Lambda})P_{m}\bar{\Sigma}_{k} \notag \\
& - \frac{2}{\eta^{2}}(\bar{\Lambda}R)(\Lambda\Gamma^{jk}\Lambda)(\bar{\Lambda}\Gamma_{cj}R)(\Lambda\Gamma^{cp}\Lambda)\bar{\Sigma}_{p}\bar{\Sigma}_{k} - \frac{2}{\eta^{3}}(\bar{\Lambda}R)(\Lambda\Gamma^{jk}\Lambda)(\bar{\Lambda}\Gamma_{cd}R)(\Lambda\Gamma^{cd}\Lambda)(\bar{\Lambda}\Gamma_{jn}\bar{\Lambda})(\Lambda\Gamma^{np}\Lambda)\bar{\Sigma}_{p}\bar{\Sigma}_{k}\notag \\
& = - \frac{1}{\eta}(\bar{\Lambda}R)(\Lambda\Gamma^{jk}\Lambda)P_{j}\bar{\Sigma}_{k} + \frac{1}{\eta}(\bar{\Lambda}R)(\Lambda\Gamma^{km}\Lambda)P_{m}\bar{\Sigma}_{k} \notag \\
& - \frac{1}{\eta^{2}}(\bar{\Lambda}R)(\Lambda\Gamma^{pk}\Lambda)(\bar{\Lambda}\Gamma^{cj}R)(\Lambda\Gamma_{cj}\Lambda)\bar{\Sigma}_{p}\bar{\Sigma}_{k} - \frac{1}{\eta^{2}}(\bar{\Lambda}R)(\bar{\Lambda}\Gamma^{cd}R)(\Lambda\Gamma_{cd}\Lambda)(\Lambda\Gamma^{kp}\Lambda)\bar{\Sigma}_{p}\bar{\Sigma}_{k} \notag \\
& = -\frac{2}{\eta}(\bar{\Lambda}R)(\Lambda\Gamma^{jk}\Lambda)P_{j}\bar{\Sigma}_{k}
\end{align}
\begin{align}
M_{5} & = -\frac{1}{\eta}(\bar{\Lambda}R)(\Lambda\Gamma^{jk}\Lambda)\bar{\Sigma}_{j}(\{Q , \bar{\Sigma}_{k}\}) \notag \\
& = \frac{1}{\eta}(\bar{\Lambda}R)(\Lambda\Gamma^{jk}\Lambda)\bar{\Sigma}_{j}P_{k} - \frac{2}{\eta^{2}}(\bar{\Lambda}R)(\Lambda\Gamma^{jk}\Lambda)\bar{\Sigma}_{j}(\Lambda\Gamma^{mb}\Lambda)(\bar{\Lambda}\Gamma_{bk}\bar{\Lambda})P_{m} \notag \\
& + \frac{2}{\eta^{2}}(\bar{\Lambda}R)(\Lambda\Gamma^{jk}\Lambda)\bar{\Sigma}_{j}(\bar{\Lambda}\Gamma_{ck}R)(\Lambda\Gamma^{cp}\Lambda)\bar{\Sigma}_{p} + \frac{2}{\eta^{3}}(\bar{\Lambda}R)(\Lambda\Gamma^{jk}\Lambda)\bar{\Sigma}_{j}(\bar{\Lambda}\Gamma^{cd}R)(\Lambda\Gamma_{cd}\Lambda)(\bar{\Lambda}\Gamma_{kn}\bar{\Lambda})(\Lambda\Gamma^{np}\Lambda)\bar{\Sigma}_{p}\notag \\
& = \frac{1}{\eta}(\bar{\Lambda}R)(\Lambda\Gamma^{kj}\Lambda)P_{j}\bar{\Sigma}_{k} - \frac{1}{\eta}(\bar{\Lambda}R)(\Lambda\Gamma^{mj}\Lambda)P_{m}\bar{\Sigma}_{j} \notag \\
& + \frac{1}{\eta^{2}}(\bar{\Lambda}R)(\Lambda\Gamma^{jp}\Lambda)\bar{\Sigma}_{j}(\bar{\Lambda}\Gamma^{ck}R)(\Lambda\Gamma_{ck}\Lambda)\bar{\Sigma}_{p} + \frac{1}{\eta^{2}}(\bar{\Lambda}R)(\Lambda\Gamma^{pj}\Lambda)\bar{\Sigma}_{j}(\bar{\Lambda}\Gamma^{cd}R)(\Lambda\Gamma_{cd}\Lambda)\bar{\Sigma}_{p}\notag \\
&= -\frac{2}{\eta}(\bar{\Lambda}R)(\Lambda\Gamma^{jk}\Lambda)P_{j}\bar{\Sigma}_{k}
\end{align}
Putting together all the terms in \eqref{eq27}:
\begin{align}
\{Q , b\} & = \sum_{i=1}^{5}M_{i} - \frac{4}{\eta^{2}}(\Lambda\Gamma^{mn}\Lambda)(\bar{\Lambda}\Gamma_{mn}R)(\bar{\Lambda}\Gamma^{aj}R)(\Lambda\Gamma_{ak}\Lambda)\bar{\Sigma}^{k}\bar{\Sigma}_{j} \notag \\
& -\frac{2}{\eta^{2}}(\Lambda\Gamma^{mn}\Lambda)(\bar{\Lambda}\Gamma_{mn}R)(\bar{\Lambda}R)(\Lambda\Gamma^{jk}\Lambda)\bar{\Sigma}_{j}\bar{\Sigma}_{k} + \frac{1}{\eta}(RR)(\Lambda\Gamma^{jk}\Lambda)\bar{\Sigma}_{j}\bar{\Sigma}_{k}\notag \\
& = -P^{2}
\end{align}
Recalling that $T = -P^{2}$ is the stress-energy tensor, we have checked that $\{Q , b\} = T$.

\subsection{$\{b,b\}$ = BRST-trivial}

In the D=10 case, the identity $\{\bar\Gamma^m, \bar \Gamma^n\}=0$ was crucial for showing that
$\{b,b\}=0$. However, in the D=11 case, it is shown in  Appendix \ref{apD3}  that 
$\{\bar\Sigma^j, \bar \Sigma^k\}$ is non-zero and is proportional to $R_\alpha$. This implies that 

\begin{equation}
\{b , b\} = R^{\alpha}G_{\alpha}(\Lambda, \bar{\Lambda}, R, W, D)
\end{equation}
for some $G_{\alpha}(\Lambda, \bar{\Lambda}, R, W, D)$.

Note that $[Q,\{b , b\}] = 0$ since $[b, T]=0$ where $T = -P_a P^a$. 
Since $Q = \Lambda^{\alpha}D_{\alpha} + R^{\alpha}\bar{W}_{\alpha}$, the quartet argument implies that the cohomology of $Q$ is independent of $R_\alpha$, which allows us to conclude that $\{b , b\} = $ BRST-trivial. It would be interesting to investigate if this BRST-triviality of $\{b,b\}$ is enough for the scattering amplitude prescription using the $b$-ghost to be consistent.

\section{Remarks}
We have succeeded in finding a considerably simpler form in \eqref{eq28} for the D=11 $b$-ghost than that of equation \eqref{eq16} which was presented in \cite{Karlsson:2014xva}. Although this simplified version is not strictly nilpotent, it satisfies the relation $\{b,b\} =$ BRST-trivial which may be good enough for consistency. 

\vspace{2mm}
It is natural to ask if the simplified $D=11$ b-ghost \eqref{eq28} is the same as the $b$-ghost presented in \eqref{eq16}. These two expressions are compared in Appendix \ref{apE} and we find that they coincide up to normal-ordering terms coming from the position of $N_{mn}$ in each expression. Note that the product of $N_{mn}$'s appears as an anticommutator in \eqref{eq16} whereas it appears as an simple ordinary product in \eqref{eq28}. However, because we have ignored normal-ordering questions in our analysis, we will not attempt to address this issue.

\section{Acknowledgments}
MG acknowledges FAPESP grant 15/23732-2 for financial support and NB acknowledges FAPESP grants 2016/01343-7 and 2014/18634-9 and CNPq grant 300256/94-9 for partial financial support.

\newpage
\begin{appendices}
\section{$D=10$ gamma matrix identities}
In $D=10$ we have chiral and antichiral spinors which will be denoted by $\chi^{\alpha}$ and $\chi_{\alpha}$ respectively. The product of two spinors can be decomposed into two forms depending on the chiralities of the spinors used:
\begin{eqnarray}
\xi_{\mu}\chi^{\nu} &=& \frac{1}{16}\delta^{\nu}_{\mu}(\xi\chi) - \frac{1}{2!16}(\gamma^{mn})^{\nu}_{\hspace{2mm}\mu}(\xi\gamma_{mn}\chi) + \frac{1}{4!16}(\gamma^{mnpq})^{\nu}_{\hspace{2mm}\mu}(\xi\gamma_{mnpq}\chi)\\
\xi^{\mu}\chi^{\nu} &=& \frac{1}{16}\gamma_{m}^{\mu\nu}(\xi\gamma^{m}\chi) + \frac{1}{3!16}(\gamma_{mnp})^{\mu\nu}(\xi\gamma^{mnp}\chi) + \frac{1}{5!32}\gamma^{\mu\nu}_{mnpqr}(\xi\gamma^{mnpqr}\chi) \label{ap4}
\end{eqnarray}
The 1-form and 5-form are symmetric, and the 3-form is antisymmetric. Furthermore, it is true that $(\gamma^{mn})^{\mu}_{\hspace{2mm}\nu} = -(\gamma^{mn})_{\nu}^{\hspace{2mm}\mu}$, $(\gamma^{mnpq})^{\mu}_{\hspace{2mm}\nu} = (\gamma^{mnpq})_{\nu}^{\hspace{2mm}\mu}$.

\vspace{2mm}
Two particularly useful identities are:
\begin{eqnarray}
(\gamma^{m})_{(\mu\nu}(\gamma_{m})_{\rho)\sigma}  &=& 0\\
(\gamma^{m})^{\mu}_{\hspace{2mm}\nu}(\gamma_{m})^{\rho}_{\hspace{2mm}\sigma} &=& 4(\gamma^{m})^{\mu\rho}(\gamma_{m})_{\nu\sigma} - 2\delta^{\mu}_{\nu}\delta^{\rho}_{\sigma} - 8\delta^{\mu}_{\sigma}\delta^{\rho}_{\nu} \label{ap1}
\end{eqnarray}
From \ref{ap1} we can deduce the following:
\begin{eqnarray}
(\gamma^{mn})^{\mu}_{\hspace{2mm}\nu}\gamma_{mnp}^{\rho\sigma} &=& 2(\gamma^{m})^{\mu\rho}(\gamma_{pm})^{\sigma}_{\hspace{2mm}\nu} + 6\gamma_{p}^{\mu\rho}\delta^{\sigma}_{\nu} - (\rho \leftrightarrow \sigma)\label{ap2} \\
(\gamma^{mn})^{\mu}_{\hspace{2mm}\nu}(\gamma_{mnp})_{\rho\sigma} &=& -2\gamma^{m}_{\nu\sigma}(\gamma_{pm})^{\mu}_{\hspace{2mm}\rho} + 6(\gamma_{p})_{\nu\sigma}\delta^{\mu}_{\rho} - (\rho \leftrightarrow \sigma)\\
\gamma_{mnp}^{\mu\nu}(\gamma^{mnp})^{\rho\sigma} &=& 12[\gamma_{m}^{\mu\sigma}(\gamma^{m})^{\nu\rho} - \gamma_{m}^{\mu\rho}(\gamma^{m})^{\nu\sigma}] \label{ap5}\\
\gamma^{\mu\nu}_{mnp}\gamma^{mnp}_{\rho\sigma} &=& 48(\delta^{\mu}_{\rho}\delta^{\nu}_{\sigma} - \delta^{\mu}_{\sigma}\delta^{\nu}_{\rho})
\end{eqnarray}
The Lorentz algebra satisfied by the ghost currents $N_{mn} = \frac{1}{2}(\lambda\gamma_{mn}w)$ is:
\begin{equation}\label{ap3}
[N_{pq}, N_{rs}] = \eta_{qs}N_{pr} - \eta_{qr}N_{ps} - \eta_{ps}N_{qr} + \eta_{pr}N_{qs}
\end{equation}
\section{$D=11$ pure spinor identities}
We list some pure spinor identities in eleven dimensions:
\begin{eqnarray}
(\bar{\Lambda}\Gamma^{ab}\bar{\Lambda})(\Gamma_{b}\bar{\Lambda})_{\alpha} &=& 0 \label{app1}\\
(\bar{\Lambda}\Gamma^{[ab}\bar{\Lambda})(\bar{\Lambda}\Gamma^{c]d}\bar{\Lambda}) &=& 0 \label{app2}\\
(\bar{\Lambda}\Gamma^{[ab}\bar{\Lambda})(\bar{\Lambda}\Gamma^{cd]}\bar{\Lambda}) &=& 0 \label{app3}\\ 
(\bar{\Lambda}\Gamma^{[ab}\bar{\Lambda})(\bar{\Lambda}\Gamma^{cd]}R) &=& 0 \label{app4}\\
(\bar{\Lambda}\Gamma_{ij}R)(\bar{\Lambda}\Gamma_{k}^{\hspace{2mm}j}R) &=& (\bar{\Lambda}\Gamma_{ik}R)(\bar{\Lambda}R) + \frac{1}{2}(\bar{\Lambda}\Gamma_{ik}\bar{\Lambda})(RR) \label{app12}\\
(\bar{\Lambda}\Gamma_{ab}R)(\bar{\Lambda}\Gamma_{cd}R)f^{ac}g^{bd} &=& 0 \label{app15}\\
(\Lambda\Gamma_{sk}\Lambda)(\Lambda\Gamma^{abcdk}\Lambda) &=& 0 \label{app11}\\
(\Gamma_{i}\Lambda)_{\alpha}(\Lambda\Gamma^{abcdi}\Lambda) &=& 6(\Gamma^{[ab}\Lambda)_{\alpha}(\Lambda\Gamma^{cd]}\Lambda) \label{app23}\\
(\Gamma_{ij}\Lambda)_{\alpha}(\Lambda\Gamma^{abcij}\Lambda) &=& -18(\Gamma^{[a}\Lambda)_{\alpha}(\Lambda\Gamma^{bc]}\Lambda)\label{app24}
\end{eqnarray}
where $f^{ac}$, $g^{bd}$ are antisymmetric in $(a,c)$, $(b,d)$ respectively. In addition, using \eqref{app4} it can be shown that
\begin{eqnarray}
L^{(1)}_{ab,cd}f^{abc} &=& (\bar{\Lambda}\Gamma_{ab}\bar{\Lambda})(\bar{\Lambda}\Gamma_{cd}R)f^{abc} \label{app5}\\
L^{(1)}_{ab,cd}f^{abc} &=& -(\bar{\Lambda}\Gamma_{cd}\bar{\Lambda})(\bar{\Lambda}\Gamma_{ab}R)f^{abc} \label{app6}\\
L^{(2)}_{ab,cd,ef}f^{abce} &=& (\bar{\Lambda}\Gamma_{ab}\bar{\Lambda})(\bar{\Lambda}\Gamma_{cd}R)(\bar{\Lambda}\Gamma_{ef}R)f^{abce} \label{app7}
\end{eqnarray}
where $f^{abc}, f^{abce}$ are antisymmetric in all of their indices.\\
Other useful identities:
\begin{eqnarray}
L^{(1)\hspace{2mm}d}_{ad,c} &=& (\bar{\Lambda}\Gamma_{ac}\bar{\Lambda})(\bar{\Lambda}R) \label{app10}\\
L^{(2)\hspace{5mm}a}_{ab,cd,e} &=& \frac{1}{3}[2(\bar{\Lambda}\Gamma_{eb}\bar{\Lambda})(\bar{\Lambda}\Gamma_{cd}R)(\bar{\Lambda}R) - (\bar{\Lambda}\Gamma_{cd}\bar{\Lambda})(\bar{\Lambda}\Gamma_{ab}R)(\bar{\Lambda}\Gamma_{e}^{\hspace{2mm}a}R)] \label{app8}\\
L^{(2)\hspace{5mm}c}_{ab,cd,e} &=& \frac{1}{3}[(\bar{\Lambda}\Gamma_{ab}\bar{\Lambda})(\bar{\Lambda}\Gamma_{cd}R)(\bar{\Lambda}\Gamma_{e}^{\hspace{2mm}c}R) - 2(\bar{\Lambda}\Gamma_{ed}\bar{\Lambda})(\bar{\Lambda}\Gamma_{ab}R)(\bar{\Lambda}R)] \label{app9}\\
(\bar{\Lambda}\Gamma_{ab}\bar{\Lambda})\bar{\Sigma}^{b} &=& 0 \label{app16}
\end{eqnarray}
Some useful commutation relations
\begin{align}
[\bar{\Sigma}^{i}, \eta] & = 0 \label{app17}\\
[\bar{\Sigma}^{j}, (\Lambda\Gamma^{mn}\Lambda)] & = \frac{2}{\eta^{2}}(\bar{\Lambda}\Gamma_{ef}\bar{\Lambda})(\bar{\Lambda}\Gamma_{gh}R)[(\Lambda\Gamma^{efgmj}\Lambda)(\Lambda\Gamma^{hn}\Lambda) - (\Lambda\Gamma^{efgnj}\Lambda)(\Lambda\Gamma^{hm}\Lambda)] \label{app18}\\
\{\bar{\Sigma}^{j}, (\bar{\Lambda}\Gamma_{mn}R)(\Lambda\Gamma^{mn}\Lambda)\} & = 0 \label{app25}\\
[Q, \eta] & = -2(\Lambda\Gamma^{mn}\Lambda)(\bar{\Lambda}\Gamma_{mn}R) \label{app19}\\
[Q, (\bar{\Lambda}\Gamma^{ab}\bar{\Lambda})] & = -2(\bar{\Lambda}\Gamma^{ab}R) \label{app20}\\
[Q, N^{hi}] & = (\Lambda\Gamma^{hi}D) \label{app21}\\
[Q, D_{\beta}] & = -2(\Gamma^{m}\Lambda)_{\beta}P_{m} \label{app22}\\
[N^{hi}, \eta] & = -2(\bar{\Lambda}\Gamma_{ab}\bar{\Lambda})[-2\eta^{ai}(\Lambda\Gamma^{bh}\Lambda) + 2\eta^{ah}(\Lambda\Gamma^{bi}\Lambda)] \notag\\
& = 4(\bar{\Lambda}\Gamma^{i}_{\hspace{2mm}b}\bar{\Lambda})(\Lambda\Gamma^{bh}\Lambda) - 4(\bar{\Lambda}\Gamma^{h}_{\hspace{2mm}b}\bar{\Lambda})(\Lambda\Gamma^{bi}\Lambda) \label{app13}\\
[N^{hi}, (\Lambda\Gamma^{lmnpq}\Lambda)] & = -2\eta^{iq}(\Lambda\Gamma^{chlmn}\Lambda) + 2\eta^{in}(\Lambda\Gamma^{chlmq}\Lambda) - 2\eta^{im}(\Lambda\Gamma^{chlnq}\Lambda)  \notag \\
& + 2\eta^{il}(\Lambda\Gamma^{chmnq}\Lambda) + 2\eta^{hq}(\Lambda\Gamma^{cilmn}\Lambda) - 2\eta^{hn}(\Lambda\Gamma^{cilmq}\Lambda) \notag \\
& + 2\eta^{hm}(\Lambda\Gamma^{cilnq}\Lambda) - 2\eta^{hl}(\Lambda\Gamma^{cimnq}\Lambda) + 2\eta^{ci}(\Lambda\Gamma^{hlmnq}\Lambda)\notag \\
& - 2\eta^{ch}(\Lambda\Gamma^{ilmnq}\Lambda) \label{app14}\\
[\Lambda\Gamma^{a}W, \Lambda\Gamma^{b}W] & = -2N^{ab}\label{app42}\\
[\Lambda\Gamma^{a}W, \Lambda\Gamma^{mns}W] & = -2\Lambda\Gamma^{amns}W\label{app43}\\
[\Lambda\Gamma^{abc}W, \Lambda\Gamma^{mnp}W] & = 4\delta^{bc}_{np}N^{am} - 4\delta^{bc}_{mp}N^{an} + 4\delta^{bc}_{mn}N^{ap} - 4\delta^{ac}_{np}N^{bm} + 4\delta^{ac}_{mp}N^{bn}\notag\\
& - 4\delta^{ac}_{mn}N^{bp} + 4\delta^{ab}_{np}N^{cm} - 4\delta^{ab}_{mp}N^{cn} + 4\delta^{ab}_{mn}N^{cp} - 2\Lambda\Gamma^{abcmnp}W\label{app41}
\end{align}
\newpage
\section{Nilpotence of D=10 $b$-ghost}\label{apI}

 The nilpotency property satisfied by this object is not obvious to see so that we will check it in detail. The first step is to show that $\{\bar{\Gamma}^{m}, \bar{\Gamma}^{n}\} = 0$. This can bee seen from the equation \eqref{eeq6} and the use of the $U(5)$ decomposition of the pure spinor variables \cite{Berkovits:2002zk}. If we choose the only non-zero component of $\bar{\lambda}_{\mu}$ to be $\bar{\lambda}_{-----} \neq 0$ then $r_{-++++}$, $r_{+-+++}$, $r_{++-++}$, $r_{+++-+}$, $r_{++++-}$ vanish as follows from the constraint $\bar{\lambda}\gamma^{m}r = 0$. This implies that the only components of $d_{\mu}$ and $N_{mn}$ appearing in \eqref{eeq6} are $d_{\nu}$ with $\nu = \{(-++++), (+-+++), (++-++), (+++-+), (++++-)\}$ and $N^{pq}$ with $p,q = \{(1-2i), (3-4i), (5-6i), (7-8i), (9-10i)\}$. Because all the components of $\bar{\lambda}_{\mu}$ with two plus signs are zero, the commutator $[N_{mn}, \bar{\lambda}_{\mu}\lambda^{\mu}]$ vanishes. Likewise the commutator $[N_{mn},N_{pq}]$ vanishes for $(p,q,m,n)$ taking the values listed above because the metric components are zero for any combination of these values (see equation \eqref{ap3}). Thus we see that $\{\bar{\Gamma}^{m},\bar{\Gamma}^{n}\} = 0$. The nilpotency property of the $b$-ghost can be seen as follows: Because the only contribution of $\lambda$ in $(\bar{\lambda}\lambda)$ is $\lambda^{+++++}$, the commutator $[\bar{\Gamma}^{m}, (\bar{\lambda}\lambda)] = 0$. Furthermore, from the constraint $(\gamma_{m}\bar{\lambda})^{\mu}\bar{\Gamma}^{m} = 0$ it is followed that the only non-zero components of $\bar{\Gamma}^{m}$ are $\bar{\Gamma}^{n}$ with $n=\{(1+2i),(3+4i),(5+6i),(7+8i),(9+10i)\}$. This implies that the term $(\lambda\gamma^{mn}r)$ in \eqref{eeq6} is non-zero only for the cases $m,n = \{(1-2i), (3-4i), (5-6i), (7-8i), (9-10i)\}$. Now, because $w_{\mu}$ can only appear with two or four plus signs in $N^{pq}$ and $\lambda^{\mu}$ appears in $(\lambda\gamma^{mn}r)$ at least with two plus signs the only relevant situation is when $w_{\mu}$ has two plus signs and $\lambda^{\mu}$ has three plus signs, however when this occurs the only components of $r_{\alpha}$ which contribute are those with one minus sign making the whole expression vanish. Therefore the commutator $[\bar{\Gamma}^{m}, (\lambda\gamma^{pq}r)]\bar{\Gamma}_{p}\bar{\Gamma}_{q} = 0$. This implies immediately that $\{b,b\} = 0$.

This result can also be shown from the following covariant computation:
\begin{eqnarray}
\{\bar{\Gamma}^{m}, \bar{\Gamma}^{n}\} &=& \{\frac{1}{2}\frac{(\bar{\lambda}\gamma^{m}d)}{(\bar{\lambda}\lambda)} - \frac{1}{8}\frac{(\bar{\lambda}\gamma^{mrs}r)N_{rs}}{(\bar{\lambda}\lambda)^{2}}, \frac{1}{2}\frac{(\bar{\lambda}\gamma^{n}d)}{(\bar{\lambda}\lambda)} - \frac{1}{8}\frac{(\bar{\lambda}\gamma^{npq}r)N_{pq}}{(\bar{\lambda}\lambda)^{2}}\}\nonumber \\
&=& -\frac{1}{4(\bar{\lambda}\lambda)^{2}}(\bar{\lambda}\gamma^{m}\gamma^{k}\gamma^{n}\bar{\lambda})P_{k} +  \frac{1}{32(\bar{\lambda}\lambda)^{4}}(\bar{\lambda}\gamma^{m}d)(\bar{\lambda}\gamma^{npq}r)(\lambda\gamma_{pq}\bar{\lambda})\nonumber \\
&& + \frac{1}{32(\bar{\lambda}\lambda)^{4}}(\bar{\lambda}\gamma^{n}d)(\bar{\lambda}\gamma^{mrs}r)(\lambda\gamma_{rs}\bar{\lambda}) + \frac{1}{16(\bar{\lambda}\lambda)^{4}}(\bar{\lambda}\gamma^{mrs}r)(\bar{\lambda}\gamma^{nps}r)N_{pr} \nonumber \\
&& + \frac{1}{64(\bar{\lambda}\lambda)^{5}}(\bar{\lambda}\gamma^{mrs}r)(\bar{\lambda}\gamma^{npq}r)(\lambda\gamma_{rs}\bar{\lambda})N_{pq}\nonumber - \frac{1}{64(\bar{\lambda}\lambda)^{5}}(\bar{\lambda}\gamma^{mrs}r)N_{rs}(\bar{\lambda}\gamma^{npq}r)(\lambda\Gamma_{pq}\bar{\lambda})
\end{eqnarray}
where we have used that $\{d_{\mu}, d_{\nu}\} = -\gamma^{m}_{\mu\nu}P_{m}$, $[N_{pq},\bar{\lambda}\lambda] = -\frac{1}{2}(\lambda\gamma_{pq}\bar{\lambda})$ and the Lorentz algebra satisfied by $N_{pq}$ given in \eqref{ap3}. The first term proportional to $P^{m}$ is zero because the pure spinor constraint and the bosonic nature of $\bar{\lambda}_{\alpha}$. From the identity \eqref{ap2} we can show that $(\bar{\lambda}\gamma^{npq}r)(\lambda\gamma_{pq}\bar{\lambda}) = 0$. Therefore the terms proportional to this expression vanish. So we are left with
\begin{equation}
\{\bar{\Gamma}^{m}, \bar{\Gamma}^{n}\} = \frac{1}{16}(\bar{\lambda}\gamma^{mrs}r)(\bar{\lambda}\gamma^{nps}r)N_{pr}
\end{equation}
The equation \eqref{ap4} allows putting this expression into the form
\begin{equation}
\{\bar{\Gamma}^{m}, \bar{\Gamma}^{n}\} = -\frac{1}{3!16^{2}}(\bar{\lambda}\gamma^{mrs}\gamma^{tuv}\gamma^{nps}\bar{\lambda})(r\gamma_{tuv}r)N_{pr}
\end{equation}
Now we can use the GAMMA package \cite{Gran:2001yh} to do gamma matrix manipulations. The expansion of this expression, the use of the pure spinor constraint and the bosonic nature of $\bar{\lambda}_{\mu}$ give us the following result
\begin{eqnarray}
\{\bar{\Gamma}^{m}, \bar{\Gamma}^{n}\} &=& -\frac{1}{3!16^{2}}[\eta^{np}(\bar{\lambda}\gamma^{tuvmr}\bar{\lambda}) + \eta^{mp}(\bar{\lambda}\gamma^{tuvnr}\bar{\lambda}) - \eta^{mn}(\bar{\lambda}\gamma^{tuvpr}\bar{\lambda})\nonumber \\
&& + 6\eta^{tn}(\bar{\lambda}\gamma^{uvmpr}\bar{\lambda}) + 6\eta^{tm}(\bar{\lambda}\gamma^{uvnpr}\bar{\lambda})](r\gamma_{tuv}r)N_{pr}
\end{eqnarray}
Using the reasons mentioned above we can write $\bar{\lambda}\gamma^{tuvmr}\bar{\lambda} = \bar{\lambda}\gamma^{tu}\gamma^{vmr}\bar{\lambda} = \bar{\lambda}\gamma^{tuv}\gamma^{mr}\bar{\lambda}$. Therefore after using the identity \eqref{ap5} and the constraint $\bar{\lambda}\gamma^{m}r = 0$ we obtain the result desired
\begin{equation}
\{\bar{\Gamma}^{m}, \bar{\Gamma}^{n}\} = 0
\end{equation}
Using this we can calculate $\{b,b\}$ directly:
\begin{eqnarray}
\{b , b\} &=& \{P^{m}\bar{\Gamma}_{m} - \frac{1}{4}\frac{(\lambda\gamma^{mn}r)}{(\bar{\lambda}\lambda)}\bar{\Gamma}_{m}\bar{\Gamma}_{n}, P^{p}\bar{\Gamma}_{p} - \frac{1}{4}\frac{(\lambda\gamma^{pq}r)}{(\bar{\lambda}\lambda)}\bar{\Gamma}_{p}\bar{\Gamma}_{q}\}\nonumber \\
&=& 0
\end{eqnarray}
where we have used $[\bar{\Gamma}^{m}, \bar{\lambda}\lambda] = \frac{1}{16(\bar{\lambda}\lambda)^{2}}(\bar{\lambda}\gamma^{mnp}r)(\lambda\gamma_{np}\bar{\lambda}) = 0$ and $\{\bar{\Gamma}^{m}, \lambda\gamma^{rs}r\}\bar{\Gamma}_{r}\bar{\Gamma}_{s} = \frac{1}{8(\bar{\lambda}\lambda)}(\bar{\lambda}\gamma^{mnp}r)(\lambda\gamma_{np}\gamma^{rs}r)\bar{\Gamma}_{r}\bar{\Gamma}_{s} = \frac{1}{2(\bar{\lambda}\lambda)}[-(\bar{\lambda}\gamma^{mns}r)(\lambda\gamma_{n}^{\hspace{2mm}r}r) + (\bar{\lambda}\gamma^{mnr}r)(\lambda\gamma_{n}^{\hspace{2mm}s}r)]\bar{\Gamma}_{r}\bar{\Gamma}_{s} = 0$ because of the constraint $(\gamma_{m}\bar{\lambda})^{\mu}\bar{\Gamma}^{m} = 0$.

\section{The $b$-ghost and $\bar{\Sigma}^{j}$ have the same $D_{\alpha}$'s}\label{apC}
We should figure out which are the $D_{\alpha}$'s appearing in the expressions for $\bar{\Sigma}^{i}$ and the $b$-ghost. For this we will decompose the eleven dimensional Lorentz group in the following way: $SO(10,1)\rightarrow SO(3,1)\times SO(7)$ and we will break the Lorentz invariance by choosing a special direction for $\bar{\Lambda}_{\alpha}$:
\begin{eqnarray}
\bar{\Lambda}\Gamma^{a}\bar{\Lambda} = 0 &\rightarrow & \mbox{We choose the only non-zero component of $\bar{\Lambda}$ to be: $\bar{\Lambda}^{++0}\neq 0$}\\
\bar{\Lambda}\Gamma^{a}R = 0 &\rightarrow & R^{+-0} = R^{-+0} = R^{--j} = 0\hspace{3mm},\hspace{3mm} \mbox{where $j=1,\ldots,7$}
\end{eqnarray}
and from the pure spinor constraint $\Lambda\Gamma^{a}\Lambda = 0$ we have:
\begin{eqnarray}
\Lambda^{-+0} &=& -\frac{\Lambda^{-+j}\Lambda^{--j}}{\Lambda^{--0}} \label{eq7}\\
\Lambda^{+-0} &=& -\frac{\Lambda^{+-j}\Lambda^{--j}}{\Lambda^{--0}}\\
\Lambda^{++j} &=& \frac{1}{\Lambda^{--0}}[\Lambda^{--j}\Lambda^{++0} - \Lambda^{-+j}\Lambda^{+-0} + \Lambda^{+-j}\Lambda^{-+0}]
\end{eqnarray}
where $j = 1, \ldots, 7$ and we have assumed that $\Lambda^{--0}\neq 0$. This allows us to expand the quadratic term in $D_{\alpha}$ in the b-ghost \eqref{eq3} in terms of these components:
\begin{eqnarray}
b_{1} &\propto & \frac{(\bar{\Lambda}^{++0}\bar{\Lambda}^{++0})(\bar{\Lambda}^{++0}R^{--0})}{(\bar{\Lambda}^{++0}\bar{\Lambda}^{++0})^{2}(\Lambda^{--0}\Lambda^{--0} + \Lambda^{--k}\Lambda^{-k})^{2}}\{[\Lambda^{--0}D^{+-0} + \Lambda^{--k}D^{+-k} + \Lambda^{+-0}D^{--0} \nonumber \\
&& + \Lambda^{+-k}D^{--k}]\times [\Lambda^{--0}D^{-+0} + \Lambda^{--k}D^{-+k} - \Lambda^{-+0}D^{--0} - \Lambda^{-+k}D^{--k}]\} \nonumber \\
&& + \frac{(\bar{\Lambda}^{++0}\bar{\Lambda}^{++0})(\bar{\Lambda}^{++0}R^{--0})}{(\bar{\Lambda}^{++0}\bar{\Lambda}^{++0})^{2}(\Lambda^{--0}\Lambda^{--0} + \Lambda^{--k}\Lambda^{--k})^{2}}\{[\Lambda^{--0}D^{-+0} + \Lambda^{--k}D^{-+k} + \Lambda^{-+0}D^{--0} \nonumber \\
&& + \Lambda^{-+k}D^{--k}]\times[\Lambda^{--0}D^{+-0} + \Lambda^{--k}D^{+-k} - \Lambda^{+-0}D^{--0} - \Lambda^{+-k}D^{--k}]\}\nonumber \\
&& + \frac{(\bar{\Lambda}^{++0}\bar{\Lambda}^{++0})(\bar{\Lambda}^{++0}R^{+-j})}{(\bar{\Lambda}^{++0}\bar{\Lambda}^{++0})^{2}(\Lambda^{--0}\Lambda^{--0} + \Lambda^{--k}\Lambda^{-k})^{2}}\{[\Lambda^{--0}D^{-+0} + \Lambda^{--k}D^{-+k} \nonumber \\ 
&& + \Lambda^{-+0}D^{--0} + \Lambda^{-+k}D^{--k}]\times[\Lambda^{--0}D^{--j} - \Lambda^{--j}D^{--0}]\}\nonumber \\
&& - \frac{(\bar{\Lambda}^{++0}\bar{\Lambda}^{++0})(\bar{\Lambda}^{++0}R^{-+j})}{(\bar{\Lambda}^{++0}\bar{\Lambda}^{++0})^{2}(\Lambda^{--0}\Lambda^{--0} + \Lambda^{--k}\Lambda^{--k})^{2}}\{[\Lambda^{--0}D^{+-0} + \Lambda^{--k}D^{+-k} \nonumber \\
&& + \Lambda^{+-0}D^{--0} + \Lambda^{+-k}D^{--k}]\times[\Lambda^{--0}D^{--j} - \Lambda^{--j}D^{--0}]\}
\end{eqnarray}
Now, we can write $\bar{\Sigma}^{i}_{0}$ in the convenient form:
\begin{eqnarray}
\bar{\Sigma}^{i}_{0} &=& \frac{1}{2\eta}[2(\bar{\Lambda}\Gamma^{ai}\bar{\Lambda})(\Lambda\Gamma_{a}D) + (\bar{\Lambda}\Gamma_{ab}\bar{\Lambda})(\Lambda\Gamma^{abi}D)]
\end{eqnarray}
After using the particular direction chosen above, $\bar{\Sigma}^{i}_{0}$ presents the following $SO(3,1)\times SO(7)$ components:
\begin{eqnarray}
\bar{\Sigma}_{0}^{1+2i} &=& 0\\
\bar{\Sigma}_{0}^{3+4i} &=& 0\\
\bar{\Sigma}_{0}^{1-2i} &\propto & \frac{\bar{\Lambda}^{++0}\bar{\Lambda}^{++0}}{(\bar{\Lambda}^{++0}\bar{\Lambda}^{++0})(\Lambda^{--0}\Lambda^{--0} + \Lambda^{--k}\Lambda^{--k})}(\Lambda^{--0}D^{+-0} + \Lambda^{--k}D^{+-k}) \label{eq4}\\
\bar{\Sigma}_{0}^{3-4i} &\propto & \frac{\bar{\Lambda}^{++0}\bar{\Lambda}^{++0}}{(\bar{\Lambda}^{++0}\bar{\Lambda}^{++0})(\Lambda^{--0}\Lambda^{--0} + \Lambda^{--k}\Lambda^{--k})}(\Lambda^{--0}D^{-+0} + \Lambda^{--k}D^{-+k}) \label{eq5}\\
\bar{\Sigma}_{0}^{j} &\propto & \frac{\bar{\Lambda}^{++0}\bar{\Lambda}^{++0}}{(\bar{\Lambda}^{++0}\bar{\Lambda}^{++0})(\Lambda^{--0}\Lambda^{--0} + \Lambda^{--k}\Lambda^{--k})}(\Lambda^{--0}D^{--j} - \Lambda^{--j}D^{--0}) \label{eq6}
\end{eqnarray}
where $k,j = 1,\ldots, 7$. Furthermore, if we multiply $\bar{\Sigma}^{j}_{0}$ by $\frac{\Lambda^{-+j}}{\Lambda^{--0}}$ we get:
\begin{equation}\label{eq8}
\frac{\Lambda^{-+j}}{\Lambda^{--0}}\bar{\Sigma}_{0}^{j} \propto \frac{\bar{\Lambda}^{++0}\bar{\Lambda}^{++0}}{(\bar{\Lambda}^{++0}\bar{\Lambda}^{++0})(\Lambda^{--0}\Lambda^{--0} + \Lambda^{--k}\Lambda^{--k})}(\Lambda^{-+j}D^{--j} + \Lambda^{-+0}D^{--0})
\end{equation}
where we used the pure spinor constraint \eqref{eq7}. In a similar way we obtain:
\begin{equation}\label{eq9}
\frac{\Lambda^{+-j}}{\Lambda^{--0}}\bar{\Sigma}_{0}^{j} \propto  \frac{\bar{\Lambda}^{++0}\bar{\Lambda}^{++0}}{(\bar{\Lambda}^{++0}\bar{\Lambda}^{++0})(\Lambda^{--0}\Lambda^{--0} + \Lambda^{--k}\Lambda^{--k})}(\Lambda^{+-j}D^{--j} + \Lambda^{+-0}D^{--0})
\end{equation}
Therefore we see that the expression for $b_{1}$ contains the same combinations of $D_{\alpha}$'s as those contained in the expression for $\bar{\Sigma}_{0}^{i}$ (\eqref{eq4}, \eqref{eq5}, \eqref{eq6}, \eqref{eq8}, \eqref{eq9}). 
\section{$D_{\alpha}$ in terms of $\bar{\Sigma}_{0}^{j}$}\label{apD}
Let us define the quantity:
\begin{equation}
H_{\alpha} = (\Lambda\Gamma_{i})_{\alpha}\bar{\Sigma}_{0}^{i} =\frac{1}{2\eta}(\Gamma_{i}\Lambda)_{\alpha}(\bar{\Lambda}\Gamma_{ab}\bar{\Lambda})(\Lambda\Gamma^{ab}\Gamma^{i}D)
\end{equation} 
Now we will assume that there exist a matrix $(M^{-1})_{\alpha}^{\hspace{2mm}\beta}$ such that:
\begin{equation}
D_{\alpha} = (M^{-1})_{\alpha}^{\hspace{2mm}\beta}H_{\beta}
\end{equation}
and let us make the following ansatz for $(M^{-1})_{\alpha}^{\hspace{2mm}\beta}$:
\begin{equation}
(M^{-1})_{\alpha}^{\hspace{2mm}\beta} = 2\delta_{\alpha}^{\hspace{1mm}\beta} + \frac{2}{\eta}(\Lambda\Gamma_{m})_{\alpha}(\bar{\Lambda}\Gamma^{mn}\bar{\Lambda})(\Lambda\Gamma_{n})^{\beta}
\end{equation}
Next we will check that this proposal for $(M^{-1})_{\alpha}^{\hspace{2mm}\beta}$ is right:
\begin{eqnarray*}
H_{\alpha} &=& \frac{1}{2\eta}(\Gamma_{i}\Lambda)_{\alpha}(\bar{\Lambda}\Gamma_{ab}\bar{\Lambda})(\Lambda\Gamma^{ab}\Gamma^{i}M^{-1}H)\\
&=& \frac{1}{2\eta}(\Gamma_{i}\Lambda)_{\alpha}(\bar{\Lambda}\Gamma_{ab}\bar{\Lambda})[2(\Lambda\Gamma^{ab}\Gamma^{i}\Gamma^{j}\Lambda)\bar{\Sigma}_{0\,j} + \frac{2}{\eta}(\Lambda\Gamma^{ab}\Gamma^{i}\Gamma^{m}\Lambda)(\bar{\Lambda}\Gamma_{mn}\bar{\Lambda})(\Lambda\Gamma^{nj}\Lambda)\bar{\Sigma}_{0\,j}]\\
&=& \frac{1}{2\eta}(\Gamma_{i}\Lambda)_{\alpha}(\bar{\Lambda}\Gamma_{ab}\bar{\Lambda})[2\eta^{ij}(\Lambda\Gamma^{ab}\Lambda) + 4\eta^{aj}(\Lambda\Gamma^{bi}\Lambda) - 4\eta^{ai}(\Lambda\Gamma^{bj}\Lambda) ]\bar{\Sigma}_{0\,j}\\
& & + \frac{1}{\eta^{2}}(\Gamma_{i}\Lambda)_{\alpha}(\bar{\Lambda}\Gamma_{ab}\bar{\Lambda})[\eta^{im}(\Lambda\Gamma^{ab}\Lambda) + 2\eta^{am}(\Lambda\Gamma^{bi}\Lambda) - 2\eta^{ai}(\Lambda\Gamma^{bm}\Lambda)](\bar{\Lambda}\Gamma_{mn}\bar{\Lambda})(\Lambda\Gamma^{nj}\Lambda)\bar{\Sigma}_{0\,j}\\
&=& (\Gamma^{j}\Lambda)_{\alpha}\bar{\Sigma}_{0\,j} - \frac{2}{\eta}(\Gamma_{i}\Lambda)_{\alpha}(\bar{\Lambda}\Gamma^{i}_{\hspace{1mm}b}\bar{\Lambda})(\Lambda\Gamma^{bj}\Lambda)\bar{\Sigma}_{0\,j} + \frac{1}{\eta^{2}}[\eta(\Gamma_{i}\Lambda)_{\alpha}(\bar{\Lambda}\Gamma^{i}_{\hspace{1mm}n}\bar{\Lambda})(\Lambda\Gamma^{nj}\Lambda)\bar{\Sigma}_{0\,j}\\
& & -2(\Gamma_{i}\Lambda)_{\alpha}(\bar{\Lambda}\Gamma^{i}_{\hspace{1mm}b}\bar{\Lambda})(\Lambda\Gamma^{b}_{\hspace{1mm}m}\Lambda)(\bar{\Lambda}\Gamma^{m}_{\hspace{2mm}n}\bar{\Lambda})(\Lambda\Gamma^{nj}\Lambda)\bar{\Sigma}_{0\,j}]\\
&=& (\Gamma^{j}\Lambda)_{\alpha}\bar{\Sigma}_{0\,j} - \frac{2}{\eta}(\Gamma_{i}\Lambda)_{\alpha}(\bar{\Lambda}\Gamma^{i}_{\hspace{1mm}b}\bar{\Lambda})(\Lambda\Gamma^{bj}\Lambda)\bar{\Sigma}_{0\,j} + \frac{1}{\eta^{2}}[\eta(\Gamma_{i}\Lambda)_{\alpha}(\bar{\Lambda}\Gamma^{i}_{\hspace{1mm}n}\bar{\Lambda})(\Lambda\Gamma^{nj}\Lambda)\bar{\Sigma}_{0\,j}\\
& & + \eta(\Gamma_{i}\Lambda)_{\alpha}(\bar{\Lambda}\Gamma^{i}_{\hspace{1mm}n}\bar{\Lambda})(\Lambda\Gamma^{nj}\Lambda)\bar{\Sigma}_{0\,j}]\\
&=& (\Gamma^{j}\Lambda)_{\alpha}\bar{\Sigma}_{0\,j}
\end{eqnarray*}
where the identity \eqref{app3} was used in the penultimate line. Therefore we have the relation:
\begin{equation}\label{app50}
D_{\alpha} = 2(\Lambda\Gamma^{c})_{\alpha}\bar{\Sigma}_{0\,c} + \frac{2}{\eta}(\Lambda\Gamma^{m})_{\alpha}(\bar{\Lambda}\Gamma_{mn}\bar{\Lambda})(\Lambda\Gamma^{nj}\Lambda)\bar{\Sigma}_{0\,j}
\end{equation}
It can be shown that $\bar{\Sigma}_{0\,i}$ can be written in terms of $H_{\alpha}$:
\begin{eqnarray*}
H_{\alpha} &=& (\Lambda\Gamma^{c})_{\alpha}\bar{\Sigma}_{0\,c}\\
(\Lambda\Gamma^{ijk}H) &=& (\Lambda\Gamma^{c}\Gamma^{ijk}\Lambda)\bar{\Sigma}_{0\,c}\\
(\bar{\Lambda}\Gamma_{ij}\bar{\Lambda})(\Lambda\Gamma^{ijk}H) &=& (\bar{\Lambda}\Gamma_{ij}\bar{\Lambda})[\eta^{ci}(\Lambda\Gamma^{jk}\Lambda) - \eta^{cj}(\Lambda\Gamma^{ik}\Lambda) + \eta^{ck}(\Lambda\Gamma^{ij}\Lambda)]\bar{\Sigma}_{0\,c}
\end{eqnarray*}
Therefore by using the constraint $(\bar{\Lambda}\Gamma_{ab}\bar{\Lambda})\bar{\Sigma}_{0}^{b} = 0$, we find
\begin{equation}
\bar{\Sigma}_{0}^{k} = \frac{1}{\eta}(\bar{\Lambda}\Gamma_{ij}\bar{\Lambda})(\Lambda\Gamma^{ijk}H) 
\end{equation}

\section{The $D_{\alpha}$'s in $\{Q, \bar{\Sigma}^{i}\}$ are gauge invariant}\label{apD1}
We will show that the $D_{\alpha}$'s appearing in \eqref{eq26} are invariant under the gauge transformations \eqref{eq400}. Therefore they are the same $D_{\alpha}$'s as those contained in the definition of $\bar{\Sigma}^{i}$. In this Appendix and the next ones we have made use of the GAMMA package \cite{Gran:2001yh} because of the heavy manipulation of gamma matrix identities which computations demanded. Let us call $I^{i}$ to the terms containing $D_{\alpha}$'s explicitly in \eqref{eq26}. The identities \eqref{app23}, \eqref{app24} allow us simplify this object:
\begin{align}
I^{i} & = -\frac{1}{\eta}(\bar{\Lambda}\Gamma_{ab}R)(\Lambda\Gamma^{abi}D) - \frac{2}{\eta}(\bar{\Lambda}\Gamma^{ai}R)(\Lambda\Gamma_{a}D) - \frac{2}{\eta^{2}}(\bar{\Lambda}\Gamma_{ab}\bar{\Lambda})(\bar{\Lambda}\Gamma_{cd}R)(\Lambda\Gamma^{abcki}\Lambda)(\Lambda\Gamma^{d}_{\hspace{1mm}k}D) \notag \\
& -\frac{2}{3\eta^{2}}(\bar{\Lambda}\Gamma_{ab}\bar{\Lambda})(\bar{\Lambda}\Gamma_{c}^{\hspace{2mm}i}R)\frac{(-18)}{6}[4(\Lambda\Gamma^{a}D)(\Lambda\Gamma^{bc}\Lambda) + 2(\Lambda\Gamma^{c}D)(\Lambda\Gamma^{ab}\Lambda)] \notag \\
& - \frac{2}{3\eta^{2}}(\bar{\Lambda}\Gamma_{ab}\bar{\Lambda})(\bar{\Lambda}R)\frac{(-18)}{6}[4(\Lambda\Gamma^{a}D)(\Lambda\Gamma^{bi}\Lambda) + 2(\Lambda\Gamma^{i}D)(\Lambda\Gamma^{ab}\Lambda)] \notag \\
& = -\frac{1}{\eta}(\bar{\Lambda}\Gamma_{ab}R)(\Lambda\Gamma^{abi}D) - \frac{2}{\eta}(\bar{\Lambda}\Gamma^{ai}R)(\Lambda\Gamma_{a}D) - \frac{2}{\eta^{2}}(\bar{\Lambda}\Gamma_{ab}\bar{\Lambda})(\bar{\Lambda}\Gamma_{cd}R)(\Lambda\Gamma^{abcik}\Lambda)(\Lambda\Gamma_{k}^{\hspace{1mm}d}D) \notag \\
& + \frac{8}{\eta^{2}}(\bar{\Lambda}\Gamma_{ab}\bar{\Lambda})(\bar{\Lambda}\Gamma_{c}^{\hspace{2mm}i}R)(\Lambda\Gamma^{bc}\Lambda)(\Lambda\Gamma^{a}D) + \frac{4}{\eta}(\bar{\Lambda}\Gamma^{ci}R)(\Lambda\Gamma_{c}D) \notag \\
& + \frac{8}{\eta^{2}}(\bar{\Lambda}\Gamma_{ab}\bar{\Lambda})(\bar{\Lambda}R)(\Lambda\Gamma^{bi}\Lambda)(\Lambda\Gamma^{a}D) + \frac{4}{\eta}(\bar{\Lambda}R)(\Lambda\Gamma^{i}D) \label{eq23}
\end{align}
The third term of this expression requires more careful manipulations, so we will do them in detail
\begin{align}
I^{*\,i} & = -\frac{2}{\eta^{2}}(\bar{\Lambda}\Gamma_{ab}\bar{\Lambda})(\bar{\Lambda}\Gamma_{cd}R)(\Lambda\Gamma^{abcik}\Lambda)(\Lambda\Gamma_{k}^{\hspace{2mm}d}D) \notag \\
& = -\frac{2}{\eta^{2}}(\bar{\Lambda}\Gamma_{ab}\bar{\Lambda})(\bar{\Lambda}\Gamma_{cd}R)[(\Lambda\Gamma^{ab}\Gamma^{d}D)(\Lambda\Gamma^{ci}\Lambda) - (\Lambda\Gamma^{ac}\Gamma^{d}D)(\Lambda\Gamma^{bi}\Lambda) + (\Lambda\Gamma^{ai}\Gamma^{d}D)(\Lambda\Gamma^{bc}\Lambda) \notag \\
& + (\Lambda\Gamma^{ci}\Gamma^{d}D)(\Lambda\Gamma^{ab}\Lambda) - (\Lambda\Gamma^{bi}\Gamma^{d}D)(\Lambda\Gamma^{ac}\Lambda) + (\Lambda\Gamma^{bc}\Gamma^{d}D)(\Lambda\Gamma^{ai}\Lambda)] \notag \\
& = -\frac{2}{\eta^{2}}(\bar{\Lambda}\Gamma_{ab}\bar{\Lambda})(\bar{\Lambda}\Gamma_{cd}R)[\eta^{bd}(\Lambda\Gamma^{a}D)(\Lambda\Gamma^{ci}\Lambda) - \eta^{ad}(\Lambda\Gamma^{b}D)(\Lambda\Gamma^{ci}\Lambda) + (\Lambda\Gamma^{abd}D)(\Lambda\Gamma^{ci}\Lambda) \notag \\
& -\eta^{cd}(\Lambda\Gamma^{a}D)(\Lambda\Gamma^{bi}\Lambda) + \eta^{ad}(\Lambda\Gamma^{c}D)(\Lambda\Gamma^{bi}\Lambda) - (\Lambda\Gamma^{acd}D)(\Lambda\Gamma^{bi}\Lambda) \notag \\
& + \eta^{id}(\Lambda\Gamma^{a}D)(\Lambda\Gamma^{bc}\Lambda) - \eta^{ad}(\Lambda\Gamma^{i}D)(\Lambda\Gamma^{bc}\Lambda) + (\Lambda\Gamma^{aid}D)(\Lambda\Gamma^{bc}\Lambda) \notag \\
& +\eta^{id}(\Lambda\Gamma^{c}D)(\Lambda\Gamma^{ab}\Lambda) - \eta^{cd}(\Lambda\Gamma^{i}D)(\Lambda\Gamma^{ab}\Lambda) + (\Lambda\Gamma^{cid}D)(\Lambda\Gamma^{ab}\Lambda) \notag \\
& -\eta^{id}(\Lambda\Gamma^{b}D)(\Lambda\Gamma^{ac}\Lambda) + \eta^{bd}(\Lambda\Gamma^{i}D)(\Lambda\Gamma^{ac}\Lambda) - (\Lambda\Gamma^{bid}D)(\Lambda\Gamma^{ac}\Lambda) \notag \\
& + \eta^{cd}(\Lambda\Gamma^{b}D)(\Lambda\Gamma^{ai}\Lambda) - \eta^{bd}(\Lambda\Gamma^{c}D)(\Lambda\Gamma^{ai}\Lambda) + (\Lambda\Gamma^{bcd}D)(\Lambda\Gamma^{ai}\Lambda)] \notag \\
& = -\frac{2}{\eta^{2}}(\bar{\Lambda}\Gamma_{ab}\bar{\Lambda})(\bar{\Lambda}\Gamma_{cd}R)[2\eta^{bd}(\Lambda\Gamma^{a}D)(\Lambda\Gamma^{ci}\Lambda) + (\Lambda\Gamma^{abd}D)(\Lambda\Gamma^{ci}\Lambda) + 2\eta^{ad}(\Lambda\Gamma^{c}D)(\Lambda\Gamma^{bi}\Lambda) \notag \\
& - 2(\Lambda\Gamma^{acd}D)(\Lambda\Gamma^{bi}\Lambda) + 2\eta^{id}(\Lambda\Gamma^{a}D)(\Lambda\Gamma^{bc}\Lambda) - 2\eta^{ad}(\Lambda\Gamma^{i}D)(\Lambda\Gamma^{bc}\Lambda) + 2(\Lambda\Gamma^{aid}D)(\Lambda\Gamma^{bc}\Lambda) \notag \\
& + \eta^{id}(\Lambda\Gamma^{c}D)(\Lambda\Gamma^{ab}\Lambda) + (\Lambda\Gamma^{cid}D)(\Lambda\Gamma^{ab}\Lambda)]\notag \\
& = -\frac{4}{\eta^{2}}(\bar{\Lambda}\Gamma_{ac}\bar{\Lambda})(\bar{\Lambda}R)(\Lambda\Gamma^{ci}\Lambda)(\Lambda\Gamma^{a}D) - \frac{2}{\eta^{2}}(\bar{\Lambda}\Gamma_{ab}\bar{\Lambda})(\bar{\Lambda}\Gamma_{cd}R)(\Lambda\Gamma^{ci}\Lambda)(\Lambda\Gamma^{abd}D) \notag \\
& + \frac{4}{\eta^{2}}(\bar{\Lambda}\Gamma_{bc}\bar{\Lambda})(\bar{\Lambda}R)(\Lambda\Gamma^{bi}\Lambda)(\Lambda\Gamma^{c}D) + \frac{4}{\eta^{2}}(\bar{\Lambda}\Gamma_{ab}\bar{\Lambda})(\bar{\Lambda}\Gamma_{cd}R)(\Lambda\Gamma^{bi}\Lambda)(\Lambda\Gamma^{acd}D) \notag \\
& - \frac{4}{\eta^{2}}(\bar{\Lambda}\Gamma_{ab}\bar{\Lambda})(\bar{\Lambda}\Gamma_{c}^{\hspace{2mm}i}R)(\Lambda\Gamma^{bc}\Lambda)(\Lambda\Gamma^{a}D) - \frac{4}{\eta}(\bar{\Lambda}R)(\Lambda\Gamma^{i}D) - \frac{4}{\eta^{2}}(\bar{\Lambda}\Gamma_{ab}\bar{\Lambda})(\bar{\Lambda}\Gamma_{cd}R)(\Lambda\Gamma^{bc}\Lambda)(\Lambda\Gamma^{aid}D) \notag \\
& - \frac{2}{\eta}(\bar{\Lambda}\Gamma^{ci}R)(\Lambda\Gamma_{c}D) - \frac{2}{\eta}(\bar{\Lambda}\Gamma_{cd}R)(\Lambda\Gamma^{cid}D) 
\end{align}
Furthermore, if we use \eqref{app4} this result can be cast as
\begin{align}
I^{*\,i} & = -\frac{8}{\eta^{2}}(\bar{\Lambda}\Gamma_{ac}\bar{\Lambda})(\bar{\Lambda}R)(\Lambda\Gamma^{ci}\Lambda)(\Lambda\Gamma^{a}D) - \frac{2}{\eta^{2}}(\bar{\Lambda}\Gamma_{ab}\bar{\Lambda})(\bar{\Lambda}\Gamma_{cd}R)(\Lambda\Gamma^{ci}\Lambda)(\Lambda\Gamma^{abd}D) \notag \\
& + \frac{4}{\eta^{2}}(\bar{\Lambda}\Gamma_{ab}\bar{\Lambda})(\bar{\Lambda}\Gamma_{cd}R)(\Lambda\Gamma^{bi}\Lambda)(\Lambda\Gamma^{acd}D) - \frac{4}{\eta^{2}}(\bar{\Lambda}\Gamma_{ab}\bar{\Lambda})(\bar{\Lambda}\Gamma_{c}^{\hspace{2mm}i}R)(\Lambda\Gamma^{bc}\Lambda)(\Lambda\Gamma^{a}D) \notag \\
& - \frac{4}{\eta}(\bar{\Lambda}R)(\Lambda\Gamma^{i}D) + \frac{1}{\eta}(\bar{\Lambda}\Gamma_{ad}R)(\Lambda\Gamma^{aid}D) + \frac{1}{\eta^{2}}(\bar{\Lambda}\Gamma_{ad}\bar{\Lambda})(\bar{\Lambda}\Gamma_{bc}R)(\Lambda\Gamma^{bc}\Lambda)(\Lambda\Gamma^{aid}\Lambda) \notag \\
& - \frac{2}{\eta}(\bar{\Lambda}\Gamma^{ci}R)(\Lambda\Gamma_{c}D) - \frac{2}{\eta}(\bar{\Lambda}\Gamma_{cd}R)(\Lambda\Gamma^{cid}D) 
\end{align}
Plugging this result into \eqref{eq23}, we find
\begin{align}
I^{i} & = \frac{4}{\eta^{2}}(\bar{\Lambda}\Gamma_{ab}\bar{\Lambda})(\bar{\Lambda}\Gamma_{c}^{\hspace{2mm}i}R)(\Lambda\Gamma^{bc}\Lambda)(\Lambda\Gamma^{a}D) + \frac{2}{\eta^{2}}(\bar{\Lambda}\Gamma_{ab}\bar{\Lambda})(\bar{\Lambda}\Gamma_{cd}R)(\Lambda\Gamma^{ci}\Lambda)(\Lambda\Gamma^{abd}D) \notag \\
& + \frac{1}{\eta^{2}}(\bar{\Lambda}\Gamma_{ad}\bar{\Lambda})(\bar{\Lambda}\Gamma_{bc}R)(\Lambda\Gamma^{bc}\Lambda)(\Lambda\Gamma^{aid}D) \label{eq25}
\end{align}
After applying the transformation \eqref{eq400} and using the identities \eqref{app2}, \eqref{app3}, \eqref{app4} one can show that this expression is invariant as mentioned above.
\vspace{2mm}

Therefore we can replace the inverse relation \eqref{app50} in \eqref{eq26}. Let us do this for each term in \eqref{eq25}:
\begin{align}
I_{1}^{i} & = \frac{8}{\eta^{2}}(\bar{\Lambda}\Gamma_{ab}\bar{\Lambda})(\bar{\Lambda}\Gamma_{c}^{\hspace{2mm}i}R)(\Lambda\Gamma^{bc}\Lambda)[(\Lambda\Gamma^{am}\Lambda)\bar{\Sigma}_{0\,m} + \frac{1}{\eta}(\Lambda\Gamma^{am}\Lambda)(\bar{\Lambda}\Gamma_{mn}\bar{\Lambda})(\Lambda\Gamma^{nk}\Lambda)\bar{\Sigma}_{0\,k}]\notag \\
& = \frac{8}{\eta^{2}}(\bar{\Lambda}\Gamma_{ab}\bar{\Lambda})(\bar{\Lambda}\Gamma_{c}^{\hspace{2mm}i}R)(\Lambda\Gamma^{bc}\Lambda)[(\Lambda\Gamma^{am}\Lambda)\bar{\Sigma}_{0\,m} - \frac{1}{2}(\Lambda\Gamma^{ak}\Lambda)\bar{\Sigma}_{0\,k}] \notag \\
& = \frac{4}{\eta^{2}}(\bar{\Lambda}\Gamma_{ab}\bar{\Lambda})(\bar{\Lambda}\Gamma_{c}^{\hspace{2mm}i}R)(\Lambda\Gamma^{bc}\Lambda)(\Lambda\Gamma^{am}\Lambda)\bar{\Sigma}_{0\,m} \notag\\
& = -\frac{2}{\eta}(\bar{\Lambda}\Gamma^{ci}R)(\Lambda\Gamma_{ck}\Lambda)\bar{\Sigma}_{0}^{k}
\end{align}

\begin{align}
I_{2}^{i} & = \frac{4}{\eta^{2}}(\bar{\Lambda}\Gamma_{ab}\bar{\Lambda})(\bar{\Lambda}\Gamma_{cd}R)(\Lambda\Gamma^{ci}\Lambda)[(\Lambda\Gamma^{abd}\Gamma^{m}\Lambda)\bar{\Sigma}_{0\,m} + \frac{1}{\eta}(\Lambda\Gamma^{abd}\Gamma^{m}\Lambda)(\bar{\Lambda}\Gamma_{mn}\bar{\Lambda})(\Lambda\Gamma^{nk}\Lambda)\bar{\Sigma}_{0\,k}] \notag \\
&= \frac{4}{\eta^{2}}(\bar{\Lambda}\Gamma_{ab}\bar{\Lambda})(\bar{\Lambda}\Gamma_{cd}R)(\Lambda\Gamma^{ci}\Lambda)[(\Lambda\Gamma^{ab}\Lambda)\bar{\Sigma}_{0}^{d} + \frac{1}{\eta}(\Lambda\Gamma^{ab}\Lambda)(\bar{\Lambda}\Gamma^{dn}\bar{\Lambda})(\Lambda\Gamma_{nk}\Lambda)\bar{\Sigma}_{0}^{k}] \notag \\
& = \frac{4}{\eta}(\bar{\Lambda}\Gamma_{cd}R)(\Lambda\Gamma^{ci}\Lambda)\bar{\Sigma}_{0}^{d} + \frac{4}{\eta^{2}}(\bar{\Lambda}\Gamma_{cd}R)(\Lambda\Gamma^{ci}\Lambda)(\bar{\Lambda}\Gamma^{dn}\bar{\Lambda})(\Lambda\Gamma_{nk}\Lambda)\bar{\Sigma}_{0}^{k}\notag \\
& = \frac{4}{\eta}(\bar{\Lambda}\Gamma_{cd}R)(\Lambda\Gamma^{ci}\Lambda)\bar{\Sigma}_{0}^{d} + \frac{2}{\eta}(\bar{\Lambda}R)(\Lambda\Gamma^{ik}\Lambda)\bar{\Sigma}_{0\,k}
\end{align}

\begin{align}
I_{3}^{i} &= -\frac{1}{\eta^{2}}(\bar{\Lambda}\Gamma_{ab}\bar{\Lambda})(\bar{\Lambda}\Gamma_{cd}R)(\Lambda\Gamma^{cd}\Lambda)(\Lambda\Gamma^{abi}D) \notag \\
&= -\frac{2}{\eta^{2}}(\bar{\Lambda}\Gamma_{ab}\bar{\Lambda})(\bar{\Lambda}\Gamma_{cd}R)(\Lambda\Gamma^{cd}\Lambda)[(\Lambda\Gamma^{abi}\Gamma^{m}\Lambda)\bar{\Sigma}_{0\,m} + \frac{1}{\eta}(\Lambda\Gamma^{abi}\Gamma^{m}\Lambda)(\bar{\Lambda}\Gamma^{mn}\bar{\Lambda})(\Lambda\Gamma_{nk}\Lambda)\bar{\Sigma}_{0}^{k}]\notag \\
&= -\frac{2}{\eta}(\bar{\Lambda}\Gamma^{cd}R)(\Lambda\Gamma_{cd}\Lambda)\bar{\Sigma}_{0}^{i} - \frac{2}{\eta^{2}}(\bar{\Lambda}\Gamma_{cd}R)(\Lambda\Gamma^{cd}\Lambda)(\bar{\Lambda}\Gamma^{in}\bar{\Lambda})(\Lambda\Gamma^{nk}\Lambda)\bar{\Sigma}_{0\,k}
\end{align}
Replacing these expressions in \eqref{eq25} and putting all together in \eqref{eq26} we obtain
\begin{align}
\{Q, \bar{\Sigma}^{i}\} & = -P^{i} - \frac{2}{\eta}[(\bar{\Lambda}\Gamma^{mb}\bar{\Lambda})(\Lambda\Gamma_{b}^{\hspace{1mm}i}\Lambda) - (\bar{\Lambda}\Gamma^{ib}\bar{\Lambda})(\Lambda\Gamma_{b}^{\hspace{1mm}m}\Lambda)]P_{m} + \frac{4}{\eta}(\bar{\Lambda}\Gamma_{mn}R)(\Lambda\Gamma^{mn}\Lambda)\bar{\Sigma}^{i} \notag \\
&- \frac{2}{\eta}(\bar{\Lambda}\Gamma_{mn}R)(\Lambda\Gamma^{mn}\Lambda)\bar{\Sigma}^{i}_{0} - \frac{2}{\eta}(\bar{\Lambda}\Gamma^{ci}R)(\Lambda\Gamma_{ck}\Lambda)\bar{\Sigma}_{0}^{k} + \frac{4}{\eta}(\bar{\Lambda}\Gamma_{cd}R)(\Lambda\Gamma^{ci}\Lambda)\bar{\Sigma}_{0}^{d} \notag \\
& + \frac{2}{\eta}(\bar{\Lambda}R)(\Lambda\Gamma^{ik}\Lambda)\bar{\Sigma}_{0\,k} -\frac{2}{\eta}(\bar{\Lambda}\Gamma^{cd}R)(\Lambda\Gamma_{cd}\Lambda)\bar{\Sigma}_{0}^{i} - \frac{2}{\eta^{2}}(\bar{\Lambda}\Gamma_{cd}R)(\Lambda\Gamma^{cd}\Lambda)(\bar{\Lambda}\Gamma^{in}\bar{\Lambda})(\Lambda\Gamma^{nk}\Lambda)\bar{\Sigma}_{0\,k} \notag \\
& - \frac{4}{\eta^{2}}(\bar{\Lambda}\Gamma_{ab}R)(\bar{\Lambda}\Gamma_{cd}R)(\Lambda\Gamma^{abcki}\Lambda)N^{d}_{\hspace{1mm}k} - \frac{4}{3\eta^{2}}(\bar{\Lambda}\Gamma_{ab}R)(\bar{\Lambda}\Gamma_{c}^{\hspace{2mm}i}R)(\Lambda\Gamma^{abcdk}\Lambda)N_{dk} \notag \\
& - \frac{4}{3\eta^{2}}(\bar{\Lambda}\Gamma_{ab}R)(\bar{\Lambda}R)(\Lambda\Gamma^{iabdk}\Lambda)N_{dk} - \frac{2}{3\eta^{2}}(\bar{\Lambda}\Gamma_{ab}\bar{\Lambda})(RR)(\Lambda\Gamma^{iabdk}\Lambda)N_{dk}
\end{align}

\section{Cancellation of all of the $N_{ab}$ contributions in the equation \eqref{eq401}}\label{apD2}

We will show this cancellation in two steps. First we will simplify the expression depending explicitly on $\bar{\Sigma}^{i}_{0}$ and then simplify the expression depending explicitly on $N_{ab}$. Finally we will see that these two expressions identically cancel out. We start with the following equation
\begin{eqnarray}
J^{i} &=& \frac{4}{\eta}(\bar{\Lambda}\Gamma^{mn}R)(\Lambda\Gamma_{mn}\Lambda)(\bar{\Sigma}^{i}-\bar{\Sigma}_{0}^{i}) - \frac{2}{\eta}(\bar{\Lambda}\Gamma^{ci}R)(\Lambda\Gamma_{ck}\Lambda)\bar{\Sigma}_{0}^{k} + \frac{4}{\eta}(\bar{\Lambda}\Gamma_{cd}R)(\Lambda\Gamma^{ci}\Lambda)\bar{\Sigma}_{0}^{d} \nonumber \\
&& + \frac{2}{\eta}(\bar{\Lambda}R)(\Lambda\Gamma^{ik}\Lambda)\bar{\Sigma}_{0\,k} - \frac{2}{\eta^{2}}(\bar{\Lambda}\Gamma^{cd}R)(\Lambda\Gamma_{cd}\Lambda)(\bar{\Lambda}\Gamma^{in}\bar{\Lambda})(\Lambda\Gamma_{nk}\Lambda)\bar{\Sigma}_{0}^{k} \label{eeq35}
\end{eqnarray}
Now let us focus on the contributions proportional to $\bar{\Lambda}R$:
\begin{eqnarray}
J_{1}^{i} &=& \frac{2}{\eta}(\bar{\Lambda}R)(\Lambda\Gamma^{ik}\Lambda)[-\frac{2}{3\eta^{2}}(\bar{\Lambda}\Gamma_{ab}\bar{\Lambda})(\bar{\Lambda}\Gamma_{ck}R)(\Lambda\Gamma^{abcqj}\Lambda)N_{qj}] \nonumber \\
&& + \frac{4}{\eta}(\bar{\Lambda}\Gamma^{mn}R)(\Lambda\Gamma_{mn}\Lambda)[\frac{2}{3\eta^{2}}(\bar{\Lambda}\Gamma_{ab}\bar{\Lambda})(\bar{\Lambda}R)(\Lambda\Gamma^{abiqj}\Lambda)N_{qj}] \nonumber \\
&& + \frac{4}{\eta}(\bar{\Lambda}\Gamma_{cd}R)(\Lambda\Gamma^{ci}\Lambda)[-\frac{4}{3\eta^{2}}(\bar{\Lambda}\Gamma_{ab}\bar{\Lambda})(\bar{\Lambda}R)(\Lambda\Gamma^{abdqj}\Lambda)N_{qj}] \nonumber \\
&=& -\frac{4}{3\eta^{3}}(\Lambda\Gamma^{ik}\Lambda)(\bar{\Lambda}R)(\bar{\Lambda}\Gamma_{ab}\bar{\Lambda})(\bar{\Lambda}\Gamma_{ck}R)(\Lambda\Gamma^{abcqj}\Lambda)N_{qj} \nonumber \\
&& + \frac{8}{3\eta^{3}}(\bar{\Lambda}\Gamma^{mn}R)(\Lambda\Gamma_{mn}\Lambda)(\bar{\Lambda}\Gamma_{ab}\bar{\Lambda})(\bar{\Lambda}R)(\Lambda\Gamma^{abiqj}\Lambda)N_{qj} \nonumber \\
&& -\frac{16}{3\eta^{3}}(\bar{\Lambda}\Gamma_{cd}R)(\Lambda\Gamma^{ci}\Lambda)(\bar{\Lambda}\Gamma_{ab}\bar{\Lambda})(\bar{\Lambda}R)(\Lambda\Gamma^{abdqj}\Lambda)N_{qj}
\end{eqnarray}
The last term can be written as
\begin{equation}
-\frac{16}{3\eta^{3}}(\bar{\Lambda}\Gamma_{cd}R)(\Lambda\Gamma^{ci}\Lambda)(\bar{\Lambda}\Gamma_{ab}\bar{\Lambda})(\bar{\Lambda}R)(\Lambda\Gamma^{abdqj}\Lambda)N_{qj} = -\frac{16}{3\eta^{3}}(\bar{\Lambda}\Gamma_{kc}R)(\Lambda\Gamma^{ki}\Lambda)(\bar{\Lambda}\Gamma_{ab}\bar{\Lambda})(\bar{\Lambda}R)(\Lambda\Gamma^{abcqj}\Lambda)N_{qj}
\end{equation}
Therefore,
\begin{eqnarray}
J_{1}^{i}&=& \frac{12}{3\eta^{3}}(\Lambda\Gamma^{ik}\Lambda)(\bar{\Lambda}R)(\bar{\Lambda}\Gamma_{ab}\bar{\Lambda})(\bar{\Lambda}\Gamma_{ck}R)(\Lambda\Gamma^{abcqj}\Lambda)N_{qj} \nonumber \\
&& + \frac{8}{3\eta^{3}}(\bar{\Lambda}\Gamma^{mn}R)(\Lambda\Gamma_{mn}\Lambda)(\bar{\Lambda}\Gamma_{ab}\bar{\Lambda})(\bar{\Lambda}R)(\Lambda\Gamma^{abiqj}\Lambda)N_{qj} \nonumber\\
&=& \frac{12}{3\eta^{3}}(\Lambda\Gamma^{ik}\Lambda)(\bar{\Lambda}R)(\bar{\Lambda}\Gamma_{ab}\bar{\Lambda})(\bar{\Lambda}\Gamma_{ck}R)(\frac{-18}{6})[4(\Lambda\Gamma^{bc}\Lambda)(\Lambda\Gamma^{a}W) + 2(\Lambda\Gamma^{ab}\Lambda)(\Lambda\Gamma^{c}W)]\nonumber \\
&& + \frac{8}{3\eta^{3}}(\bar{\Lambda}\Gamma^{mn}R)(\Lambda\Gamma_{mn}\Lambda)(\bar{\Lambda}\Gamma_{ab}\bar{\Lambda})(\bar{\Lambda}R)(\frac{-18}{6})[4(\Lambda\Gamma^{bi}\Lambda)(\Lambda\Gamma^{a}W) + 2(\Lambda\Gamma^{ab}\Lambda)(\Lambda\Gamma^{i}W)]\nonumber \\
&=& -\frac{48}{\eta^{3}}(\Lambda\Gamma^{ik}\Lambda)(\bar{\Lambda}R)(\bar{\Lambda}\Gamma_{ab}\bar{\Lambda})(\bar{\Lambda}\Gamma_{ck}R)(\Lambda\Gamma^{bc}\Lambda)(\Lambda\Gamma^{a}W) \nonumber \\
&& - \frac{24}{\eta^{2}}(\Lambda\Gamma^{ik}\Lambda)(\bar{\Lambda}R)(\bar{\Lambda}\Gamma_{ck}R)(\Lambda\Gamma^{c}W) \nonumber \\
&& -\frac{32}{\eta^{3}}(\bar{\Lambda}\Gamma^{mn}R)(\Lambda\Gamma_{mn}\Lambda)(\bar{\Lambda}\Gamma_{ab}\bar{\Lambda})(\bar{\Lambda}R)(\Lambda\Gamma^{bi}\Lambda)(\Lambda\Gamma^{a}W) \nonumber \\
&& - \frac{16}{\eta^{2}}(\bar{\Lambda}\Gamma^{mn}R)(\Lambda\Gamma_{mn}\Lambda)(\bar{\Lambda}R)(\Lambda\Gamma^{i}W)\nonumber \\
&=& -\frac{24}{\eta^{3}}(\Lambda\Gamma^{bi}\Lambda)(\bar{\Lambda}R)(\bar{\Lambda}\Gamma_{ab}\bar{\Lambda})(\bar{\Lambda}\Gamma_{ck}R)(\Lambda\Gamma^{ck}\Lambda)(\Lambda\Gamma^{a}W) \nonumber \\
&& - \frac{24}{\eta^{2}}(\Lambda\Gamma^{ik}\Lambda)(\bar{\Lambda}R)(\bar{\Lambda}\Gamma_{ck}R)(\Lambda\Gamma^{c}W) \nonumber \\
&& -\frac{32}{\eta^{3}}(\bar{\Lambda}\Gamma^{mn}R)(\Lambda\Gamma_{mn}\Lambda)(\bar{\Lambda}\Gamma_{ab}\bar{\Lambda})(\bar{\Lambda}R)(\Lambda\Gamma^{bi}\Lambda)(\Lambda\Gamma^{a}W) \nonumber \\
&& - \frac{16}{\eta^{2}}(\bar{\Lambda}\Gamma^{mn}R)(\Lambda\Gamma_{mn}\Lambda)(\bar{\Lambda}R)(\Lambda\Gamma^{i}W)
\end{eqnarray}
As a result, we get
\begin{eqnarray}
J_{1}^{i} &=& \frac{8}{\eta^{3}}(\Lambda\Gamma^{bi}\Lambda)(\bar{\Lambda}R)(\bar{\Lambda}\Gamma_{ab}\bar{\Lambda})(\bar{\Lambda}\Gamma_{ck}R)(\Lambda\Gamma^{ck}\Lambda)(\Lambda\Gamma^{a}W) \nonumber \\
&& - \frac{24}{\eta^{2}}(\Lambda\Gamma^{ik}\Lambda)(\bar{\Lambda}R)(\bar{\Lambda}\Gamma_{ck}R)(\Lambda\Gamma^{c}W) \nonumber \\
&& - \frac{16}{\eta^{2}}(\bar{\Lambda}\Gamma^{mn}R)(\Lambda\Gamma_{mn}\Lambda)(\bar{\Lambda}R)(\Lambda\Gamma^{i}W)
\end{eqnarray}
Now let us focus on the term proportional to $(\bar{\Lambda}\Gamma^{ci}R)$:
\begin{eqnarray}
J_{2}^{i} &=& \frac{4}{\eta}(\bar{\Lambda}\Gamma^{mn}R)(\Lambda\Gamma_{mn}\Lambda)[\frac{2}{3\eta^{2}}(\bar{\Lambda}\Gamma_{ab}\bar{\Lambda})(\bar{\Lambda}\Gamma_{c}^{\hspace{2mm}i}R)(\Lambda\Gamma^{abcqj}\Lambda)N_{qj}] \nonumber \\
&& -\frac{2}{\eta}(\bar{\Lambda}\Gamma^{ci}R)(\Lambda\Gamma_{ck}\Lambda)[-\frac{2}{3\eta^{2}}(\bar{\Lambda}\Gamma_{ab}\bar{\Lambda})(\bar{\Lambda}\Gamma_{f}^{\hspace{2mm}k}R)(\Lambda\Gamma^{abfqj}\Lambda)N_{qj}] \nonumber \\
&& -\frac{2}{\eta^{2}}(\bar{\Lambda}\Gamma^{cd}R)(\Lambda\Gamma_{cd}\Lambda)(\bar{\Lambda}\Gamma^{in}\bar{\Lambda})(\Lambda\Gamma_{nk}\Lambda)[-\frac{2}{3\eta^{2}}(\bar{\Lambda}\Gamma_{ab}\bar{\Lambda})(\bar{\Lambda}\Gamma_{f}^{\hspace{2mm}k}R)(\Lambda\Gamma^{abfqj}\Lambda)N_{qj}] \nonumber 
\end{eqnarray}

\begin{eqnarray}
&=& \frac{8}{3\eta^{3}}(\bar{\Lambda}\Gamma^{mn}R)(\Lambda\Gamma_{mn}\Lambda)\bar{\Lambda}\Gamma_{ab}\bar{\Lambda})(\bar{\Lambda}\Gamma_{c}^{\hspace{2mm}i}R)(\Lambda\Gamma^{abcqj}\Lambda)N_{qj} \nonumber \\
&& + \frac{4}{3\eta^{3}}(\bar{\Lambda}\Gamma^{ci}R)(\Lambda\Gamma_{ck}\Lambda)(\bar{\Lambda}\Gamma_{ab}\bar{\Lambda})(\bar{\Lambda}\Gamma_{f}^{\hspace{2mm}k}R)(\Lambda\Gamma^{abfqj}\Lambda)N_{qj} \nonumber \\
&& + \frac{4}{3\eta^{4}}(\bar{\Lambda}\Gamma^{cd}R)(\Lambda\Gamma_{cd}\Lambda)(\bar{\Lambda}\Gamma^{in}\bar{\Lambda})(\Lambda\Gamma_{nk}\Lambda)(\bar{\Lambda}\Gamma_{ab}\bar{\Lambda})(\bar{\Lambda}\Gamma_{f}^{\hspace{2mm}k}R)(\Lambda\Gamma^{abfqj}\Lambda)N_{qj} \nonumber \\
&=& \frac{8}{3\eta^{3}}(\bar{\Lambda}\Gamma^{mn}R)(\Lambda\Gamma_{mn}\Lambda)\bar{\Lambda}\Gamma_{ab}\bar{\Lambda})(\bar{\Lambda}\Gamma_{c}^{\hspace{2mm}i}R)(\Lambda\Gamma^{abcqj}\Lambda)N_{qj} \nonumber \\
&& + \frac{4}{3\eta^{3}}(\bar{\Lambda}\Gamma^{ci}R)(\Lambda\Gamma_{ck}\Lambda)(\bar{\Lambda}\Gamma_{ab}\bar{\Lambda})(\bar{\Lambda}\Gamma_{f}^{\hspace{2mm}k}R)(\Lambda\Gamma^{abfqj}\Lambda)N_{qj} \nonumber \\
&& + \frac{2}{3\eta^{4}}(\bar{\Lambda}\Gamma^{cd}R)(\Lambda\Gamma_{cd}\Lambda)(\bar{\Lambda}\Gamma_{ab}\bar{\Lambda})(\bar{\Lambda}\Gamma^{i}_{\hspace{1mm}f}R)(\Lambda\Gamma^{abfqj}\Lambda)N_{qj} \nonumber \\
&=& \frac{6}{3\eta^{3}}(\bar{\Lambda}\Gamma^{mn}R)(\Lambda\Gamma_{mn}\Lambda)\bar{\Lambda}\Gamma_{ab}\bar{\Lambda})(\bar{\Lambda}\Gamma_{c}^{\hspace{2mm}i}R)(\Lambda\Gamma^{abcqj}\Lambda)N_{qj} \nonumber \\
&& + \frac{4}{3\eta^{3}}(\bar{\Lambda}\Gamma^{ci}R)(\Lambda\Gamma_{ck}\Lambda)(\bar{\Lambda}\Gamma_{ab}\bar{\Lambda})(\bar{\Lambda}\Gamma_{f}^{\hspace{2mm}k}R)(\Lambda\Gamma^{abfqj}\Lambda)N_{qj}
\end{eqnarray}
Now we use the identity \eqref{app24}:
\begin{eqnarray}
J_{2}^{i} &=& \frac{6}{3\eta^{3}}(\bar{\Lambda}\Gamma^{mn}R)(\Lambda\Gamma_{mn}\Lambda)\bar{\Lambda}\Gamma_{ab}\bar{\Lambda})(\bar{\Lambda}\Gamma_{c}^{\hspace{2mm}i}R)(-\frac{18}{6})[4(\Lambda\Gamma^{bc}\Lambda)(\Lambda\Gamma^{a}W) + 2(\Lambda\Gamma^{ab}\Lambda)(\Lambda\Gamma^{c}W)]\nonumber \\
&& +  \frac{4}{3\eta^{3}}(\bar{\Lambda}\Gamma^{ci}R)(\Lambda\Gamma_{ck}\Lambda)(\bar{\Lambda}\Gamma_{ab}\bar{\Lambda})(\bar{\Lambda}\Gamma_{f}^{\hspace{2mm}k}R)(-\frac{18}{6})[4(\Lambda\Gamma^{bf}\Lambda)(\Lambda\Gamma^{a}W) + 2(\Lambda\Gamma^{ab}\Lambda)(\Lambda\Gamma^{f}W)]\nonumber \\
&=& -\frac{24}{\eta^{3}}(\bar{\Lambda}\Gamma^{mn}R)(\Lambda\Gamma_{mn}\Lambda)(\bar{\Lambda}\Gamma_{ab}\bar{\Lambda})(\bar{\Lambda}\Gamma_{c}^{\hspace{2mm}i}R)(\Lambda\Gamma^{bc}\Lambda)(\Lambda\Gamma^{a}W) - \frac{12}{\eta^{2}}(\bar{\Lambda}\Gamma^{mn}R)(\Lambda\Gamma_{mn}\Lambda)(\bar{\Lambda}\Gamma_{c}^{\hspace{2mm}i}R)(\Lambda\Gamma^{c}W)\nonumber \\
&& -\frac{16}{\eta^{3}}(\bar{\Lambda}\Gamma^{ci}R)(\Lambda\Gamma_{ck}\Lambda)(\bar{\Lambda}\Gamma_{ab}\bar{\Lambda})(\bar{\Lambda}\Gamma_{f}^{\hspace{2mm}k}R)(\Lambda\Gamma^{bf}\Lambda)(\Lambda\Gamma^{a}W) - \frac{8}{\eta^{2}}(\bar{\Lambda}\Gamma^{ci}R)(\Lambda\Gamma_{ck}\Lambda)(\bar{\Lambda}\Gamma_{f}^{\hspace{2mm}k}R)(\Lambda\Gamma^{f}W)\nonumber \\
&=& -\frac{24}{\eta^{3}}(\bar{\Lambda}\Gamma^{mn}R)(\Lambda\Gamma_{mn}\Lambda)(\bar{\Lambda}\Gamma_{ab}\bar{\Lambda})(\bar{\Lambda}\Gamma_{c}^{\hspace{2mm}i}R)(\Lambda\Gamma^{bc}\Lambda)(\Lambda\Gamma^{a}W) - \frac{12}{\eta^{2}}(\bar{\Lambda}\Gamma^{mn}R)(\Lambda\Gamma_{mn}\Lambda)(\bar{\Lambda}\Gamma_{c}^{\hspace{2mm}i}R)(\Lambda\Gamma^{c}W)\nonumber \\
&& -\frac{8}{\eta^{3}}(\bar{\Lambda}\Gamma^{ci}R)(\Lambda\Gamma^{b}_{\hspace{2mm}c}\Lambda)(\bar{\Lambda}\Gamma_{ab}\bar{\Lambda})(\bar{\Lambda}\Gamma_{fk}R)(\Lambda\Gamma^{fk}\Lambda)(\Lambda\Gamma^{a}W) - \frac{8}{\eta^{2}}(\bar{\Lambda}\Gamma^{ci}R)(\Lambda\Gamma_{ck}\Lambda)(\bar{\Lambda}\Gamma_{f}^{\hspace{2mm}k}R)(\Lambda\Gamma^{f}W)\nonumber
\end{eqnarray}
Therefore,
\begin{eqnarray}
J_{2}^{i} &=& -\frac{16}{\eta^{3}}(\bar{\Lambda}\Gamma^{mn}R)(\Lambda\Gamma_{mn}\Lambda)(\bar{\Lambda}\Gamma_{ab}\bar{\Lambda})(\bar{\Lambda}\Gamma_{c}^{\hspace{2mm}i}R)(\Lambda\Gamma^{bc}\Lambda)(\Lambda\Gamma^{a}W) - \frac{12}{\eta^{2}}(\bar{\Lambda}\Gamma^{mn}R)(\Lambda\Gamma_{mn}\Lambda)(\bar{\Lambda}\Gamma_{c}^{\hspace{2mm}i}R)(\Lambda\Gamma^{c}W)\nonumber \\
&& - \frac{8}{\eta^{2}}(\bar{\Lambda}\Gamma^{ci}R)(\Lambda\Gamma_{ck}\Lambda)(\bar{\Lambda}\Gamma_{f}^{\hspace{2mm}k}R)(\Lambda\Gamma^{f}W) 
\end{eqnarray}
Now let us simplify the remaining terms in \eqref{eeq35}:
\begin{eqnarray}
J_{3}^{i} &=& \frac{4}{\eta}(\bar{\Lambda}\Gamma^{mn}R)(\Lambda\Gamma_{mn}\Lambda)[\frac{2}{\eta^{2}}(\bar{\Lambda}\Gamma_{ab}\bar{\Lambda})(\bar{\Lambda}\Gamma_{cd}R)(\Lambda\Gamma^{abcki}\Lambda)N^{d}_{\hspace{2mm}k}]\nonumber \\
&& + \frac{4}{\eta}(\bar{\Lambda}\Gamma_{cd}R)(\Lambda\Gamma^{ci}\Lambda)[-\frac{2}{\eta^{2}}(\bar{\Lambda}\Gamma_{ab}\bar{\Lambda})(\bar{\Lambda}\Gamma_{ef}R)(\Lambda\Gamma^{abekd}\Lambda)N^{f}_{\hspace{2mm}k}]\nonumber \\
&=& \frac{8}{\eta^{3}}(\bar{\Lambda}\Gamma^{mn}R)(\Lambda\Gamma_{mn}\Lambda)(\bar{\Lambda}\Gamma_{ab}\bar{\Lambda})(\bar{\Lambda}\Gamma_{cd}R)(\Lambda\Gamma^{abcki}\Lambda)N^{d}_{\hspace{2mm}k} \nonumber \\
&& - \frac{8}{\eta^{3}}(\bar{\Lambda}\Gamma_{cd}R)(\Lambda\Gamma^{ci}\Lambda)(\bar{\Lambda}\Gamma_{ab}\bar{\Lambda})(\bar{\Lambda}\Gamma_{ef}R)(\Lambda\Gamma^{abekd}\Lambda)N^{f}_{\hspace{2mm}k}
\end{eqnarray}
Now we apply the identity \eqref{app23} to each term:
\begin{eqnarray}
J_{3}^{i\,(1)} &=& \frac{8}{\eta^{3}}(\bar{\Lambda}\Gamma^{mn}R)(\Lambda\Gamma_{mn}\Lambda)(\bar{\Lambda}\Gamma_{ab}\bar{\Lambda})(\bar{\Lambda}\Gamma_{cd}R)(\frac{6}{24})[4(\Lambda\Gamma^{ci}\Lambda)(\Lambda\Gamma^{ab}\Gamma^{d}W) - 4(\Lambda\Gamma^{ca}\Lambda)(\Lambda\Gamma^{ib}\Gamma^{d}W) \nonumber \\
&& + 4(\Lambda\Gamma^{cb}\Lambda)(\Lambda\Gamma^{ia}\Gamma^{d}W) + 4(\Lambda\Gamma^{ab}\Lambda)(\Lambda\Gamma^{ci}\Gamma^{d}W) - 4(\Lambda\Gamma^{ib}\Lambda)(\Lambda\Gamma^{ca}\Gamma^{d}W) + 4(\Lambda\Gamma^{ia}\Lambda)(\Lambda\Gamma^{cb}\Gamma^{d}W)]\nonumber \\
&=&\frac{8}{\eta^{3}}(\bar{\Lambda}\Gamma^{mn}R)(\Lambda\Gamma_{mn}\Lambda)(\bar{\Lambda}\Gamma_{ab}\bar{\Lambda})(\bar{\Lambda}\Gamma_{cd}R)[2\eta^{bd}(\Lambda\Gamma^{ci}\Lambda)(\Lambda\Gamma^{a}W) + (\Lambda\Gamma^{ci}\Lambda)(\Lambda\Gamma^{abd}W)\nonumber \\
&& +2\eta^{di}(\Lambda\Gamma^{ca}\Lambda)(\Lambda\Gamma^{b}W) - 2\eta^{bd}(\Lambda\Gamma^{ca}\Lambda)(\Lambda\Gamma^{i}W) - 2(\Lambda\Gamma^{ca}\Lambda)(\Lambda\Gamma^{bdi}W) + \eta^{di}(\Lambda\Gamma^{ab}\Lambda)(\Lambda\Gamma^{c}W) \nonumber\\
&& - \eta^{cd}(\Lambda\Gamma^{ab}\Lambda)(\Lambda\Gamma^{i}W) - (\Lambda\Gamma^{ab}\Lambda)(\Lambda\Gamma^{cdi}W) + 2\eta^{cd}(\Lambda\Gamma^{ib}\Lambda)(\Lambda\Gamma^{a}W) - 2\eta^{ad}(\Lambda\Gamma^{ib}\Lambda)(\Lambda\Gamma^{c}W) \nonumber\\
&& + 2(\Lambda\Gamma^{ib}\Lambda)(\Lambda\Gamma^{acd}W)]\nonumber \\
&=& \frac{16}{\eta^{3}}(\bar{\Lambda}\Gamma^{mn}R)(\Lambda\Gamma_{mn}\Lambda)(\bar{\Lambda}\Gamma_{ac}\bar{\Lambda})(\bar{\Lambda}R)(\Lambda\Gamma^{ci}\Lambda)(\Lambda\Gamma^{a}W)\nonumber\\
&& + \frac{8}{\eta^{3}}(\bar{\Lambda}\Gamma^{mn}R)(\Lambda\Gamma_{mn}\Lambda)(\bar{\Lambda}\Gamma_{ab}\bar{\Lambda})(\bar{\Lambda}\Gamma_{cd}R)(\Lambda\Gamma^{ci}\Lambda)(\Lambda\Gamma^{abd}W)\nonumber \\
&& +\frac{16}{\eta^{3}}(\bar{\Lambda}\Gamma^{mn}R)(\Lambda\Gamma_{mn}\Lambda)(\bar{\Lambda}\Gamma_{ab}\bar{\Lambda})(\bar{\Lambda}\Gamma_{c}^{\hspace{2mm}i}R)(\Lambda\Gamma^{ca}\Lambda)(\Lambda\Gamma^{b}W)\nonumber\\
&& + \frac{16}{\eta^{2}}(\bar{\Lambda}\Gamma^{mn}R)(\Lambda\Gamma_{mn}\Lambda)(\bar{\Lambda}R)(\Lambda\Gamma^{i}W)\nonumber \\
&& -\frac{16}{\eta^{3}}(\bar{\Lambda}\Gamma^{mn}R)(\Lambda\Gamma_{mn}\Lambda)(\bar{\Lambda}\Gamma_{ab}\bar{\Lambda})(\bar{\Lambda}\Gamma_{cd}R)(\Lambda\Gamma^{ca}\Lambda)(\Lambda\Gamma^{bdi}W)\nonumber \\
&& + \frac{8}{\eta^{2}}(\bar{\Lambda}\Gamma^{mn}R)(\Lambda\Gamma_{mn}\Lambda)(\bar{\Lambda}\Gamma_{c}^{\hspace{2mm}i}R)(\Lambda\Gamma^{c}W)\nonumber\\
&& -\frac{8}{\eta^{2}}(\bar{\Lambda}\Gamma^{mn}R)(\Lambda\Gamma_{mn}\Lambda)(\bar{\Lambda}\Gamma_{cd}R)(\Lambda\Gamma^{cdi}W)\nonumber\\
&& -\frac{16}{\eta^{2}}(\bar{\Lambda}\Gamma^{mn}R)(\Lambda\Gamma_{mn}\Lambda)(\bar{\Lambda}\Gamma_{bc}\bar{\Lambda})(\bar{\Lambda}R)(\Lambda\Gamma^{bi}\Lambda)(\Lambda\Gamma^{c}W)\nonumber \\
&& + \frac{16}{\eta^{3}}(\bar{\Lambda}\Gamma^{mn}R)(\Lambda\Gamma_{mn}\Lambda)(\bar{\Lambda}\Gamma_{ab}\bar{\Lambda})(\bar{\Lambda}\Gamma_{cd}R)(\Lambda\Gamma^{ib}\Lambda)(\Lambda\Gamma^{acd}W)
\end{eqnarray}
\begin{eqnarray}
J_{3}^{i\,(2)} &=& - \frac{8}{\eta^{3}}(\bar{\Lambda}\Gamma_{cd}R)(\Lambda\Gamma^{ci}\Lambda)(\bar{\Lambda}\Gamma_{ab}\bar{\Lambda})(\bar{\Lambda}\Gamma_{ef}R)(\frac{6}{24})[4(\Lambda\Gamma^{ed}\Lambda)(\Lambda\Gamma^{ab}\Gamma^{f}W) - 4(\Lambda\Gamma^{ea}\Lambda)(\Lambda\Gamma^{db}\Gamma^{f}W)\nonumber \\
&& + 4(\Lambda\Gamma^{eb}\Lambda)(\Lambda\Gamma^{da}\Gamma^{f}W) + 4(\Lambda\Gamma^{ab}\Lambda)(\Lambda\Gamma^{ed}\Gamma^{f}W) - 4(\Lambda\Gamma^{db}\Lambda)(\Lambda\Gamma^{ea}\Gamma^{f}W) + 4(\Lambda\Gamma^{da}\Lambda)(\Lambda\Gamma^{eb}\Gamma^{f}W)]\nonumber
\end{eqnarray}
\begin{eqnarray}
&=&  - \frac{8}{\eta^{3}}(\bar{\Lambda}\Gamma_{cd}R)(\Lambda\Gamma^{ci}\Lambda)(\bar{\Lambda}\Gamma_{ab}\bar{\Lambda})(\bar{\Lambda}\Gamma_{ef}R)[2\eta^{bf}(\Lambda\Gamma^{ed}\Lambda)(\Lambda\Gamma^{a}W) + (\Lambda\Gamma^{ed}\Lambda)(\Lambda\Gamma^{abf}W)\nonumber \\
&& +2\eta^{fd}(\Lambda\Gamma^{ea}\Lambda)(\Lambda\Gamma^{b}W) - 2\eta^{bf}(\Lambda\Gamma^{ea}\Lambda)(\Lambda\Gamma^{d}W) - 2(\Lambda\Gamma^{ea}\Lambda)(\Lambda\Gamma^{bfd}W) + \eta^{fd}(\Lambda\Gamma^{ab}\Lambda)(\Lambda\Gamma^{e}W) \nonumber\\
&& - \eta^{ef}(\Lambda\Gamma^{ab}\Lambda)(\Lambda\Gamma^{d}W) - (\Lambda\Gamma^{ab}\Lambda)(\Lambda\Gamma^{efd}W) + 2\eta^{ef}(\Lambda\Gamma^{db}\Lambda)(\Lambda\Gamma^{a}W) - 2\eta^{af}(\Lambda\Gamma^{db}\Lambda)(\Lambda\Gamma^{e}W) \nonumber\\
&& + 2(\Lambda\Gamma^{db}\Lambda)(\Lambda\Gamma^{aef}W)]\nonumber \\
&=& -\frac{8}{\eta^{3}}(\bar{\Lambda}\Gamma_{cd}R)(\Lambda\Gamma^{cd}\Lambda)(\bar{\Lambda}\Gamma_{ae}\bar{\Lambda})(\bar{\Lambda}R)(\Lambda\Gamma^{ei}\Lambda)(\Lambda\Gamma^{a}W)\nonumber \\
&& -\frac{4}{\eta^{3}}(\bar{\Lambda}\Gamma_{cd}R)(\Lambda\Gamma^{cd}\Lambda)(\bar{\Lambda}\Gamma_{ab}\bar{\Lambda})(\bar{\Lambda}\Gamma_{ef}R)(\Lambda\Gamma^{ei}\Lambda)(\Lambda\Gamma^{abf}W)\nonumber\\
&& - \frac{8}{\eta^{3}}(\bar{\Lambda}\Gamma_{ce}R)(\Lambda\Gamma^{ce}\Lambda)(\bar{\Lambda}\Gamma_{ab}\bar{\Lambda})(\bar{\Lambda}R)(\Lambda\Gamma^{ia}\Lambda)(\Lambda\Gamma^{b}W) - \frac{4}{\eta^{2}}(\bar{\Lambda}\Gamma_{ab}\bar{\Lambda})(RR)(\Lambda\Gamma^{ia}\Lambda)(\Lambda\Gamma^{b}W)\nonumber \\
&& +\frac{16}{\eta^{3}}(\bar{\Lambda}\Gamma_{cd}R)(\Lambda\Gamma^{ci}\Lambda)(\bar{\Lambda}\Gamma_{ab}\bar{\Lambda})(\bar{\Lambda}\Gamma_{ef}R)(\Lambda\Gamma^{ea}\Lambda)(\Lambda\Gamma^{bfd}W)\nonumber\\
&& - \frac{16}{\eta^{2}}(\bar{\Lambda}\Gamma_{cd}R)(\Lambda\Gamma^{ci}\Lambda)(\bar{\Lambda}R)(\Lambda\Gamma^{d}W)\nonumber \\
&&-\frac{8}{\eta^{2}}(\bar{\Lambda}\Gamma_{ce}R)(\Lambda\Gamma^{ci}\Lambda)(\bar{\Lambda}R)(\Lambda\Gamma^{e}W) - \frac{4}{\eta^{2}}(\bar{\Lambda}\Gamma_{ce}\bar{\Lambda})(\Lambda\Gamma^{ci}\Lambda)(RR)(\Lambda\Gamma^{e}W)\nonumber\\
&& + \frac{8}{\eta^{2}}(\bar{\Lambda}\Gamma_{cd}R)(\Lambda\Gamma^{ci}\Lambda)(\bar{\Lambda}\Gamma_{ef}R)(\Lambda\Gamma^{efd}W)\nonumber\\
&& + \frac{8}{\eta^{3}}(\bar{\Lambda}\Gamma_{cd}R)(\Lambda\Gamma^{cd}\Lambda)(\bar{\Lambda}\Gamma_{eb}\bar{\Lambda})(\bar{\Lambda}R)(\Lambda\Gamma^{ib}\Lambda)(\Lambda\Gamma^{e}W)\nonumber \\
&& -\frac{8}{\eta^{3}}(\bar{\Lambda}\Gamma_{cd}R)(\Lambda\Gamma^{cd}\Lambda)(\bar{\Lambda}\Gamma_{ab}\bar{\Lambda})(\bar{\Lambda}\Gamma_{ef}R)(\Lambda\Gamma^{ib}\Lambda)(\Lambda\Gamma^{aef}W)
\end{eqnarray}
Therefore $J_{3}^{i}$ takes the form
\begin{eqnarray}
J_{3}^{i} &=& \frac{8}{\eta^{3}}(\Lambda\Gamma^{mn}R)(\Lambda\Gamma_{mn}\Lambda)(\bar{\Lambda}\Gamma_{ac}\bar{\Lambda})(\bar{\Lambda}R)(\Lambda\Gamma^{ci}\Lambda)(\Lambda\Gamma^{a}W)\nonumber\\ && - \frac{4}{\eta^{3}}(\bar{\Lambda}\Gamma_{mn}R)(\Lambda\Gamma^{mn}\Lambda)(\bar{\Lambda}\Gamma_{ab}\bar{\Lambda})(\bar{\Lambda}\Gamma_{cd}R)(\Lambda\Gamma^{ci}\Lambda)(\Lambda\Gamma^{abd}W)\nonumber\\
&& + \frac{16}{\eta^{3}}(\bar{\Lambda}\Gamma^{mn}R)(\Lambda\Gamma_{mn}\Lambda)(\bar{\Lambda}\Gamma_{ab}\bar{\Lambda})(\bar{\Lambda}\Gamma_{c}^{\hspace{2mm}i}R)(\Lambda\Gamma^{ca}\Lambda)(\Lambda\Gamma^{b}W)\nonumber \\
&& + \frac{16}{\eta^{2}}(\bar{\Lambda}\Gamma^{mn}R)(\Lambda\Gamma_{mn}\Lambda)(\bar{\Lambda}R)(\Lambda\Gamma^{i}W)\nonumber \\
&& - \frac{16}{\eta^{3}}(\bar{\Lambda}\Gamma^{mn}R)(\Lambda\Gamma_{mn}\Lambda)(\bar{\Lambda}\Gamma_{ab}\bar{\Lambda})(\bar{\Lambda}\Gamma_{cd}R)(\Lambda\Gamma^{ca}\Lambda)(\Lambda\Gamma^{bdi}W) \nonumber\\
&& + \frac{8}{\eta^{2}}(\bar{\Lambda}\Gamma^{mn}R)(\Lambda\Gamma_{mn}\Lambda)(\bar{\Lambda}\Gamma_{c}^{\hspace{2mm}i}R)(\Lambda\Gamma^{c}W)\nonumber\\
&& - \frac{8}{\eta^{2}}(\bar{\Lambda}\Gamma^{mn}R)(\Lambda\Gamma_{mn}\Lambda)(\bar{\Lambda}\Gamma_{cd}R)(\Lambda\Gamma^{cdi}W)\nonumber\\
&& - \frac{24}{\eta^{2}}(\bar{\Lambda}\Gamma_{cd}R)(\Lambda\Gamma^{ci}\Lambda)(\bar{\Lambda}R)(\Lambda\Gamma^{d}W)\nonumber\\
&& + \frac{16}{\eta^{3}}(\bar{\Lambda}\Gamma_{cd}R)(\Lambda\Gamma^{ci}\Lambda)(\bar{\Lambda}\Gamma_{ab}\bar{\Lambda})(\bar{\Lambda}\Gamma_{ef}R)(\Lambda\Gamma^{ea}\Lambda)(\Lambda\Gamma^{bfd}W)\nonumber\\
&& +\frac{8}{\eta^{2}}(\bar{\Lambda}\Gamma_{cd}R)(\Lambda\Gamma^{ci}\Lambda)(\bar{\Lambda}\Gamma_{ef}R)(\Lambda\Gamma^{efd}W)
\end{eqnarray}
And we also have
\begin{eqnarray}
J_{1}^{i} + J_{2}^{i} &=& -\frac{16}{\eta^{3}}(\bar{\Lambda}\Gamma^{mn}R)(\Lambda\Gamma_{mn}\Lambda)(\bar{\Lambda}\Gamma_{ab}\bar{\Lambda})(\bar{\Lambda}\Gamma_{c}^{\hspace{2mm}i}R)(\Lambda\Gamma^{bc}\Lambda)(\Lambda\Gamma^{a}W)\nonumber\\
&&-\frac{12}{\eta^{2}}(\bar{\Lambda}\Gamma^{mn}R)(\Lambda\Gamma_{mn}\Lambda)(\bar{\Lambda}\Gamma_{c}^{\hspace{2mm}i}R)(\Lambda\Gamma^{c}W)\nonumber\\
&&-\frac{8}{\eta^{2}})(\bar{\Lambda}\Gamma_{c}^{\hspace{2mm}i}R)(\Lambda\Gamma^{ck}\Lambda)(\bar{\Lambda}\Gamma_{fk}R)(\Lambda\Gamma^{f}W)\nonumber\\
&&-\frac{8}{\eta^{3}}(\bar{\Lambda}\Gamma^{mn}R)(\Lambda\Gamma_{mn}\Lambda)(\bar{\Lambda}R)(\bar{\Lambda}\Gamma_{ab}\bar{\Lambda})(\Lambda\Gamma^{bi}\Lambda)(\Lambda\Gamma^{a}W)\nonumber\\
&&+\frac{24}{\eta^{2}}(\bar{\Lambda}\Gamma_{cd}R)(\bar{\Lambda}R)(\Lambda\Gamma_{c}^{\hspace{2mm}i}\Lambda)(\Lambda\Gamma^{c}W)\nonumber \\
&&-\frac{16}{\eta^{2}}(\bar{\Lambda}\Gamma^{mn}R)(\Lambda\Gamma_{mn}\Lambda)(\bar{\Lambda}R)(\Lambda\Gamma^{i}W)
\end{eqnarray}
The sum of these quantities gives us the following result
\begin{eqnarray}
J^{i} &=& -\frac{4}{\eta^{3}}(\bar{\Lambda}\Gamma^{mn}R)(\Lambda\Gamma_{mn}\Lambda)(\bar{\Lambda}\Gamma_{ab}\bar{\Lambda})(\bar{\Lambda}\Gamma_{cd}R)(\Lambda\Gamma^{ci}\Lambda)(\Lambda\Gamma^{abd}W)\nonumber \\
&& + \frac{4}{\eta^{2}}(\bar{\Lambda}\Gamma^{mn}R)(\Lambda\Gamma_{mn}\Lambda)(\bar{\Lambda}\Gamma_{bd}R)(\Lambda\Gamma^{bdi}W)\nonumber \\
&&-\frac{4}{\eta^{2}}(\bar{\Lambda}\Gamma^{mn}R)(\Lambda\Gamma_{mn}\Lambda)(\bar{\Lambda}\Gamma_{c}^{\hspace{2mm}i}R)(\Lambda\Gamma^{c}W)\nonumber\\
&& -\frac{8}{\eta^{2}}(\bar{\Lambda}\Gamma^{mn}R)(\Lambda\Gamma_{mn}\Lambda)(\bar{\Lambda}\Gamma_{cd}R)(\Lambda\Gamma^{cdi}W)\nonumber \\
&& + \frac{16}{\eta^{3}}(\bar{\Lambda}\Gamma_{cd}R)(\Lambda\Gamma^{ci}\Lambda)(\bar{\Lambda}\Gamma_{ab}\bar{\Lambda})(\bar{\Lambda}\Gamma_{ef}R)(\Lambda\Gamma^{ea}\Lambda)(\Lambda\Gamma^{bfd}W)\nonumber\\
&& + \frac{8}{\eta^{2}}(\bar{\Lambda}\Gamma_{cd}R)(\Lambda\Gamma^{ci}\Lambda)(\bar{\Lambda}\Gamma_{ef}R)(\Lambda\Gamma^{efd}W)\nonumber\\
&&-\frac{8}{\eta^{2}}(\bar{\Lambda}\Gamma_{c}^{\hspace{2mm}i}R)(\Lambda\Gamma^{ck}\Lambda)(\bar{\Lambda}\Gamma_{fk}R)(\Lambda\Gamma^{f}W)
\end{eqnarray}
where we have used that $(\bar{\Lambda}\Gamma^{mn}R)(\Lambda\Gamma_{mn}\Lambda)(\bar{\Lambda}\Gamma^{ab}R)(\Lambda\Gamma_{ab}\Lambda) = 0$. After using the identity \eqref{app4} this expression simplifies to
\begin{eqnarray}
J^{i} &=& -\frac{4}{\eta^{2}}(\bar{\Lambda}\Gamma^{mn}R)(\Lambda\Gamma_{mn}\Lambda)(\bar{\Lambda}\Gamma_{bd}R)(\Lambda\Gamma^{bdi}W) + \frac{4}{\eta^{2}}(\bar{\Lambda}\Gamma_{cd}R)(\Lambda\Gamma^{ci}\Lambda)(\bar{\Lambda}\Gamma_{ef}R)(\Lambda\Gamma^{efd}W)\nonumber\\
&& -\frac{4}{\eta^{2}}(\bar{\Lambda}\Gamma^{mn}R)(\Lambda\Gamma_{mn}\Lambda)(\bar{\Lambda}\Gamma_{c}^{\hspace{2mm}i}R)(\Lambda\Gamma^{c}W)- \frac{8}{\eta^{2}}(\bar{\Lambda}\Gamma_{c}^{\hspace{2mm}i}R)(\Lambda\Gamma^{ck}\Lambda)(\bar{\Lambda}\Gamma_{fk}R)(\Lambda\Gamma^{f}W)\nonumber \\
\end{eqnarray}
Now we will simplify the expressions containing $N_{mn}$ explicitly:
\begin{eqnarray}
S^{i} &=& -\frac{4}{\eta^{2}}(\bar{\Lambda}\Gamma_{ab}R)(\bar{\Lambda}\Gamma_{cd}R)(\Lambda\Gamma^{abcki}\Lambda)N^{d}_{\hspace{2mm}k} - \frac{4}{3\eta^{2}}(\bar{\Lambda}\Gamma_{ab}R)(\bar{\Lambda}\Gamma_{c}^{\hspace{2mm}i}R)(\Lambda\Gamma^{abcdk}\Lambda)N_{dk}\nonumber \\
&& - \frac{4}{3\eta^{2}}(\bar{\Lambda}\Gamma_{ab}R)(\bar{\Lambda}R)(\Lambda\Gamma^{iabdk}\Lambda)N_{dk} - \frac{2}{3\eta^{2}}(\bar{\Lambda}\Gamma_{ab}\bar{\Lambda})(RR)(\Lambda\Gamma^{iabdk}\Lambda)N_{dk}
\end{eqnarray}
For convenience let us focus first on the last three terms:
\begin{eqnarray}
S_{2}^{i} &=& -\frac{4}{3\eta^{2}}(\bar{\Lambda}\Gamma_{ab}R)(\bar{\Lambda}\Gamma_{c}^{\hspace{2mm}i}R)(-\frac{18}{6})[4(\Lambda\Gamma^{bc}\Lambda)(\Lambda\Gamma^{a}W) + 2(\Lambda\Gamma^{ab}\Lambda)(\Lambda\Gamma^{c}W)] \nonumber \\
&& - \frac{4}{3\eta^{2}}(\bar{\Lambda}\Gamma_{ab}R)(\bar{\Lambda}R)(-\frac{18}{6})[2(\Lambda\Gamma^{ab}\Lambda)(\Lambda\Gamma^{i}W) + 4(\Lambda\Gamma^{bi}\Lambda)(\Lambda\Gamma^{a}W)]\nonumber\\
&& - \frac{2}{3\eta^{2}}(\bar{\Lambda}\Gamma_{ab}\bar{\Lambda})(RR)(-\frac{18}{6})[2(\Lambda\Gamma^{ab}\Lambda)(\Lambda\Gamma^{i}W) + 4(\Lambda\Gamma^{bi}\Lambda)(\Lambda\Gamma^{a}W)]\nonumber \\
&=& \frac{16}{\eta^{2}}(\bar{\Lambda}\Gamma_{ab}R)(\bar{\Lambda}\Gamma_{c}^{\hspace{2mm}i}R)(\Lambda\Gamma^{bc}\Lambda)(\Lambda\Gamma^{a}W) + \frac{8}{\eta^{2}}(\bar{\Lambda}\Gamma_{ab}R)(\Lambda\Gamma^{ab}\Lambda)(\bar{\Lambda}\Gamma_{c}^{\hspace{2mm}i}R)(\Lambda\Gamma^{c}W)\nonumber\\
&& +\frac{8}{\eta^{2}}(\bar{\Lambda}\Gamma_{ab}R)(\Lambda\Gamma^{ab}\Lambda)(\bar{\Lambda}R)(\Lambda\Gamma^{i}W) + \frac{16}{\eta^{2}}(\bar{\Lambda}\Gamma_{ab}R)(\bar{\Lambda}R)(\Lambda\Gamma^{bi}\Lambda)(\Lambda\Gamma^{a}W)\nonumber\\
&& + \frac{4}{\eta}(RR)(\Lambda\Gamma^{i}W) + \frac{8}{\eta^{2}}(\bar{\Lambda}\Gamma_{ab}\bar{\Lambda})(RR)(\Lambda\Gamma^{bi}\Lambda)(\Lambda\Gamma^{a}W)
\end{eqnarray}
The same manipulations for the first term of $S^{i}$ give us
\begin{eqnarray}
S_{1}^{i} &=& -\frac{4}{\eta^{2}}(\bar{\Lambda}\Gamma_{ab}R)(\bar{\Lambda}\Gamma_{cd}R)(\frac{6}{24})[4(\Lambda\Gamma^{ci}\Lambda)(\Lambda\Gamma^{ab}\Gamma^{d}W) - 4(\Lambda\Gamma^{ca}\Lambda)(\Lambda\Gamma^{ib}\Gamma^{d}W)\nonumber \\
&& + 4(\Lambda\Gamma^{cb}\Lambda)(\Lambda\Gamma^{ia}\Gamma^{d}W) + 4(\Lambda\Gamma^{ab}\Lambda)(\Lambda\Gamma^{ci}\Gamma^{d}W) - 4(\Lambda\Gamma^{ib}\Lambda)(\Lambda\Gamma^{ca}\Gamma^{d}W) + 4(\Lambda\Gamma^{ia}\Lambda)(\Lambda\Gamma^{cb}\Gamma^{d}W)]\nonumber\\
&=& -\frac{4}{\eta^{2}}(\bar{\Lambda}\Gamma_{ab}R)(\bar{\Lambda}\Gamma_{cd}R)[(\Lambda\Gamma^{ci}\Lambda)(\Lambda\Gamma^{ab}\Gamma^{d}W) - 2(\Lambda\Gamma^{ca}\Lambda)(\Lambda\Gamma^{ib}\Gamma^{d}W) + (\Lambda\Gamma^{ab}\Lambda)(\Lambda\Gamma^{ci}\Gamma^{d}W)\nonumber\\
&& + 2(\Lambda\Gamma^{ia}\Lambda)(\Lambda\Gamma^{cb}\Gamma^{d}W)]\nonumber\\
&=& -\frac{4}{\eta^{2}}(\bar{\Lambda}\Gamma_{ab}R)(\bar{\Lambda}\Gamma_{cd}R)[2\eta^{bd}(\Lambda\Gamma^{ci}\Lambda)(\Lambda\Gamma^{a}W) + (\Lambda\Gamma^{ci}\Lambda)(\Lambda\Gamma^{abd}W) + 2\eta^{di}(\Lambda\Gamma^{ca}\Lambda)(\Lambda\Gamma^{b}W)\nonumber\\
&& -2\eta^{bd}(\Lambda\Gamma^{ca}\Lambda)(\Lambda\Gamma^{i}W) - 2(\Lambda\Gamma^{ca}\Lambda)(\Lambda\Gamma^{bdi}W) + \eta^{di}(\Lambda\Gamma^{ab}\Lambda)(\Lambda\Gamma^{c}W) - \eta^{cd}(\Lambda\Gamma^{ab}\Lambda)(\Lambda\Gamma^{i}W) \nonumber\\
&& - (\Lambda\Gamma^{ab}\Lambda)(\Lambda\Gamma^{cdi}W) -\eta^{cd}(\Lambda\Gamma^{ia}\Lambda)(\Lambda\Gamma^{b}W) + 2\eta^{bd}(\Lambda\Gamma^{ia}\Lambda)(\Lambda\Gamma^{c}W) -2(\Lambda\Gamma^{ia}\Lambda)(\Lambda\Gamma^{bcd}W)]\nonumber \\
\end{eqnarray}
Therefore we get
\begin{eqnarray}
S_{1}^{i} &=& -\frac{8}{\eta^{2}}(\bar{\Lambda}\Gamma_{ac}R)(\bar{\Lambda}R)(\Lambda\Gamma^{ci}\Lambda)(\Lambda\Gamma^{a}W) - \frac{4}{\eta^{2}}(\bar{\Lambda}\Gamma_{ac}\bar{\Lambda})(RR)(\Lambda\Gamma^{ci}\Lambda)(\Lambda\Gamma^{a}W)\nonumber \\
&& -\frac{4}{\eta^{2}}(\bar{\Lambda}\Gamma_{ab}R)(\bar{\Lambda}\Gamma_{cd}R)(\Lambda\Gamma^{ci}\Lambda)(\Lambda\Gamma^{abd}W) - \frac{8}{\eta^{2}}(\bar{\Lambda}\Gamma_{ab}\bar{\Lambda})(\bar{\Lambda}\Gamma_{c}^{\hspace{2mm}i}R)(\Lambda\Gamma^{ca}\Lambda)(\Lambda\Gamma^{b}W)\\
&& +\frac{8}{\eta^{2}}(\bar{\Lambda}\Gamma_{ca}R)(\bar{\Lambda}R)(\Lambda\Gamma^{ca}\Lambda)(\Lambda\Gamma^{i}W) - \frac{4}{\eta}(RR)(\Lambda\Gamma^{i}W)\nonumber\\
&& -\frac{4}{\eta^{2}}(\bar{\Lambda}\Gamma^{ab}R)(\Lambda\Gamma_{ab}\Lambda)(\bar{\Lambda}\Gamma_{c}^{\hspace{2mm}i}R)(\Lambda\Gamma^{c}W) + \frac{4}{\eta^{2}}(\bar{\Lambda}\Gamma^{ab}R)(\Lambda\Gamma_{ab}\Lambda)(\bar{\Lambda}\Gamma_{cd}R)(\Lambda\Gamma^{cdi}W)\nonumber\\
&& - \frac{8}{\eta^{2}}(\bar{\Lambda}\Gamma_{ac}R)(\bar{\Lambda}R)(\Lambda\Gamma^{ia}\Lambda)(\Lambda\Gamma^{c}W) - \frac{4}{\eta^{2}}(\bar{\Lambda}\Gamma_{ac}\bar{\Lambda})(RR)(\Lambda\Gamma^{ia}\Lambda)(\Lambda\Gamma^{c}W)\nonumber\\
&& + \frac{8}{\eta^{2}}(\bar{\Lambda}\Gamma_{ab}R)(\bar{\Lambda}\Gamma_{cd}R)(\Lambda\Gamma^{ia}\Lambda)(\Lambda\Gamma^{bcd}W)
\end{eqnarray}
When summing $S_{1}^{1} + S_{2}^{i}$ we obtain
\begin{eqnarray}
S^{i} &=& \frac{4}{\eta^{2}}(\bar{\Lambda}\Gamma^{mn}R)(\Lambda\Gamma_{mn}\Lambda)(\bar{\Lambda}\Gamma_{bd}R)(\Lambda\Gamma^{bdi}W) + \frac{4}{\eta^{2}}(\bar{\Lambda}\Gamma_{cd}R)(\Lambda\Gamma^{ci}\Lambda)(\bar{\Lambda}\Gamma_{ef}R)(\Lambda\Gamma^{efd}W)\nonumber\\
&& + \frac{4}{\eta^{2}}(\bar{\Lambda}\Gamma^{mn}R)(\Lambda\Gamma_{mn}\Lambda)(\bar{\Lambda}\Gamma_{c}^{\hspace{2mm}i}R)(\Lambda\Gamma^{c}W) + \frac{8}{\eta^{2}}(\bar{\Lambda}\Gamma_{c}^{\hspace{2mm}i}R)(\Lambda\Gamma^{ck}\Lambda)(\bar{\Lambda}\Gamma_{fk}R)(\Lambda\Gamma^{f}W)\nonumber \\
\end{eqnarray}
Thus we have a full cancellation $J^{i} + S^{i} = 0$.

\section{Calculation of $\{\bar{\Sigma}^{i}, {\bar{\Sigma}}^{j}\}$}\label{apD3}
The object $\bar{\Sigma}^{i}$ has a part depending on $D_{\alpha}$ and other part depending on $N_{mn}$, as it can be seen in \eqref{eeq22}. The part depending on $N_{mn}$ will be called $\bar{\Sigma}^{i}_{1}$ and as before we use $\bar{\Sigma}^{i}_{0}$ to denote the part depending on $D_{\alpha}$. Therefore
\begin{equation}
\bar{\Sigma}^{i} = \bar{\Sigma}^{i}_{0} + \bar{\Sigma}^{i}_{1}
\end{equation}
It is easy to see that $\{\bar{\Sigma}^{i}_{0}, \bar{\Sigma}^{j}_{0}\} = 0$:
\begin{eqnarray}
\{\bar{\Sigma}^{i}_{0}, \bar{\Sigma}^{j}_{0}\} &=& \{\frac{1}{2\eta}(\bar{\Lambda}\Gamma_{ab}\bar{\Lambda})(\Lambda\Gamma^{ab}\Gamma^{i}D), \frac{1}{2\eta}(\bar{\Lambda}\Gamma_{cd}\bar{\Lambda})(\Lambda\Gamma^{cd}\Gamma^{j}D)\}\nonumber\\
&=& \frac{1}{4\eta^{2}}(\bar{\Lambda}\Gamma_{ab}\bar{\Lambda})(\bar{\Lambda}\Gamma_{cd}\bar{\Lambda})(\Lambda\Gamma^{ab}\Gamma^{i})^{\alpha}(\Lambda\Gamma^{cd}\Gamma^{j})^{\beta}\{D_{\alpha}, D_{\beta}\}\nonumber\\
&=& -\frac{1}{2\eta^{2}}(\bar{\Lambda}\Gamma_{ab}\bar{\Lambda})(\bar{\Lambda}\Gamma_{cd}\bar{\Lambda})(\Lambda\Gamma^{ab}\Gamma^{i}\Gamma^{m}\Gamma^{j}\Gamma^{cd}\Lambda)P_{m}\nonumber\\
&=& 0
\end{eqnarray}
using the identity \eqref{app3}. 

\vspace{2mm}
The next step is to compute the anticommutator $\{\bar{\Sigma}_{0}^{i}, \bar{\Sigma}_{1}^{j}\}$. To this end let us write $\bar{\Sigma}_{1}^{j}$ explicitly:
\begin{eqnarray}
\bar{\Sigma}_{1}^{j} &=& \frac{2}{\eta^{2}}(\bar{\Lambda}\Gamma_{ab}\bar{\Lambda})(\bar{\Lambda}\Gamma_{cd}R)(\Lambda\Gamma^{abckj}\Lambda)N^{d}_{\hspace{2mm}k} + \frac{2}{3\eta^{2}}(\bar{\Lambda}\Gamma_{ab}\bar{\Lambda})(\bar{\Lambda}\Gamma_{c}^{\hspace{2mm}j}R)(\Lambda\Gamma^{abcdk}\Lambda)N_{dk} \nonumber\\
&& + \frac{2}{3\eta^{2}}(\bar{\Lambda}\Gamma_{ab}\bar{\Lambda})(\bar{\Lambda}R)(\Lambda\Gamma^{jabdk}\Lambda)N_{dk}
\end{eqnarray}
and denote each term by $\bar{\Sigma}^{j(1)}_{1}$, $\bar{\Sigma}^{j(2)}_{1}$, $\bar{\Sigma}^{j(3)}_{1}$, respectively. It can be shown that $\{\bar{\Sigma}^{i}_{0}, \bar{\Sigma}_{1}^{j(1)}\} = 0$. Now we rewrite $\bar{\Sigma}_{1}^{j(2)}$, $\bar{\Sigma}_{1}^{j(3)}$ in a more convenient way:
\begin{eqnarray}
\bar{\Sigma}^{j(2)}_{1} &=& -\frac{4}{\eta}(\bar{\Lambda}\Gamma^{cj}R)(\Lambda\Gamma_{c}W) - \frac{8}{\eta^{2}}(\bar{\Lambda}\Gamma_{ab}\bar{\Lambda})(\bar{\Lambda}\Gamma_{c}^{\hspace{2mm}j}R)(\Lambda\Gamma_{ca}\Lambda)(\Lambda\Gamma^{b}W)\\
\bar{\Sigma}^{j(3)}_{1} &=& -\frac{4}{\eta}(\bar{\Lambda}R)(\Lambda\Gamma^{j}W) - \frac{8}{\eta^{2}}(\bar{\Lambda}\Gamma_{ab}\bar{\Lambda})(\bar{\Lambda}R)(\Lambda\Gamma^{ja}\Lambda)(\Lambda\Gamma^{b}W)
\end{eqnarray}
after using the identity \eqref{app24}. Therefore
\begin{eqnarray}
\{\bar{\Sigma}^{i}_{0}, \bar{\Sigma}^{j(2)}_{1}\} &=& \frac{2}{\eta^{2}}(\bar{\Lambda}\Gamma_{mn}\bar{\Lambda})(\bar{\Lambda}\Gamma_{c}^{\hspace{2mm}j}R)(\Lambda\Gamma^{c}\Gamma^{mn}\Gamma^{i}D) + \frac{4}{\eta^{3}}(\bar{\Lambda}\Gamma_{mn}\bar{\Lambda})(\bar{\Lambda}\Gamma_{ab}\bar{\Lambda})(\bar{\Lambda}\Gamma_{c}^{\hspace{2mm}j}R)(\Lambda\Gamma^{ca}\Lambda)(\Lambda\Gamma^{b}\Gamma^{mn}\Gamma^{i}D)\nonumber\\
&=& -\frac{8}{\eta^{3}}(\bar{\Lambda}\Gamma_{ab}\bar{\Lambda})(\bar{\Lambda}\Gamma_{c}^{\hspace{2mm}j}R)(\Lambda\Gamma^{ca}\Lambda)(\bar{\Lambda}\Gamma^{in}\bar{\Lambda})(\Lambda\Gamma^{b}_{\hspace{2mm}n}D)\nonumber\\
&& + \frac{4}{\eta^{3}}(\bar{\Lambda}\Gamma_{a}^{\hspace{2mm}i}\bar{\Lambda})(\bar{\Lambda}\Gamma_{c}^{\hspace{2mm}j}R)(\Lambda\Gamma^{ca}\Lambda)(\bar{\Lambda}\Gamma_{mn}\bar{\Lambda})(\Lambda\Gamma^{mn}D)\nonumber\\
&& + \frac{4}{\eta^{2}}(\bar{\Lambda}\Gamma_{c}^{\hspace{2mm}j}R)(\bar{\Lambda}\Gamma^{ci}\bar{\Lambda})(\Lambda D) - \frac{4}{\eta^{2}}(\bar{\Lambda}\Gamma^{cj}R)(\bar{\Lambda}\Gamma^{in}\bar{\Lambda})(\Lambda\Gamma_{cn}D) \nonumber \\
&& - \frac{4}{\eta^{2}}(\bar{\Lambda}\Gamma^{cj}R)(\bar{\Lambda}\Gamma_{cn}\bar{\Lambda})(\Lambda\Gamma^{in}D) + \frac{2}{\eta^{2}}(\bar{\Lambda}\Gamma^{ij}R)(\bar{\Lambda}\Gamma^{mn}\bar{\Lambda})(\Lambda\Gamma_{mn}D)\nonumber \\
&& + \frac{2}{\eta^{2}}(\bar{\Lambda}\Gamma_{c}^{\hspace{2mm}j}R)(\bar{\Lambda}\Gamma_{mn}\bar{\Lambda})(\Lambda\Gamma^{cimn}D)\nonumber\\
&=& \frac{4}{\eta^{2}}(\bar{\Lambda}R)(\bar{\Lambda}\Gamma^{ij}\bar{\Lambda})(\Lambda D)- \frac{4}{\eta^{2}}(\bar{\Lambda}\Gamma^{cj}R)(\bar{\Lambda}\Gamma^{in}\bar{\Lambda})(\Lambda\Gamma_{cn}D) + \frac{4}{\eta^{2}}(\bar{\Lambda}R)(\bar{\Lambda}\Gamma^{j}_{\hspace{2mm}n}\bar{\Lambda})(\Lambda\Gamma^{in}D)\nonumber\\
&& + \frac{2}{\eta^{2}}(\bar{\Lambda}\Gamma^{ij}R)(\bar{\Lambda}\Gamma_{mn}\bar{\Lambda})(\Lambda\Gamma^{mn}D) + \frac{2}{\eta^{2}}(\bar{\Lambda}\Gamma_{c}^{\hspace{2mm}j}R)(\bar{\Lambda}\Gamma_{mn}\bar{\Lambda})(\Lambda\Gamma^{cimn}D)
\end{eqnarray}
and by doing the same for $\bar{\Sigma}^{j(3)}_{1}$ we obtain
\begin{eqnarray}
\{\bar{\Sigma}_{0}^{i}, \bar{\Sigma}^{j(3)}_{1}\} &=& \frac{2}{\eta^{2}}(\bar{\Lambda}R)(\bar{\Lambda}\Gamma_{mn}\bar{\Lambda})(\Lambda\Gamma^{j}\Gamma^{mn}\Gamma^{i}D) + \frac{4}{\eta^{3}}(\bar{\Lambda}\Gamma_{ab}\bar{\Lambda})(\bar{\Lambda}R)(\Lambda\Gamma^{ja}\Lambda)(\bar{\Lambda}\Gamma_{mn}\bar{\Lambda})(\Lambda\Gamma^{b}\Gamma^{mn}\Gamma^{i}D)\nonumber\\
&=& - \frac{8}{\eta^{3}}(\bar{\Lambda}\Gamma_{ab}\bar{\Lambda})(\bar{\Lambda}R)(\Lambda\Gamma^{ja}\Lambda)(\bar{\Lambda}\Gamma^{i}_{\hspace{2mm}n}\bar{\Lambda})(\Lambda\Gamma^{bn}D)\nonumber\\
&& + \frac{4}{\eta^{3}}(\bar{\Lambda}\Gamma_{a}^{\hspace{2mm}i}\bar{\Lambda})(\bar{\Lambda}R)(\Lambda\Gamma^{ja}\Lambda)(\bar{\Lambda}\Gamma_{mn}\bar{\Lambda})(\Lambda\Gamma^{mn}D)\nonumber\\
&& -\frac{4}{\eta^{2}}(\bar{\Lambda}R)(\bar{\Lambda}\Gamma^{ij}\bar{\Lambda})(\Lambda D) - \frac{4}{\eta^{2}}(\bar{\Lambda}R)(\bar{\Lambda}\Gamma^{j}_{\hspace{2mm}n}\bar{\Lambda})(\Lambda\Gamma^{in}D)\nonumber\\
&& - \frac{4}{\eta^{2}}(\bar{\Lambda}R)(\bar{\Lambda}\Gamma^{i}_{\hspace{2mm}n}\bar{\Lambda})(\Lambda\Gamma^{jn}D) + \frac{4}{\eta^{2}}\eta^{ij}(\bar{\Lambda}R)(\bar{\Lambda}\Gamma_{mn}\bar{\Lambda})(\Lambda\Gamma^{mn}D)\nonumber\\
&& -\frac{4}{\eta^{2}}(\bar{\Lambda}R)(\bar{\Lambda}\Gamma_{mn}\bar{\Lambda})(\Lambda\Gamma^{ijmn}D)
\end{eqnarray}
Hence we get
\begin{eqnarray}
\{\bar{\Sigma}^{i}_{0}, \bar{\Sigma}^{j}_{1}\} &=& -\frac{4}{\eta^{2}}(\bar{\Lambda}\Gamma_{cj}R)(\bar{\Lambda}\Gamma^{in}\bar{\Lambda})(\Lambda\Gamma^{cn}D) + \frac{2}{\eta^{2}}(\bar{\Lambda}\Gamma^{ij}R)(\bar{\Lambda}\Gamma_{mn}\bar{\Lambda})(\Lambda\Gamma^{mn}D)\nonumber\\
&& + \frac{2}{\eta^{2}}(\bar{\Lambda}\Gamma_{c}^{\hspace{2mm}j}R)(\bar{\Lambda}\Gamma_{mn}\bar{\Lambda})(\Lambda\Gamma^{cimn}D) - \frac{4}{\eta^{2}}(\bar{\Lambda}R)(\bar{\Lambda}\Gamma^{i}_{\hspace{2mm}n}\bar{\Lambda})(\Lambda\Gamma^{jn}D)\nonumber\\
&& + \frac{2}{\eta^{2}}\eta^{ij}(\bar{\Lambda}R)(\bar{\Lambda}\Gamma_{mn}\bar{\Lambda})(\Lambda\Gamma^{mn}D) - \frac{2}{\eta}(\bar{\Lambda}R)(\bar{\Lambda}\Gamma_{mn}\bar{\Lambda})(\Lambda\Gamma^{ijmn}D)\label{eq500a}
\end{eqnarray}
Analogously we obtain
\begin{eqnarray}
\{\bar{\Sigma}^{i}_{1}, \bar{\Sigma}^{j}_{0}\} &=& -\frac{4}{\eta^{2}}(\bar{\Lambda}\Gamma_{ci}R)(\bar{\Lambda}\Gamma^{jn}\bar{\Lambda})(\Lambda\Gamma^{cn}D) + \frac{2}{\eta^{2}}(\bar{\Lambda}\Gamma^{ji}R)(\bar{\Lambda}\Gamma_{mn}\bar{\Lambda})(\Lambda\Gamma^{mn}D)\nonumber\\
&& + \frac{2}{\eta^{2}}(\bar{\Lambda}\Gamma_{c}^{\hspace{2mm}i}R)(\bar{\Lambda}\Gamma_{mn}\bar{\Lambda})(\Lambda\Gamma^{cjmn}D) - \frac{4}{\eta^{2}}(\bar{\Lambda}R)(\bar{\Lambda}\Gamma^{j}_{\hspace{2mm}n}\bar{\Lambda})(\Lambda\Gamma^{in}D)\nonumber\\
&& + \frac{2}{\eta^{2}}\eta^{ji}(\bar{\Lambda}R)(\bar{\Lambda}\Gamma_{mn}\bar{\Lambda})(\Lambda\Gamma^{mn}D) - \frac{2}{\eta}(\bar{\Lambda}R)(\bar{\Lambda}\Gamma_{mn}\bar{\Lambda})(\Lambda\Gamma^{jimn}D) \label{eq501}
\end{eqnarray}
and thus the sum of \eqref{eq500a} and \eqref{eq501} is
\begin{eqnarray}
\{\bar{\Sigma}^{i}_{0}, \bar{\Sigma}^{j}_{1}\} + \{\bar{\Sigma}^{i}_{1}, \bar{\Sigma}^{j}_{0}\} &=& -\frac{4}{\eta^{2}}(\bar{\Lambda}\Gamma^{in}\bar{\Lambda})(\bar{\Lambda}\Gamma^{cj}R)(\Lambda\Gamma_{cn}D) -\frac{4}{\eta^{2}}(\bar{\Lambda}\Gamma^{jn}\bar{\Lambda})(\bar{\Lambda}\Gamma^{ci}R)(\Lambda\Gamma_{cn}D) \nonumber \\
&& + \frac{2}{\eta^{2}}(\bar{\Lambda}\Gamma_{mn}\bar{\Lambda})(\bar{\Lambda}\Gamma_{c}^{\hspace{2mm}j}R)(\Lambda\Gamma^{cimn}D) + \frac{2}{\eta^{2}}(\bar{\Lambda}\Gamma_{mn}\bar{\Lambda})(\bar{\Lambda}\Gamma_{c}^{\hspace{2mm}i}R)(\Lambda\Gamma^{cjmn}D)\nonumber \\
&&-\frac{4}{\eta^{2}}(\bar{\Lambda}\Gamma^{i}_{\hspace{2mm}n}\bar{\Lambda})(\bar{\Lambda}R)(\Lambda\Gamma^{jn}D) -\frac{4}{\eta^{2}}(\bar{\Lambda}\Gamma^{j}_{\hspace{2mm}n}\bar{\Lambda})(\bar{\Lambda}R)(\Lambda\Gamma^{in}D)\nonumber\\
&& +\frac{4}{\eta^{2}}\eta^{ij}(\bar{\Lambda}\Gamma^{mn}\bar{\Lambda})(\bar{\Lambda}R)(\Lambda\Gamma_{mn}D)
\end{eqnarray}
Now we will simplify the expression corresponding to $\bar{\Sigma}_{1}^{j}$:
\begin{eqnarray}
\bar{\Sigma}^{j}_{1} &=& \frac{2}{\eta^{2}}(\bar{\Lambda}\Gamma_{ab}\bar{\Lambda})(\bar{\Lambda}\Gamma_{cd}R)(\Lambda\Gamma^{abckj}\Lambda)N^{d}_{\hspace{2mm}k} - \frac{8}{\eta^{2}}(\bar{\Lambda}\Gamma_{ab}\bar{\Lambda})(\bar{\Lambda}\Gamma_{c}^{\hspace{2mm}j}R)(\Lambda\Gamma^{bc}\Lambda)(\Lambda\Gamma^{a}W) \nonumber\\
&&-\frac{4}{\eta}(\bar{\Lambda}\Gamma^{cj}R)(\Lambda\Gamma_{c}W) - \frac{4}{\eta}(\bar{\Lambda}R)(\Lambda\Gamma^{j}W) - \frac{8}{\eta^{2}}(\bar{\Lambda}\Gamma_{ab}\bar{\Lambda})(\bar{\Lambda}R)(\Lambda\Gamma^{ja}\Lambda)(\Lambda\Gamma^{b}W)\nonumber \\
\label{eq502}
\end{eqnarray}
Lets us call $Y_{1}^{j}$ to the first term of this expression and expand it as follows
\begin{eqnarray}
Y_{1}^{j} &=& (\frac{6}{24})(\bar{\Lambda}\Gamma_{ab}\bar{\Lambda})(\bar{\Lambda}\Gamma_{cd}R)[4(\Lambda\Gamma^{cj}\Lambda)(\Lambda\Gamma^{ab}\Gamma^{d}W) - 4(\Lambda\Gamma^{ca}\Lambda)(\Lambda\Gamma^{jb}\Gamma^{d}W) + 4(\Lambda\Gamma^{cb}\Lambda)(\Lambda\Gamma^{ja}\Gamma^{d}W)\nonumber \\
&& + 4(\Lambda\Gamma^{ab}\Lambda)(\Lambda\Gamma^{cj}\Gamma^{d}W) - 4(\Lambda\Gamma^{jb}\Lambda)(\Lambda\Gamma^{ca}\Gamma^{d}W) + 4(\Lambda\Gamma^{ja}\Lambda)(\Lambda\Gamma^{cb}\Gamma^{d}W)] \nonumber \\
&=& \frac{2}{\eta^{2}}(\bar{\Lambda}\Gamma_{ab}\bar{\Lambda})(\bar{\Lambda}\Gamma_{cd}R)[(\Lambda\Gamma^{cj}\Lambda)(\Lambda\Gamma^{ab}\Gamma^{d}W) - 2(\Lambda\Gamma^{ca}\Lambda)(\Lambda\Gamma^{jb}\Gamma^{d}W) + (\Lambda\Gamma^{ab}\Lambda)(\Lambda\Gamma^{cj}\Gamma^{d}W)\nonumber\\
&& - 2(\Lambda\Gamma^{jb}\Lambda)(\Lambda\Gamma^{ca}\Gamma^{d}W)]\nonumber\\
&=& \frac{2}{\eta^{2}}(\bar{\Lambda}\Gamma_{ab}\bar{\Lambda})(\bar{\Lambda}\Gamma_{cd}R)[2\eta^{bd}(\Lambda\Gamma^{cj}\Lambda)(\Lambda\Gamma^{a}W) + (\Lambda\Gamma^{cj}\Lambda)(\Lambda\Gamma^{abd}W) + 2\eta^{dj}(\Lambda\Gamma^{ca}\Lambda)(\Lambda\Gamma^{b}W) \nonumber\\
&& -2\eta^{bd}(\Lambda\Gamma^{ca}\Lambda)(\Lambda\Gamma^{j}W) - 2(\Lambda\Gamma^{ca}\Lambda)(\Lambda\Gamma^{bdj}W) + \eta^{dj}(\Lambda\Gamma^{ab}\Lambda)(\Lambda\Gamma^{c}W) -\eta^{cd}(\Lambda\Gamma^{ab}\Lambda)(\Lambda\Gamma^{j}W) \nonumber\\
&& - (\Lambda\Gamma^{ab}\Lambda)(\Lambda\Gamma^{cdj}W) + 2\eta^{cd}(\Lambda\Gamma^{jb}\Lambda)(\Lambda\Gamma^{a}W) -2\eta^{ad}(\Lambda\Gamma^{jb}\Lambda)(\Lambda\Gamma^{c}W) + 2(\Lambda\Gamma^{jb}\Lambda)(\Lambda\Gamma^{acd}W)]\nonumber\\
&=&\frac{4}{\eta^{2}}(\bar{\Lambda}\Gamma_{ac}\bar{\Lambda})(\bar{\Lambda}R)(\Lambda\Gamma^{cj}\Lambda)(\Lambda\Gamma^{a}W) + \frac{2}{\eta^{2}}(\bar{\Lambda}\Gamma_{ab}\bar{\Lambda})(\bar{\Lambda}\Gamma_{cd}R)(\Lambda\Gamma^{cj}\Lambda)(\Lambda\Gamma^{abd}W) \nonumber\\
&& + \frac{4}{\eta^{2}}(\bar{\Lambda}\Gamma^{ab}\bar{\Lambda})(\bar{\Lambda}\Gamma^{cj}R)(\Lambda\Gamma_{ca}\Lambda)(\Lambda\Gamma_{b}W) + \frac{4}{\eta}(\bar{\Lambda}R)(\Lambda\Gamma^{j}W) - \frac{4}{\eta^{2}}(\bar{\Lambda}\Gamma_{ab}\bar{\Lambda})(\bar{\Lambda}\Gamma_{cd}R)(\Lambda\Gamma^{ca}\Lambda)(\Lambda\Gamma^{bdj}W)\nonumber\\
&& + \frac{2}{\eta}(\bar{\Lambda}\Gamma^{cj}R)(\Lambda\Gamma_{c}W) - \frac{2}{\eta}(\bar{\Lambda}\Gamma_{cd}R)(\Lambda\Gamma^{cdj}W) - \frac{4}{\eta^{2}}(\bar{\Lambda}\Gamma_{ab}\bar{\Lambda})(\bar{\Lambda}R)(\Lambda\Gamma^{jb}\Lambda)(\Lambda\Gamma^{c}W)\nonumber\\
&& + \frac{4}{\eta^{2}}(\bar{\Lambda}\Gamma_{ab}\bar{\Lambda})(\bar{\Lambda}\Gamma_{cd}R)(\Lambda\Gamma^{jb}\Lambda)(\Lambda\Gamma^{acd}W)\nonumber\\
&=&\frac{4}{\eta^{2}}(\bar{\Lambda}\Gamma_{ac}\bar{\Lambda})(\bar{\Lambda}R)(\Lambda\Gamma^{cj}\Lambda)(\Lambda\Gamma^{a}W) + \frac{2}{\eta^{2}}(\bar{\Lambda}\Gamma_{ab}\bar{\Lambda})(\bar{\Lambda}\Gamma_{cd}R)(\Lambda\Gamma^{cj}\Lambda)(\Lambda\Gamma^{abd}W) \nonumber\\
&& + \frac{2}{\eta^{2}}(\bar{\Lambda}\Gamma^{ac}\bar{\Lambda})(\bar{\Lambda}\Gamma^{bj}R)(\Lambda\Gamma_{ca}\Lambda)(\Lambda\Gamma_{b}W) +  \frac{2}{\eta^{2}}(\bar{\Lambda}\Gamma^{bj}\bar{\Lambda})(\bar{\Lambda}\Gamma^{ac}R)(\Lambda\Gamma_{ca}\Lambda)(\Lambda\Gamma_{b}W) \nonumber\\
&& + \frac{4}{\eta^{2}}(\bar{\Lambda}\Gamma^{cj}\bar{\Lambda})(\bar{\Lambda}\Gamma^{ba}R)(\Lambda\Gamma_{ca}\Lambda)(\Lambda\Gamma_{b}W) + \frac{4}{\eta}(\bar{\Lambda}R)(\Lambda\Gamma^{j}W) - \frac{1}{\eta^{2}}(\bar{\Lambda}\Gamma_{ac}\bar{\Lambda})(\bar{\Lambda}\Gamma_{bd}R)(\Lambda\Gamma^{ca}\Lambda)(\Lambda\Gamma^{bdj}W)\nonumber \\
&& - \frac{1}{\eta^{2}}(\bar{\Lambda}\Gamma_{bd}\bar{\Lambda})(\bar{\Lambda}\Gamma_{ac}R)(\Lambda\Gamma^{ca}\Lambda)(\Lambda\Gamma^{bdj}W) + \frac{2}{\eta}(\bar{\Lambda}\Gamma^{cj}R)(\Lambda\Gamma_{c}W) - \frac{2}{\eta}(\bar{\Lambda}\Gamma_{cd}R)(\Lambda\Gamma^{cdj}W)\nonumber\\
&& -\frac{4}{\eta^{2}}(\bar{\Lambda}\Gamma_{cb}\bar{\Lambda})(\bar{\Lambda}R)(\Lambda\Gamma^{jb}\Lambda)(\Lambda\Gamma^{c}W) - \frac{4}{\eta^{2}}(\bar{\Lambda}\Gamma_{ab}\bar{\Lambda})(\bar{\Lambda}\Gamma_{cd}R)(\Lambda\Gamma^{cj}\Lambda)(\Lambda\Gamma^{abd}W)\nonumber\\
&=& \frac{8}{\eta^{2}}(\bar{\Lambda}\Gamma_{ac}\bar{\Lambda})(\bar{\Lambda}R)(\Lambda\Gamma^{cj}\Lambda)(\Lambda\Gamma^{a}W) - \frac{2}{\eta^{2}}(\bar{\Lambda}\Gamma_{ab}\bar{\Lambda})(\bar{\Lambda}\Gamma_{cd}R)(\Lambda\Gamma^{cj}\Lambda)(\Lambda\Gamma^{abd}W) - \frac{1}{\eta}(\bar{\Lambda}\Gamma_{cd}R)(\Lambda\Gamma^{cdj}W)\nonumber\\
&& + \frac{2}{\eta^{2}}(\bar{\Lambda}\Gamma^{bj}\bar{\Lambda})(\bar{\Lambda}\Gamma^{ac}R)(\Lambda\Gamma_{ca}\Lambda)(\Lambda\Gamma_{b}W) + \frac{4}{\eta^{2}}(\bar{\Lambda}\Gamma^{cj}\bar{\Lambda})(\bar{\Lambda}\Gamma^{ba}R)(\Lambda\Gamma_{ca}\Lambda)(\Lambda\Gamma_{b}W)\nonumber\\
&& + \frac{4}{\eta}(\bar{\Lambda}R)(\Lambda\Gamma^{j}W) - \frac{1}{\eta^{2}}(\bar{\Lambda}\Gamma_{bd}\bar{\Lambda})(\bar{\Lambda}\Gamma_{ac}R)(\Lambda\Gamma^{ca}\Lambda)(\Lambda\Gamma^{bdj}W)
\end{eqnarray}
Plugging this result into the equation \eqref{eq502}
\begin{eqnarray}
\bar{\Sigma}_{1}^{j} &=& - \frac{2}{\eta^{2}}(\bar{\Lambda}\Gamma_{ab}\bar{\Lambda})(\bar{\Lambda}\Gamma_{cd}R)(\Lambda\Gamma^{cj}\Lambda)(\Lambda\Gamma^{abd}W) - \frac{1}{\eta}(\bar{\Lambda}\Gamma_{cd}R)(\Lambda\Gamma^{cdj}W)\nonumber\\
&& + \frac{2}{\eta^{2}}(\bar{\Lambda}\Gamma^{bj}\bar{\Lambda})(\bar{\Lambda}\Gamma^{ac}R)(\Lambda\Gamma_{ca}\Lambda)(\Lambda\Gamma_{b}W) + \frac{4}{\eta^{2}}(\bar{\Lambda}\Gamma^{cj}\bar{\Lambda})(\bar{\Lambda}\Gamma^{ba}R)(\Lambda\Gamma_{ca}\Lambda)(\Lambda\Gamma_{b}W)\nonumber\\
&& - \frac{1}{\eta^{2}}(\bar{\Lambda}\Gamma_{bd}\bar{\Lambda})(\bar{\Lambda}\Gamma_{ac}R)(\Lambda\Gamma^{ca}\Lambda)(\Lambda\Gamma^{bdj}W)  - \frac{8}{\eta^{2}}(\bar{\Lambda}\Gamma_{ab}\bar{\Lambda})(\bar{\Lambda}\Gamma_{c}^{\hspace{2mm}j}R)(\Lambda\Gamma^{bc}\Lambda)(\Lambda\Gamma^{a}W) \nonumber\\
&&-\frac{4}{\eta}(\bar{\Lambda}\Gamma^{cj}R)(\Lambda\Gamma_{c}W)\nonumber\\
&=& - \frac{2}{\eta^{2}}(\bar{\Lambda}\Gamma_{ab}\bar{\Lambda})(\bar{\Lambda}\Gamma_{cd}R)(\Lambda\Gamma^{cj}\Lambda)(\Lambda\Gamma^{abd}W) - \frac{1}{\eta}(\bar{\Lambda}\Gamma_{cd}R)(\Lambda\Gamma^{cdj}W)\nonumber\\
&& + \frac{2}{\eta^{2}}(\bar{\Lambda}\Gamma^{bj}\bar{\Lambda})(\bar{\Lambda}\Gamma^{ac}R)(\Lambda\Gamma_{ca}\Lambda)(\Lambda\Gamma_{b}W) + \frac{4}{\eta^{2}}(\bar{\Lambda}\Gamma^{cj}\bar{\Lambda})(\bar{\Lambda}\Gamma^{ba}R)(\Lambda\Gamma_{ca}\Lambda)(\Lambda\Gamma_{b}W)\nonumber\\
&& - \frac{1}{\eta^{2}}(\bar{\Lambda}\Gamma_{bd}\bar{\Lambda})(\bar{\Lambda}\Gamma_{ac}R)(\Lambda\Gamma^{ca}\Lambda)(\Lambda\Gamma^{bdj}W) + \frac{4}{\eta^{2}}(\bar{\Lambda}\Gamma^{nr}\bar{\Lambda})(\bar{\Lambda}\Gamma^{mj}R)(\Lambda\Gamma_{nr}\Lambda)(\Lambda\Gamma^{m}W) \nonumber\\
&& + \frac{4}{\eta^{2}}(\bar{\Lambda}\Gamma^{mj}\bar{\Lambda})(\bar{\Lambda}\Gamma^{nr}R)(\Lambda\Gamma_{nr}\Lambda)(\Lambda\Gamma_{m}W)  + \frac{8}{\eta^{2}}(\bar{\Lambda}\Gamma^{rj}\bar{\Lambda})(\bar{\Lambda}\Gamma^{mn}R)(\Lambda\Gamma_{nr}\Lambda)(\Lambda\Gamma_{m}W)\nonumber\\
&& - \frac{4}{\eta}(\bar{\Lambda}\Gamma^{cj}R)(\Lambda\Gamma_{c}W)\nonumber\\
&=& - \frac{2}{\eta^{2}}(\bar{\Lambda}\Gamma_{ab}\bar{\Lambda})(\bar{\Lambda}\Gamma_{cd}R)(\Lambda\Gamma^{cj}\Lambda)(\Lambda\Gamma^{abd}W) - \frac{1}{\eta}(\bar{\Lambda}\Gamma_{cd}R)(\Lambda\Gamma^{cdj}W)\nonumber\\
&& + \frac{2}{\eta^{2}}(\bar{\Lambda}\Gamma^{bj}\bar{\Lambda})(\bar{\Lambda}\Gamma^{ca}R)(\Lambda\Gamma_{ca}\Lambda)(\Lambda\Gamma_{b}W) 
- \frac{4}{\eta^{2}}(\bar{\Lambda}\Gamma^{cj}\bar{\Lambda})(\bar{\Lambda}\Gamma^{ba}R)(\Lambda\Gamma_{ca}\Lambda)(\Lambda\Gamma_{b}W)\nonumber\\
&& + \frac{1}{\eta^{2}}(\bar{\Lambda}\Gamma_{bd}\bar{\Lambda})(\bar{\Lambda}\Gamma_{ac}R)(\Lambda\Gamma^{ac}\Lambda)(\Lambda\Gamma^{bdj}W)
\end{eqnarray}
This expression is invariant under the gauge symmetry generated by the pure spinor constraint, as it should. Now let us make the following definitions:
\begin{eqnarray}
W_{1}^{j} &=& -\frac{2}{\eta^{2}}(\bar{\Lambda}\Gamma_{ab}\bar{\Lambda})(\bar{\Lambda}\Gamma_{cd}R)(\Lambda\Gamma^{cj}\Lambda)(\Lambda\Gamma^{abd}W)\\
W_{2}^{j} &=& -\frac{1}{\eta}(\bar{\Lambda}\Gamma_{cd}R)(\Lambda\Gamma^{cdj}W)\\
W_{3}^{j} &=& \frac{2}{\eta^{2}}(\bar{\Lambda}\Gamma^{bj}\bar{\Lambda})(\bar{\Lambda}\Gamma^{ca}R)(\Lambda\Gamma_{ca}\Lambda)(\Lambda\Gamma_{b}W)\\
W_{4}^{j} &=& -\frac{4}{\eta^{2}}(\bar{\Lambda}\Gamma^{cj}\bar{\Lambda})(\bar{\Lambda}\Gamma^{ba}R)(\Lambda\Gamma_{ca}\Lambda)(\Lambda\Gamma_{b}W)\\
W_{5}^{j} &=& \frac{1}{\eta^{2}}(\bar{\Lambda}\Gamma_{bd}\bar{\Lambda})(\bar{\Lambda}\Gamma_{ac}R)(\Lambda\Gamma^{ac}\Lambda)(\Lambda\Gamma^{bdj}W)
\end{eqnarray}
Consequently to compute $\{\bar{\Sigma}_{1}^{i}, \bar{\Sigma}_{1}^{j}\}$ we should calculate the anticommutator between each pair of these $W_{\{1,2,3,4,5\}}^{j}$ variables. Explicitly this computation works as follows
\begin{eqnarray}
\{W_{1}^{i}, W_{1}^{j}\} &=& \{-\frac{2}{\eta^{2}}(\bar{\Lambda}\Gamma_{mn}\bar{\Lambda})(\bar{\Lambda}\Gamma_{rs}R)(\Lambda\Gamma^{ri}\Lambda)(\Lambda\Gamma^{mns}W),-\frac{2}{\eta^{2}}(\bar{\Lambda}\Gamma_{ab}\bar{\Lambda})(\bar{\Lambda}\Gamma_{cd}R)(\Lambda\Gamma^{cj}\Lambda)(\Lambda\Gamma^{abd}W)\} \nonumber \\
&=& \frac{4}{\eta^{4}}(\bar{\Lambda}\Gamma_{mn}\bar{\Lambda})(\bar{\Lambda}\Gamma_{rs}R)(\bar{\Lambda}\Gamma_{ab}\bar{\Lambda})(\bar{\Lambda}\Gamma_{cd}R)[(\Lambda\Gamma^{ri}\Lambda)(\Lambda\Gamma^{mns}W), (\Lambda\Gamma^{cj}\Lambda)(\Lambda\Gamma^{abd}W)] \nonumber \\
&=& \frac{4}{\eta^{4}}(\bar{\Lambda}\Gamma_{mn}\bar{\Lambda})(\bar{\Lambda}\Gamma_{rs}R)(\bar{\Lambda}\Gamma_{ab}\bar{\Lambda})(\bar{\Lambda}\Gamma_{cd}R)\{(\Lambda\Gamma^{ri}\Lambda)(\Lambda\Gamma^{cj}\Lambda)[(\Lambda\Gamma^{mns}W), (\Lambda\Gamma^{abd}W)] \nonumber\\
&& + (\Lambda\Gamma^{ri}\Lambda)[(\Lambda\Gamma^{mns}W),(\Lambda\Gamma^{cj}\Lambda)](\Lambda\Gamma^{abd}W) + (\Lambda\Gamma^{cj}\Lambda)[(\Lambda\Gamma^{ri}\Lambda),(\Lambda\Gamma^{abd}W)](\Lambda\Gamma^{mns}W)\}\nonumber\\
&=& \frac{64}{\eta^{4}}(\bar{\Lambda}\Gamma_{mn}\bar{\Lambda})(\bar{\Lambda}\Gamma_{rs}R)(\bar{\Lambda}\Gamma_{ab}\bar{\Lambda})(\bar{\Lambda}\Gamma_{cd}R)(\Lambda\Gamma^{ri}\Lambda)(\Lambda\Gamma^{cj}\Lambda)[\delta^{ms}_{ad}N^{nd}]\nonumber\\
&=& -\frac{32}{\eta^{4}}(\bar{\Lambda}\Gamma_{dn}\bar{\Lambda})(\bar{\Lambda}\Gamma_{ra}R)(\bar{\Lambda}\Gamma_{ab}\bar{\Lambda})(\bar{\Lambda}\Gamma_{cd}R)(\Lambda\Gamma^{ri}\Lambda)(\Lambda\Gamma^{cj}\Lambda)N^{nd}\nonumber\\
&=& 0
\end{eqnarray}
because of the identities \eqref{app41} and $(\bar{\Lambda}R)(\bar{\Lambda}R)=0$.
\begin{eqnarray}
\{W_{2}^{i}, W_{2}^{j}\} &=& \{-\frac{1}{\eta}(\bar{\Lambda}\Gamma_{mn}R)(\Lambda\Gamma^{mni}W), -\frac{1}{\eta}(\bar{\Lambda}\Gamma_{cd}R)(\Lambda\Gamma^{cdj}W)\}\nonumber\\
&=& \frac{1}{\eta^{2}}(\bar{\Lambda}\Gamma_{mn}R)(\bar{\Lambda}\Gamma_{cd}R)[(\Lambda\Gamma^{mni}W), (\Lambda\Gamma^{cdj}W)]\nonumber\\
&=& \frac{1}{\eta^{2}}(\bar{\Lambda}\Gamma_{mn}R)(\bar{\Lambda}\Gamma_{cd}R)(4\delta^{mn}_{cd}N^{ij} - 8\delta^{mn}_{cj}N^{id} - 8\delta^{mi}_{cd}N^{nj} - 16\delta^{ni}_{cj}N^{md})\nonumber\\
&=& -\frac{8}{\eta^{2}}(\bar{\Lambda}\Gamma^{cj}R)(\bar{\Lambda}\Gamma_{cd}R)N^{id} - \frac{8}{\eta^{2}}(\bar{\Lambda}\Gamma_{mn}R)(\bar{\Lambda}\Gamma^{mi}R)N^{nj}\nonumber\\
&& -\frac{8}{\eta^{2}}(\bar{\Lambda}\Gamma_{m}^{\hspace{3mm}c}R)(\bar{\Lambda}\Gamma_{cd}R)\eta^{ij} + \frac{8}{\eta^{2}}(\bar{\Lambda}\Gamma^{mj}R)(\bar{\Lambda}\Gamma^{id}R)N_{md}
\end{eqnarray}
The use of the identity \eqref{app12} allows us to write
\begin{eqnarray}
\{W_{2}^{i}, W_{2}^{j}\} &=& -\frac{8}{\eta^{2}}[(\bar{\Lambda}\Gamma^{jd}R)(\bar{\Lambda}R) + \frac{1}{2}(\bar{\Lambda}\Gamma^{jd}\bar{\Lambda})(RR)]N^{i}_{\hspace{2mm}d}\nonumber \\
&& +\frac{8}{\eta^{2}}[(\bar{\Lambda}\Gamma^{ni}R)(\bar{\Lambda}R) + \frac{1}{2}(\bar{\Lambda}\Gamma^{ni}\bar{\Lambda})(RR)]N^{j}_{\hspace{2mm}n}\nonumber \\
&& +\frac{8}{\eta^{2}}\eta^{ij}[(\bar{\Lambda}\Gamma^{md}R)(\bar{\Lambda}R) + \frac{1}{2}(\bar{\Lambda}\Gamma^{md}\bar{\Lambda})(RR)]N_{md}\nonumber \\
&& + \frac{8}{\eta^{2}}(\bar{\Lambda}\Gamma^{mj}R)(\bar{\Lambda}\Gamma^{id}R)N_{md}\nonumber\\
&=& -\frac{8}{\eta^{2}}(\bar{\Lambda}\Gamma^{jd}R)(\bar{\Lambda}R)N^{i}_{\hspace{2mm}d} - \frac{4}{\eta^{2}}(\bar{\Lambda}\Gamma^{jd}\bar{\Lambda})(RR)N^{i}_{\hspace{2mm}d}\nonumber\\
&& +\frac{8}{\eta^{2}}(\bar{\Lambda}\Gamma^{di}R)(\bar{\Lambda}R)N^{j}_{\hspace{2mm}d} + \frac{4}{\eta^{2}}(\bar{\Lambda}\Gamma^{di}\bar{\Lambda})(RR)N^{j}_{\hspace{2mm}d}\nonumber\\
&& + \frac{8}{\eta^{2}}\eta^{ij}(\bar{\Lambda}\Gamma^{md}R)(\bar{\Lambda}R)N_{md} + \frac{4}{\eta^{2}}\eta^{ij}(\bar{\Lambda}\Gamma^{md}\bar{\Lambda})(RR)N_{md}\nonumber\\
&& + \frac{8}{\eta^{2}}(\bar{\Lambda}\Gamma^{mj}R)(\bar{\Lambda}\Gamma^{id}R)N_{md}
\end{eqnarray}

\begin{eqnarray}
\{W_{3}^{i}, W_{3}^{j}\} &=& \{\frac{2}{\eta^{2}}(\bar{\Lambda}\Gamma^{ni}\bar{\Lambda})(\bar{\Lambda}\Gamma^{rm}R)(\Lambda\Gamma_{rm}\Lambda)(\Lambda\Gamma_{n}W), \frac{2}{\eta^{2}}(\bar{\Lambda}\Gamma^{bj}\bar{\Lambda})(\bar{\Lambda}\Gamma^{ca}R)(\Lambda\Gamma_{ca}\Lambda)(\Lambda\Gamma_{b}W)\}\nonumber\\
&=& -\frac{8}{\eta^{2}}(\bar{\Lambda}\Gamma^{ni}\bar{\Lambda})(\bar{\Lambda}\Gamma^{rm}R)(\bar{\Lambda}\Gamma^{bj})\bar{\Lambda}(\bar{\Lambda}\Gamma^{ca}R)(\Lambda\Gamma_{rm}\Lambda)(\Lambda\Gamma_{ca}\Lambda)N_{nb}\nonumber\\
&=& 0
\end{eqnarray}
because of the identities \eqref{app42} and $(\Lambda\Gamma_{mn}\Lambda)(\bar{\Lambda}\Gamma^{mn}R)(\Lambda\Gamma_{ab}\Lambda)(\bar{\Lambda}\Gamma^{ab}R) = 0$.

\begin{eqnarray}
\{W_{4}^{i}, W_{4}^{j}\} &=& \{-\frac{4}{\eta^{2}}(\bar{\Lambda}\Gamma^{ri}\bar{\Lambda})(\bar{\Lambda}\Gamma^{nm}R)(\Lambda\Gamma_{rm}\Lambda)(\Lambda\Gamma_{a}W), -\frac{4}{\eta^{2}}(\bar{\Lambda}\Gamma^{cj}\bar{\Lambda})(\bar{\Lambda}\Gamma^{ba}R)(\Lambda\Gamma_{ca}\Lambda)(\Lambda\Gamma_{b}W)\}\nonumber\\
&=& -\frac{32}{\eta^{4}}(\bar{\Lambda}\Gamma^{ri}\bar{\Lambda})(\bar{\Lambda}\Gamma^{nm}R)(\bar{\Lambda}\Gamma^{cj}\bar{\Lambda})(\bar{\Lambda}\Gamma^{ba}R)(\Lambda\Gamma_{rm}\Lambda)(\Lambda\Gamma_{ca}\Lambda)N_{nb}
\end{eqnarray}
which follows directly from the identity \eqref{app42}
\begin{eqnarray}
\{W_{5}^{i}, W_{5}^{j}\} &=& \{\frac{1}{\eta^{2}}(\bar{\Lambda}\Gamma_{ns}\bar{\Lambda})(\bar{\Lambda}\Gamma^{mr}R)(\Lambda\Gamma_{mr}\Lambda)(\Lambda\Gamma^{nsi}W), \frac{1}{\eta^{2}}(\bar{\Lambda}\Gamma_{bd}\bar{\Lambda})(\bar{\Lambda}\Gamma^{ac}R)(\Lambda\Gamma_{ac}\Lambda)(\Lambda\Gamma^{bdj}W)\}\nonumber\\
&=& \frac{1}{\eta^{4}}(\bar{\Lambda}\Gamma_{ns}\bar{\Lambda})(\bar{\Lambda}\Gamma_{bd}\bar{\Lambda})(\bar{\Lambda}\Gamma^{mr}R)(\bar{\Lambda}\Gamma^{ac}R)(\Lambda\Gamma_{mr}\Lambda)(\Lambda\Gamma_{ac}\Lambda)[(\Lambda\Gamma^{nsi}W), (\Lambda\Gamma^{bdj}W)]\nonumber\\ 
&=& \frac{8}{\eta^{4}}(\bar{\Lambda}\Gamma^{nj}\bar{\Lambda})(\bar{\Lambda}\Gamma^{id}\bar{\Lambda})(\bar{\Lambda}\Gamma^{mr}R)(\bar{\Lambda}\Gamma^{ac}R)(\Lambda\Gamma_{mr}\Lambda)(\Lambda\Gamma_{ac}\Lambda)N_{nd}\nonumber\\
&=& 0
\end{eqnarray}
because of the identity $(\Lambda\Gamma_{mn}\Lambda)(\bar{\Lambda}\Gamma^{mn}R)(\Lambda\Gamma_{ab}\Lambda)(\bar{\Lambda}\Gamma^{ab}R) = 0$.

\begin{eqnarray}
\{W_{1}^{i}, W_{2}^{j}\} &=& \{-\frac{2}{\eta^{2}}(\bar{\Lambda}\Gamma_{mn}\bar{\Lambda})(\bar{\Lambda}\Gamma_{rs}R)(\Lambda\Gamma^{ri}\Lambda)(\Lambda\Gamma^{mns}W),-\frac{1}{\eta}(\bar{\Lambda}\Gamma_{cd}R)(\Lambda\Gamma^{cdj}W)\}\nonumber\\
&=& \frac{2}{\eta^{3}}(\bar{\Lambda}\Gamma_{mn}\bar{\Lambda})(\bar{\Lambda}\Gamma_{rs}R)(\bar{\Lambda}\Gamma_{cd}R)[(\Lambda\Gamma^{ri}\Lambda)(\Lambda\Gamma^{mns}W), (\Lambda\Gamma^{cdj}W)]\nonumber\\
&=& \frac{2}{\eta^{3}}(\bar{\Lambda}\Gamma_{mn}\bar{\Lambda})(\bar{\Lambda}\Gamma_{rs}R)(\bar{\Lambda}\Gamma_{cd}R)\{(\Lambda\Gamma^{ri}\Lambda)[(\Lambda\Gamma^{mns}W), (\Lambda\Gamma^{cdj}W)] \nonumber\\
&& + [(\Lambda\Gamma^{ri}\Lambda), (\Lambda\Gamma^{cdj}W)](\Lambda\Gamma^{mns}W)\}\nonumber\\
&=& \frac{2}{\eta^{3}}(\bar{\Lambda}\Gamma_{mn}\bar{\Lambda})(\bar{\Lambda}\Gamma_{rs}R)(\bar{\Lambda}\Gamma_{cd}R)[-8\delta^{mn}_{cj} + 8\delta^{ns}_{cd} - 16\delta^{ns}_{cj}N^{md}]\nonumber\\
&& +\frac{4}{\eta^{3}}(\bar{\Lambda}\Gamma_{mn}\bar{\Lambda})(\bar{\Lambda}\Gamma_{rs}R)(\bar{\Lambda}\Gamma_{cd}R)(\Lambda\Gamma^{cdjri}\Lambda)(\Lambda\Gamma^{mns}W)\nonumber\\
&=& -\frac{16}{\eta^{3}}(\bar{\Lambda}\Gamma^{cj}\bar{\Lambda})(\bar{\Lambda}\Gamma_{rs}R)(\bar{\Lambda}\Gamma_{cd}R)(\Lambda\Gamma^{ri}\Lambda)N^{sd}\nonumber\\
&& +\frac{16}{\eta^{3}}(\bar{\Lambda}\Gamma_{mn}\bar{\Lambda})(\bar{\Lambda}\Gamma_{rs}R)(\bar{\Lambda}\Gamma^{ns}R)(\Lambda\Gamma^{ri}\Lambda)N^{mj}\nonumber\\
&& - \frac{16}{\eta^{3}}[(\bar{\Lambda}\Gamma_{mc}\bar{\Lambda})(\bar{\Lambda}\Gamma_{r}^{\hspace{2mm}j}R)(\bar{\Lambda}\Gamma^{cd}R)(\Lambda\Gamma^{ri}\Lambda) - (\bar{\Lambda}\Gamma_{m}^{\hspace{3mm}j}\bar{\Lambda})(\bar{\Lambda}\Gamma_{rc}R)(\bar{\Lambda}\Gamma^{cd}R)(\Lambda\Gamma^{ri}\Lambda)]N^{m}_{\hspace{2mm}d}\nonumber \\
&& +\frac{4}{\eta^{3}}(\bar{\Lambda}\Gamma_{mn}\bar{\Lambda})(\bar{\Lambda}\Gamma_{rs}R)(\bar{\Lambda}\Gamma_{cd}R)(\Lambda\Gamma^{cdjri}\Lambda)(\Lambda\Gamma^{mns}W)\nonumber\\
&=& \frac{32}{\eta^{3}}(\bar{\Lambda}\Gamma_{d}^{\hspace{2mm}j}\bar{\Lambda})(\bar{\Lambda}\Gamma_{rs}R)(\bar{\Lambda}R)(\Lambda\Gamma^{ri}\Lambda)N^{sd} + \frac{16}{\eta^{3}}(\bar{\Lambda}\Gamma_{sd}\bar{\Lambda})(\bar{\Lambda}\Gamma^{rj}R)(\bar{\Lambda}R)(\Lambda\Gamma_{r}^{\hspace{2mm}i}\Lambda)N^{sd}\nonumber\\
&& -\frac{16}{\eta^{3}}(\bar{\Lambda}\Gamma_{d}^{\hspace{2mm}j}\bar{\Lambda})(\bar{\Lambda}\Gamma_{rs}\bar{\Lambda})(RR)(\Lambda\Gamma^{ri}\Lambda)N^{ds}\nonumber \\
&& +\frac{4}{\eta^{3}}(\bar{\Lambda}\Gamma_{mn}\bar{\Lambda})(\bar{\Lambda}\Gamma_{rs}R)(\bar{\Lambda}\Gamma_{cd}R)(\Lambda\Gamma^{cdjri}\Lambda)(\Lambda\Gamma^{mns}W)
\end{eqnarray}

\begin{eqnarray}
\{W_{1}^{i}, W_{3}^{j}\} &=& \{-\frac{2}{\eta^{2}}(\bar{\Lambda}\Gamma_{mn}\bar{\Lambda})(\bar{\Lambda}\Gamma_{rs}R)(\Lambda\Gamma^{ri}\Lambda)(\Lambda\Gamma^{mns}W), \frac{2}{\eta^{2}}(\bar{\Lambda}\Gamma^{bj}\bar{\Lambda})(\bar{\Lambda}\Gamma^{ca}R)(\Lambda\Gamma_{ca}\Lambda)(\Lambda\Gamma_{b}W)\}\nonumber\\
&=& \frac{8}{\eta^{4}}(\bar{\Lambda}\Gamma_{mn}\bar{\Lambda})(\bar{\Lambda}\Gamma_{rs}R)(\Lambda\Gamma^{ri}\Lambda)(\bar{\Lambda}\Gamma^{bj}\bar{\Lambda})(\bar{\Lambda}\Gamma^{ca}R)(\Lambda\Gamma_{ca}\Lambda)(\Lambda\Gamma^{mns}_{\hspace{6mm}b}W)\nonumber\\
&=& 0
\end{eqnarray}
where we have used the identities \eqref{app2}, \eqref{app43}.

\begin{eqnarray}
\{W_{1}^{i}, W_{4}^{j}\} &=& \{-\frac{2}{\eta^{2}}(\bar{\Lambda}\Gamma_{mn}\bar{\Lambda})(\bar{\Lambda}\Gamma_{rs}R)(\Lambda\Gamma^{ri}\Lambda)(\Lambda\Gamma^{mns}W), -\frac{4}{\eta^{2}}(\bar{\Lambda}\Gamma^{cj}\bar{\Lambda})(\bar{\Lambda}\Gamma^{ba}R)(\Lambda\Gamma_{ca}\Lambda)(\Lambda\Gamma_{b}W)\}\nonumber\\
&=& \frac{8}{\eta^{4}}(\bar{\Lambda}\Gamma_{mn}\bar{\Lambda})(\bar{\Lambda}\Gamma_{rs}R)(\bar{\Lambda}\Gamma^{cj}\bar{\Lambda})(\bar{\Lambda}\Gamma^{ba}R)\{(\Lambda\Gamma^{ri}\Lambda)[(\Lambda\Gamma^{mns}W),(\Lambda\Gamma_{ca}\Lambda)](\Lambda\Gamma_{b}W)\nonumber\\
&& + (\Lambda\Gamma_{ca}\Lambda)[(\Lambda\Gamma^{mns}W),(\Lambda\Gamma_{b}W)]\}\nonumber\\
&=& -\frac{16}{\eta^{4}}(\bar{\Lambda}\Gamma_{mn}\bar{\Lambda})(\bar{\Lambda}\Gamma_{rs}R)(\bar{\Lambda}\Gamma^{cj}\bar{\Lambda})(\bar{\Lambda}\Gamma^{ba}R)(\Lambda\Gamma^{ri}\Lambda)(\Lambda\Gamma^{mns}_{\hspace{6mm}ca}\Lambda)(\Lambda\Gamma_{b}W)\nonumber\\ 
&&-\frac{16}{\eta^{3}}(\bar{\Lambda}\Gamma_{mn}\bar{\Lambda})(\bar{\Lambda}\Gamma_{rs}R)(\bar{\Lambda}\Gamma^{bj}R)(\Lambda\Gamma^{ri}\Lambda)(\Lambda\Gamma^{mns}_{\hspace{6mm}b}W)\nonumber\\
&=&-\frac{16}{\eta^{3}}(\bar{\Lambda}\Gamma_{mn}\bar{\Lambda})(\bar{\Lambda}\Gamma_{rs}R)(\bar{\Lambda}\Gamma^{bj}R)(\Lambda\Gamma^{ri}\Lambda)(\Lambda\Gamma^{mns}_{\hspace{6mm}b}W)
\end{eqnarray}
because of the identity \eqref{app2}.

\begin{eqnarray}
\{W_{1}^{i}, W_{5}^{j}\} &=& \{-\frac{2}{\eta}(\bar{\Lambda}\Gamma_{mn}\bar{\Lambda})(\bar{\Lambda}\Gamma_{rs}R)(\Lambda\Gamma^{ri}\Lambda)(\Lambda\Gamma^{mns}W), \frac{1}{\eta^{2}}(\bar{\Lambda}\Gamma_{bd}\bar{\Lambda})(\bar{\Lambda}\Gamma_{ac}R)(\Lambda\Gamma^{ac}\Lambda)(\Lambda\Gamma^{bdj}W)\}\nonumber\\
&=& -\frac{2}{\eta^{4}}(\bar{\Lambda}\Gamma_{mn}\bar{\Lambda})(\bar{\Lambda}\Gamma_{rs}R)(\bar{\Lambda}\Gamma_{bd}\bar{\Lambda})(\bar{\Lambda}\Gamma_{ac}R)[(\Lambda\Gamma^{ri}\Lambda)(\Lambda\Gamma^{mns}W),(\Lambda\Gamma^{ac}\Lambda)(\Lambda\Gamma^{bdj}W)]\nonumber\\
&=& -\frac{2}{\eta^{4}}(\bar{\Lambda}\Gamma_{mn}\bar{\Lambda})(\bar{\Lambda}\Gamma_{rs}R)(\bar{\Lambda}\Gamma_{bd}\bar{\Lambda})(\bar{\Lambda}\Gamma_{ac}R)\{(\Lambda\Gamma^{ri}\Lambda)(\Lambda\Gamma^{ac}\Lambda)[(\Lambda\Gamma^{mns}W),(\Lambda\Gamma^{bdj}W)] \nonumber\\
&& + (\Lambda\Gamma^{ac}\Lambda)[(\Lambda\Gamma^{ri}\Lambda),(\Lambda\Gamma^{bdj}W)](\Lambda\Gamma^{mns}W)\}\nonumber\\
&=& -\frac{4}{\eta^{4}}(\bar{\Lambda}\Gamma_{mn}\bar{\Lambda})(\bar{\Lambda}\Gamma_{rs}R)(\bar{\Lambda}\Gamma_{bd}\bar{\Lambda})(\bar{\Lambda}\Gamma_{ac}R)(\Lambda\Gamma^{ac}\Lambda)(\Lambda\Gamma^{bdjri}\Lambda)(\Lambda\Gamma^{mns}W)\nonumber\\
&& - \frac{2}{\eta^{4}}(\bar{\Lambda}\Gamma_{mn}\bar{\Lambda})(\bar{\Lambda}\Gamma_{rs}R)(\bar{\Lambda}\Gamma_{bd}\bar{\Lambda})(\bar{\Lambda}\Gamma_{ac}R)(\Lambda\Gamma^{ri}\Lambda)(\Lambda\Gamma^{ac}\Lambda)[16\delta^{ms}_{bj}N^{nd}]\nonumber\\
&=& \frac{16}{\eta^{4}}(\bar{\Lambda}\Gamma^{jn}\bar{\Lambda})(\bar{\Lambda}\Gamma_{rb}R)(\bar{\Lambda}\Gamma^{bd}\bar{\Lambda})(\bar{\Lambda}\Gamma^{ac}R)(\Lambda\Gamma_{ac}\Lambda)(\Lambda\Gamma^{ri}\Lambda)N_{nd}\nonumber\\
&=& \frac{8}{\eta^{4}}(\bar{\Lambda}\Gamma^{j}_{\hspace{2mm}r}\bar{\Lambda})(\bar{\Lambda}\Gamma^{nd}\bar{\Lambda})(\bar{\Lambda}R)(\bar{\Lambda}\Gamma^{ac}R)(\Lambda\Gamma_{ac}\Lambda)(\Lambda\Gamma^{ri}\Lambda)N_{nd}
\end{eqnarray}

\begin{eqnarray}
\{W_{2}^{i}, W_{3}^{j}\} &=& \{-\frac{1}{\eta}(\bar{\Lambda}\Gamma_{mn}R)(\Lambda\Gamma^{mni}W) , \frac{2}{\eta^{2}}(\bar{\Lambda}\Gamma_{b}^{\hspace{2mm}j}\bar{\Lambda})(\bar{\Lambda}\Gamma^{ca}R)(\Lambda\Gamma_{ca}\Lambda)(\Lambda\Gamma^{b}W)\}\nonumber\\
&=& \frac{4}{\eta^{3}}(\bar{\Lambda}\Gamma_{mn}R)(\bar{\Lambda}\Gamma_{b}^{\hspace{2mm}j}\bar{\Lambda})(\bar{\Lambda}\Gamma^{ca}R)(\Lambda\Gamma_{ca}\Lambda)(\Lambda\Gamma^{mnib}W)
\end{eqnarray}

\begin{eqnarray}
\{W_{2}^{i}, W_{4}^{j}\} &=& \{-\frac{1}{\eta}(\bar{\Lambda}\Gamma_{mn}R)(\Lambda\Gamma^{mni}W), -\frac{4}{\eta^{2}}(\bar{\Lambda}\Gamma^{cj}\bar{\Lambda})(\bar{\Lambda}\Gamma^{ba}R)(\Lambda\Gamma_{ca}\Lambda)(\Lambda\Gamma_{b}W)\}\nonumber\\
&=&\frac{4}{\eta^{3}}(\bar{\Lambda}\Gamma_{mn}R)(\bar{\Lambda}\Gamma_{c}^{\hspace{2mm}j}\bar{\Lambda})(\bar{\Lambda}\Gamma_{ba}R)[-2(\Lambda\Gamma^{ca}\Lambda)(\Lambda\Gamma^{mnib}W) - 2(\Lambda\Gamma^{mnica}\Lambda)(\Lambda\Gamma^{b}W)]\nonumber\\
&=& -\frac{8}{\eta^{3}}(\bar{\Lambda}\Gamma_{mn}R)(\bar{\Lambda}\Gamma_{c}^{\hspace{2mm}j}\bar{\Lambda})(\bar{\Lambda}\Gamma_{ba}R)(\Lambda\Gamma^{ca}\Lambda)(\Lambda\Gamma^{mnib}W)\nonumber\\
&&  -\frac{8}{\eta^{3}}(\bar{\Lambda}\Gamma_{mn}R)(\bar{\Lambda}\Gamma_{c}^{\hspace{2mm}j}\bar{\Lambda})(\bar{\Lambda}\Gamma_{ba}R)(\Lambda\Gamma^{mnica}\Lambda)(\Lambda\Gamma^{b}W)
\end{eqnarray}

\begin{eqnarray}
\{W_{2}^{i}, W_{5}^{j}\} &=& \{-\frac{1}{\eta}(\bar{\Lambda}\Gamma_{mn}R)(\Lambda\Gamma^{mni}W), \frac{1}{\eta^{2}}(\bar{\Lambda}\Gamma_{bd}\bar{\Lambda})(\bar{\Lambda}\Gamma_{ac}R)(\Lambda\Gamma^{ac}\Lambda)(\Lambda\Gamma^{bdj}W)\}\nonumber\\
&=& - \frac{1}{\eta^{3}}(\bar{\Lambda}\Gamma_{mn}R)(\bar{\Lambda}\Gamma_{bd}\bar{\Lambda})(\bar{\Lambda}\Gamma_{ac}R)(\Lambda\Gamma^{ac}\Lambda)[(\Lambda\Gamma^{mni}W), (\Lambda\Gamma^{bdj}W)]\nonumber\\
&=& - \frac{1}{\eta^{3}}(\bar{\Lambda}\Gamma_{mn}R)(\bar{\Lambda}\Gamma_{bd}\bar{\Lambda})(\bar{\Lambda}\Gamma_{ac}R)(\Lambda\Gamma^{ac}\Lambda)[-8\delta^{mn}_{bj} - 8\delta^{mi}_{bd} + 16\delta^{mi}_{bj}]\nonumber\\
&=& \frac{8}{\eta^{3}}(\bar{\Lambda}\Gamma^{bj}R)(\bar{\Lambda}\Gamma_{bd}\bar{\Lambda})(\bar{\Lambda}\Gamma_{ac}R)(\Lambda\Gamma^{ac}\Lambda)N^{id}\nonumber\\
&& + \frac{8}{\eta^{3}}(\bar{\Lambda}\Gamma_{mn}R)(\bar{\Lambda}\Gamma^{mi}\bar{\Lambda})(\bar{\Lambda}\Gamma_{ac}R)(\Lambda\Gamma^{ac}\Lambda)N^{nj}\nonumber\\
&& -\frac{16}{\eta^{3}}(\frac{1}{2})[\eta^{ij}(\bar{\Lambda}\Gamma^{bn}R)(\bar{\Lambda}\Gamma_{b}^{\hspace{2mm}d}\bar{\Lambda}) - (\bar{\Lambda}\Gamma^{jn}R)(\bar{\Lambda}\Gamma^{id}\bar{\Lambda})](\bar{\Lambda}\Gamma_{ac}R)(\Lambda\Gamma^{ac}\Lambda)N_{nd}\nonumber\\
&=& -\frac{8}{\eta^{3}}(\bar{\Lambda}\Gamma^{jd}\bar{\Lambda})(\bar{\Lambda}R)(\bar{\Lambda}\Gamma_{ac}R)(\Lambda\Gamma^{ac}\Lambda)N^{i}_{\hspace{2mm}d}\nonumber\\
&& -\frac{8}{\eta^{3}}(\bar{\Lambda}\Gamma^{id}\bar{\Lambda})(\bar{\Lambda}R)(\bar{\Lambda}\Gamma_{ac}R)(\Lambda\Gamma^{ac}\Lambda)N^{j}_{\hspace{2mm}d}\nonumber\\
&& + \frac{8}{\eta^{3}}(\bar{\Lambda}\Gamma^{nd}\bar{\Lambda})(\bar{\Lambda}R)(\bar{\Lambda}\Gamma_{ac}R)(\Lambda\Gamma^{ac}\Lambda)\eta^{ij}N_{nd}\nonumber\\
&&+\frac{8}{\eta{3}}(\bar{\Lambda}\Gamma^{jn}R)(\bar{\Lambda}\Gamma^{id}\bar{\Lambda})(\bar{\Lambda}\Gamma_{ac}R)(\Lambda\Gamma^{ac}\Lambda)N_{nd}
\end{eqnarray}

\begin{eqnarray}
\{W_{3}^{i}, W_{4}^{j}\} &=& \{\frac{2}{\eta^{2}}(\bar{\Lambda}\Gamma^{ni}\bar{\Lambda})(\bar{\Lambda}\Gamma^{rm}R)(\Lambda\Gamma_{rm}\Lambda)(\Lambda\Gamma_{n}W),-\frac{4}{\eta^{2}}(\bar{\Lambda}\Gamma^{cj}\bar{\Lambda})(\bar{\Lambda}\Gamma^{ba}R)(\Lambda\Gamma_{ca}\Lambda)(\Lambda\Gamma_{b}W)\}\nonumber\\
&=& \frac{16}{\eta^{4}}(\bar{\Lambda}\Gamma^{ni}\bar{\Lambda})(\bar{\Lambda}\Gamma^{rm}R)(\Lambda\Gamma_{rm}\Lambda)(\bar{\Lambda}\Gamma^{cj}\bar{\Lambda})(\bar{\Lambda}\Gamma^{ba}R)(\Lambda\Gamma_{ca}\Lambda)N_{nb}
\end{eqnarray}

\begin{eqnarray}
\{W_{3}^{i}, W_{5}^{j}\} &=& \{\frac{2}{\eta^{2}}(\bar{\Lambda}\Gamma^{ni}\bar{\Lambda})(\bar{\Lambda}\Gamma^{rm}R)(\Lambda\Gamma_{rm}\Lambda)(\Lambda\Gamma_{n}W), \frac{1}{\eta^{2}}(\bar{\Lambda}\Gamma_{bd}\bar{\Lambda})(\bar{\Lambda}\Gamma_{ac}R)(\Lambda\Gamma^{ac}\Lambda)(\Lambda\Gamma^{bdj}W)\}\nonumber\\
&=& 0
\end{eqnarray}

\begin{eqnarray}
\{W_{4}^{i}, W_{5}^{j}\} &=& \{-\frac{4}{\eta^{2}}(\bar{\Lambda}\Gamma^{ri}\bar{\Lambda})(\bar{\Lambda}\Gamma^{nm}R)(\Lambda\Gamma_{rm}\Lambda)(\Lambda\Gamma_{n}W), \frac{1}{\eta^{2}}(\bar{\Lambda}\Gamma_{bd}\bar{\Lambda})(\bar{\Lambda}\Gamma_{ac}R)(\Lambda\Gamma^{ac}\Lambda)(\Lambda\Gamma^{bdj}W)\}\nonumber\\
&=& -\frac{8}{\eta^{4}}(\bar{\Lambda}\Gamma^{ri}\bar{\Lambda})(\bar{\Lambda}\Gamma_{n}^{\hspace{2mm}m}R)(\bar{\Lambda}\Gamma_{bd}\bar{\Lambda})(\bar{\Lambda}\Gamma_{ac}R)(\Lambda\Gamma_{rm}\Lambda)(\Lambda\Gamma^{ac}\Lambda)(\Lambda\Gamma^{bdjn}W)\nonumber\\
&=& -\frac{4}{\eta^{3}}(\bar{\Lambda}\Gamma_{n}^{\hspace{2mm}i}R)(\bar{\Lambda}\Gamma_{bd}\bar{\Lambda})(\bar{\Lambda}\Gamma_{ac}R)(\Lambda\Gamma^{ac}\Lambda)(\Lambda\Gamma^{bdjn}W)
\end{eqnarray}
Putting all together, the result is
\begin{eqnarray}
\{\bar{\Sigma}^{i}, \bar{\Sigma}^{j}\} &=& -\frac{4}{\eta^{2}}(\bar{\Lambda}\Gamma^{in}\bar{\Lambda})(\bar{\Lambda}\Gamma^{cj}R)(\Lambda\Gamma_{cn}D) -\frac{4}{\eta^{2}}(\bar{\Lambda}\Gamma^{jn}\bar{\Lambda})(\bar{\Lambda}\Gamma^{ci}R)(\Lambda\Gamma_{cn}D)\nonumber\\
&& + \frac{2}{\eta^{2}}(\bar{\Lambda}\Gamma_{mn}\bar{\Lambda})(\bar{\Lambda}\Gamma_{c}^{\hspace{2mm}j}R)(\Lambda\Gamma^{cimn}D) + \frac{2}{\eta^{2}}(\bar{\Lambda}\Gamma_{mn}\bar{\Lambda})(\bar{\Lambda}\Gamma_{c}^{\hspace{2mm}i}R)(\Lambda\Gamma^{cjmn}D)\nonumber \\
&&-\frac{4}{\eta^{2}}(\bar{\Lambda}\Gamma^{i}_{\hspace{2mm}n}\bar{\Lambda})(\bar{\Lambda}R)(\Lambda\Gamma^{jn}D) -\frac{4}{\eta^{2}}(\bar{\Lambda}\Gamma^{j}_{\hspace{2mm}n}\bar{\Lambda})(\bar{\Lambda}R)(\Lambda\Gamma^{in}D) +\frac{4}{\eta^{2}}\eta^{ij}(\bar{\Lambda}\Gamma^{mn}\bar{\Lambda})(\bar{\Lambda}R)(\Lambda\Gamma_{mn}D)\nonumber\\
&& -\frac{8}{\eta^{2}}(\bar{\Lambda}\Gamma^{jd}R)(\bar{\Lambda}R)N^{i}_{\hspace{2mm}d} - \frac{4}{\eta^{2}}(\bar{\Lambda}\Gamma^{jd}\bar{\Lambda})(RR)N^{i}_{\hspace{2mm}d}\nonumber\\
&& +\frac{8}{\eta^{2}}(\bar{\Lambda}\Gamma^{di}R)(\bar{\Lambda}R)N^{j}_{\hspace{2mm}d} + \frac{4}{\eta^{2}}(\bar{\Lambda}\Gamma^{di}\bar{\Lambda})(RR)N^{j}_{\hspace{2mm}d}\nonumber\\
&& + \frac{8}{\eta^{2}}\eta^{ij}(\bar{\Lambda}\Gamma^{md}R)(\bar{\Lambda}R)N_{md} + \frac{4}{\eta^{2}}\eta^{ij}(\bar{\Lambda}\Gamma^{md}\bar{\Lambda})(RR)N_{md} + \frac{8}{\eta^{2}}(\bar{\Lambda}\Gamma^{mj}R)(\bar{\Lambda}\Gamma^{id}R)N_{md}\nonumber \\
&& + \frac{32}{\eta^{3}}(\bar{\Lambda}\Gamma_{d}^{\hspace{2mm}j}\bar{\Lambda})(\bar{\Lambda}\Gamma_{rs}R)(\bar{\Lambda}R)(\Lambda\Gamma^{ri}\Lambda)N^{sd} + \frac{16}{\eta^{3}}(\bar{\Lambda}\Gamma_{sd}\bar{\Lambda})(\bar{\Lambda}\Gamma^{rj}R)(\bar{\Lambda}R)(\Lambda\Gamma_{r}^{\hspace{2mm}i}\Lambda)N^{sd}\nonumber\\
&& -\frac{16}{\eta^{3}}(\bar{\Lambda}\Gamma_{d}^{\hspace{2mm}j}\bar{\Lambda})(\bar{\Lambda}\Gamma_{rs}\bar{\Lambda})(RR)(\Lambda\Gamma^{ri}\Lambda)N^{ds} + \frac{32}{\eta^{3}}(\bar{\Lambda}\Gamma_{d}^{\hspace{2mm}i}\bar{\Lambda})(\bar{\Lambda}\Gamma_{rs}R)(\bar{\Lambda}R)(\Lambda\Gamma^{rj}\Lambda)N^{sd}\nonumber\\
&& + \frac{16}{\eta^{3}}(\bar{\Lambda}\Gamma_{sd}\bar{\Lambda})(\bar{\Lambda}\Gamma^{ri}R)(\bar{\Lambda}R)(\Lambda\Gamma_{r}^{\hspace{2mm}j}\Lambda)N^{sd} -\frac{16}{\eta^{3}}(\bar{\Lambda}\Gamma_{d}^{\hspace{2mm}i}\bar{\Lambda})(\bar{\Lambda}\Gamma_{rs}\bar{\Lambda})(RR)(\Lambda\Gamma^{rj}\Lambda)N^{ds}\nonumber\\
&& -\frac{16}{\eta^{3}}(\bar{\Lambda}\Gamma_{mn}\bar{\Lambda})(\bar{\Lambda}\Gamma_{rs}R)(\bar{\Lambda}\Gamma^{bj}R)(\Lambda\Gamma^{ri}\Lambda)(\Lambda\Gamma^{mns}_{\hspace{6mm}b}W)\nonumber\\
&& -\frac{16}{\eta^{3}}(\bar{\Lambda}\Gamma_{mn}\bar{\Lambda})(\bar{\Lambda}\Gamma_{rs}R)(\bar{\Lambda}\Gamma^{bi}R)(\Lambda\Gamma^{rj}\Lambda)(\Lambda\Gamma^{mns}_{\hspace{6mm}b}W)\nonumber\\
&& + \frac{8}{\eta^{4}}(\bar{\Lambda}\Gamma^{j}_{\hspace{2mm}r}\bar{\Lambda})(\bar{\Lambda}\Gamma^{nd}\bar{\Lambda})(\bar{\Lambda}R)(\bar{\Lambda}\Gamma^{ac}R)(\Lambda\Gamma_{ac}\Lambda)(\Lambda\Gamma^{ri}\Lambda)N_{nd}\nonumber\\
&& + \frac{8}{\eta^{4}}(\bar{\Lambda}\Gamma^{i}_{\hspace{2mm}r}\bar{\Lambda})(\bar{\Lambda}\Gamma^{nd}\bar{\Lambda})(\bar{\Lambda}R)(\bar{\Lambda}\Gamma^{ac}R)(\Lambda\Gamma_{ac}\Lambda)(\Lambda\Gamma^{rj}\Lambda)N_{nd}\nonumber\\
&& + \frac{4}{\eta^{3}}(\bar{\Lambda}\Gamma_{mn}R)(\bar{\Lambda}\Gamma_{b}^{\hspace{2mm}j}\bar{\Lambda})(\bar{\Lambda}\Gamma^{ca}R)(\Lambda\Gamma_{ca}\Lambda)(\Lambda\Gamma^{mnib}W)\nonumber\\
&& + \frac{4}{\eta^{3}}(\bar{\Lambda}\Gamma_{mn}R)(\bar{\Lambda}\Gamma_{b}^{\hspace{2mm}i}\bar{\Lambda})(\bar{\Lambda}\Gamma^{ca}R)(\Lambda\Gamma_{ca}\Lambda)(\Lambda\Gamma^{mnjb}W)\nonumber\\
&&  -\frac{8}{\eta^{3}}(\bar{\Lambda}\Gamma_{mn}R)(\bar{\Lambda}\Gamma_{c}^{\hspace{2mm}j}\bar{\Lambda})(\bar{\Lambda}\Gamma_{ba}R)(\Lambda\Gamma^{ca}\Lambda)(\Lambda\Gamma^{mnib}W)\nonumber\\
&&  -\frac{8}{\eta^{3}}(\bar{\Lambda}\Gamma_{mn}R)(\bar{\Lambda}\Gamma_{c}^{\hspace{2mm}j}\bar{\Lambda})(\bar{\Lambda}\Gamma_{ba}R)(\Lambda\Gamma^{mnica}\Lambda)(\Lambda\Gamma^{b}W)\nonumber\\
&&  -\frac{8}{\eta^{3}}(\bar{\Lambda}\Gamma_{mn}R)(\bar{\Lambda}\Gamma_{c}^{\hspace{2mm}i}\bar{\Lambda})(\bar{\Lambda}\Gamma_{ba}R)(\Lambda\Gamma^{ca}\Lambda)(\Lambda\Gamma^{mnjb}W)\nonumber\\
&&  -\frac{8}{\eta^{3}}(\bar{\Lambda}\Gamma_{mn}R)(\bar{\Lambda}\Gamma_{c}^{\hspace{2mm}i}\bar{\Lambda})(\bar{\Lambda}\Gamma_{ba}R)(\Lambda\Gamma^{mnjca}\Lambda)(\Lambda\Gamma^{b}W)\nonumber\\
&& -\frac{16}{\eta^{3}}(\bar{\Lambda}\Gamma^{jd}\bar{\Lambda})(\bar{\Lambda}R)(\bar{\Lambda}\Gamma_{ac}R)(\Lambda\Gamma^{ac}\Lambda)N^{i}_{\hspace{2mm}d} -\frac{16}{\eta^{3}}(\bar{\Lambda}\Gamma^{id}\bar{\Lambda})(\bar{\Lambda}R)(\bar{\Lambda}\Gamma_{ac}R)(\Lambda\Gamma^{ac}\Lambda)N^{j}_{\hspace{2mm}d}\nonumber\\
&& + \frac{16}{\eta^{3}}(\bar{\Lambda}\Gamma^{nd}\bar{\Lambda})(\bar{\Lambda}R)(\bar{\Lambda}\Gamma_{ac}R)(\Lambda\Gamma^{ac}\Lambda)\eta^{ij}N_{nd} +\frac{16}{\eta{3}}(\bar{\Lambda}\Gamma^{jn}R)(\bar{\Lambda}\Gamma^{id}\bar{\Lambda})(\bar{\Lambda}\Gamma_{ac}R)(\Lambda\Gamma^{ac}\Lambda)N_{nd}\nonumber\\ 
&& +\frac{16}{\eta^{4}}(\bar{\Lambda}\Gamma^{ni}\bar{\Lambda})(\bar{\Lambda}\Gamma^{rm}R)(\Lambda\Gamma_{rm}\Lambda)(\bar{\Lambda}\Gamma^{cj}\bar{\Lambda})(\bar{\Lambda}\Gamma^{ba}R)(\Lambda\Gamma_{ca}\Lambda)N_{nb}\nonumber\\
&& + \frac{16}{\eta^{4}}(\bar{\Lambda}\Gamma^{nj}\bar{\Lambda})(\bar{\Lambda}\Gamma^{rm}R)(\Lambda\Gamma_{rm}\Lambda)(\bar{\Lambda}\Gamma^{ci}\bar{\Lambda})(\bar{\Lambda}\Gamma^{ba}R)(\Lambda\Gamma_{ca}\Lambda)N_{nb}\nonumber\\
&& -\frac{4}{\eta^{3}}(\bar{\Lambda}\Gamma_{n}^{\hspace{2mm}i}R)(\bar{\Lambda}\Gamma_{bd}\bar{\Lambda})(\bar{\Lambda}\Gamma_{ac}R)(\Lambda\Gamma^{ac}\Lambda)(\Lambda\Gamma^{bdjn}W)\nonumber\\
&& -\frac{4}{\eta^{3}}(\bar{\Lambda}\Gamma_{n}^{\hspace{2mm}j}R)(\bar{\Lambda}\Gamma_{bd}\bar{\Lambda})(\bar{\Lambda}\Gamma_{ac}R)(\Lambda\Gamma^{ac}\Lambda)(\Lambda\Gamma^{bdin}W)
\end{eqnarray}
One of the useful things that can be extracted from this result is the fact that $\{\bar{\Sigma}^{i}, \bar{\Sigma}^{j}\}$ depends linearly and quadratically on $R_{\alpha}$. This allows us to find the $R_{\alpha}$-dependence of $\{b , b\}$ which it turns out to be of the form:
\begin{equation}\label{eq503}
\{b,b\} = R^{\alpha}f^{(1)}_{\alpha} + \ldots + R^{\alpha}R^{\beta}R^{\delta}R^{\sigma}R^{\rho}R^{\lambda}f^{(6)}_{\alpha\beta\delta\sigma\rho\lambda}
\end{equation}
where $f^{(i)}_{\alpha_{1}\ldots\alpha_{i}}$ for $i=1,\ldots ,6$ are functions of pure spinor variables $\Lambda^{\alpha}, \bar{\Lambda}_{\alpha}, W_{\alpha}$ and the fermionic constraints $D_{\alpha}$.

This can be used to check that $\{b,b\} = Q\Omega$ where $\Omega$ is an arbitrary function of pure spinor variables and the constraints $D_{\alpha}$. To see this let us expand $\Omega$ in terms of $R^{\alpha}$:
\begin{equation}
\Omega = \Omega^{(0)} + R^{\alpha}\Omega_{\alpha}^{(1)} + R^{\alpha\beta}\Omega^{(2)}_{\alpha\beta} + \ldots + R^{\alpha_{1}\ldots\alpha_{23}}\Omega^{(23)}_{\alpha_{1}\ldots\alpha_{23}} 
\end{equation}
Thus the action of the BRST operator $Q = Q_{0} + R^{\alpha}\bar{W}_{\alpha}$ on $\Omega$ gives us
\begin{eqnarray}
Q\Omega = Q_{0}\Omega^{(0)} + R^{\alpha}(\frac{\partial}{\partial\bar{\Lambda}^{\alpha}}\Omega^{(0)} + Q_{0}\Omega^{(1)}_{\alpha}) + R^{\alpha}R^{\beta}(\frac{\partial}{\partial\bar{\Lambda}^{\alpha}}\Omega_{\beta}^{(1)} + Q_{0}\Omega_{\alpha\beta}^{(2)}) + \ldots
\end{eqnarray} 
The comparison of this result with the equation \eqref{eq503} determines the functions $\Omega^{(k)}$ for $k=1,\ldots,23$:
\begin{eqnarray}
0 &=& Q_{0}\Omega^{(0)}\\
f^{(1)}_{\alpha} &=& \frac{\partial}{\partial \bar{\Lambda}^{\alpha}}\Omega^{(0)} + Q_{0}\Omega^{(1)}_{\alpha}\\
f^{(2)}_{\alpha\beta} &=& \frac{\partial}{\partial \bar{\Lambda}^{\alpha}}\Omega^{(1)}_{\beta} + Q_{0}\Omega^{(2)}_{\alpha\beta}\\
&\vdots & \nonumber
\end{eqnarray}
Therefore if we make the following definitions:
\begin{eqnarray}
\Omega^{(0)} &=& \bar{\Lambda}^{\alpha}f_{\alpha}^{(1)}\\
\Omega^{(1)}_{\beta} &=& \bar{\Lambda}^{\alpha}f_{\alpha\beta}^{(2)}\\
\Omega^{(2)}_{\beta\delta} &=& \bar{\Lambda}^{\alpha}f_{\alpha\beta\delta}^{(3)}\\
&\vdots & \nonumber\\
\Omega^{(5)}_{\beta\delta\sigma\rho\lambda} &=& \bar{\Lambda}^{\alpha}f_{\alpha\beta\delta\sigma\rho\lambda}^{(6)}\\
\Omega^{(6)}_{\beta\delta\sigma\rho\lambda\gamma} &=& 0\\
&\vdots & \nonumber\\
\Omega^{(23)} &=& 0
\end{eqnarray}
the equations above are automatically solved.
\section{Expanding the simplified $D=11$ $b$-ghost}\label{apE}
In this Appendix we will reproduce the terms contained in $O(\bar{\Sigma}^{2})$ in the expression for the simplified $D=11$ $b$-ghost. First we will reproduce the quadratic term in $D_{\alpha}$ in the expression for the $b$-ghost \eqref{eq16}
\begin{equation}\label{eq3}
\eta^{-2}L^{(1)}_{ab,cd}(\Lambda\Gamma^{a}D)(\Lambda\Gamma^{bcd}D)
\end{equation}
We will work with the expression
\begin{equation}\label{eq14}
b_{simpl} = P^{i}\bar{\Sigma}_{i} -\frac{4}{\eta^{2}}(\bar{\Lambda}\Gamma_{ab}\bar{\Lambda})(\bar{\Lambda}\Gamma_{cd}R)(\Lambda\Gamma^{aj}\Lambda)\bar{\Sigma}_{j}[(\Lambda\Gamma^{bd}\Lambda)\bar{\Sigma}^{c} + \frac{1}{\eta}(\Lambda\Gamma^{bd}\Lambda)(\bar{\Lambda}\Gamma^{cs}\bar{\Lambda})(\Lambda\Gamma_{sk}\Lambda)\bar{\Sigma}^{k}]
\end{equation}
It is useful to write $\bar{\Sigma}^{i}_{0}$ in the convenient way:
\begin{equation}
\bar{\Sigma}^{i}_{0} = \frac{1}{2\eta}[(\bar{\Lambda}\Gamma_{ab}\bar{\Lambda})(\Lambda\Gamma^{i}\Gamma^{ab}D) + 4(\bar{\Lambda}\Gamma^{ai}\bar{\Lambda})(\Lambda\Gamma_{a}D)]
\end{equation}
Therefore we have
\begin{equation}\label{eq13}
(\Lambda\Gamma_{ij}\Lambda)\bar{\Sigma}^{j}_{0} = \frac{2}{\eta}(\Lambda\Gamma_{ij}\Lambda)(\bar{\Lambda}\Gamma^{aj}\bar{\Lambda})(\Lambda\Gamma_{a}D)
\end{equation}
which is a direct consequence of the identity \eqref{app1}. Now we will expand $\bar{\Sigma}^{i}$ as it was done in \eqref{eeq22}:
\begin{align}\label{eq22}
\bar{\Sigma}^{i} & = \bar{\Sigma}_{0}^{i} + \frac{2}{\eta^{2}}(\bar{\Lambda}\Gamma_{ab}\bar{\Lambda})(\bar{\Lambda}\Gamma_{cd}R)(\Lambda\Gamma^{abcki}\Lambda)N^{d}_{\hspace{2mm}k} + \frac{2}{3\eta^{2}}(\bar{\Lambda}\Gamma_{ab}\bar{\Lambda})(\bar{\Lambda}\Gamma_{c}^{\hspace{2mm}i}R)(\Lambda\Gamma^{abcqj}\Lambda)N_{qj} \notag \\
&  - \frac{2}{3\eta^{2}}(\bar{\Lambda}\Gamma_{ac}\bar{\Lambda})(\bar{\Lambda}R)(\Lambda\Gamma^{aicqj}\Lambda)N_{qj}
\end{align}
Using this equation we can write $\bar{\Sigma}^{j}$, $\bar{\Sigma}^{c}$, $\bar{\Sigma}^{k}$ in the following way:
\begin{align}
\bar{\Sigma}^{j} & = \bar{\Sigma}_{0}^{j} + \frac{2}{\eta^{2}}(\bar{\Lambda}\Gamma_{ef}\bar{\Lambda})(\bar{\Lambda}\Gamma_{gh}R)(\Lambda\Gamma^{efgij}\Lambda)N^{h}_{\hspace{2mm}i} + \frac{2}{3\eta^{2}}(\bar{\Lambda}\Gamma_{ef}\bar{\Lambda})(\bar{\Lambda}\Gamma_{g}^{\hspace{2mm}j}R)(\Lambda\Gamma^{efghi}\Lambda)N_{hi} \notag \\
&  - \frac{2}{3\eta^{2}}(\bar{\Lambda}\Gamma_{eg}\bar{\Lambda})(\bar{\Lambda}R)(\Lambda\Gamma^{ejghi}\Lambda)N_{hi}
\end{align}
\begin{align}
\bar{\Sigma}^{c} & = \bar{\Sigma}_{0}^{c} + \frac{2}{\eta^{2}}(\bar{\Lambda}\Gamma_{lm}\bar{\Lambda})(\bar{\Lambda}\Gamma_{np}R)(\Lambda\Gamma^{lmnqc}\Lambda)N^{p}_{\hspace{2mm}q} + \frac{2}{3\eta^{2}}(\bar{\Lambda}\Gamma_{lm}\bar{\Lambda})(\bar{\Lambda}\Gamma_{n}^{\hspace{2mm}c}R)(\Lambda\Gamma^{lmnpq}\Lambda)N_{pq} \notag \\
&  - \frac{2}{3\eta^{2}}(\bar{\Lambda}\Gamma_{ln}\bar{\Lambda})(\bar{\Lambda}R)(\Lambda\Gamma^{lcnpq}\Lambda)N_{pq}
\end{align}
\begin{align}
\bar{\Sigma}^{k} & = \bar{\Sigma}_{0}^{k} + \frac{2}{\eta^{2}}(\bar{\Lambda}\Gamma_{rt}\bar{\Lambda})(\bar{\Lambda}\Gamma_{uw}R)(\Lambda\Gamma^{rtuyk}\Lambda)N^{w}_{\hspace{2mm}y} + \frac{2}{3\eta^{2}}(\bar{\Lambda}\Gamma_{rt}\bar{\Lambda})(\bar{\Lambda}\Gamma_{u}^{\hspace{2mm}k}R)(\Lambda\Gamma^{rtuwy}\Lambda)N_{wy} \notag \\
&  - \frac{2}{3\eta^{2}}(\bar{\Lambda}\Gamma_{ru}\bar{\Lambda})(\bar{\Lambda}R)(\Lambda\Gamma^{rkuwy}\Lambda)N_{wy}
\end{align}
Replacing these expressions in \eqref{eq14} we have
\begin{align}\label{eq19}
b_{simpl} & = P^{i}\bar{\Sigma}_{i} - \frac{4}{\eta^{2}}(\bar{\Lambda}\Gamma_{ab}\bar{\Lambda})(\bar{\Lambda}\Gamma_{cd}R)(\Lambda\Gamma^{bd}\Lambda)(\Lambda\Gamma^{a}_{\hspace{2mm}j}\Lambda)[\bar{\Sigma}_{0}^{j} +  \frac{2}{\eta^{2}}(\bar{\Lambda}\Gamma_{ef}\bar{\Lambda})(\bar{\Lambda}\Gamma_{gh}R)(\Lambda\Gamma^{efgij}\Lambda)N^{h}_{\hspace{2mm}i} \notag \\
& + \frac{2}{3\eta^{2}}(\bar{\Lambda}\Gamma_{ef}\bar{\Lambda})(\bar{\Lambda}\Gamma_{g}^{\hspace{2mm}j}R)(\Lambda\Gamma^{efghi}\Lambda)N_{hi} \notag - \frac{2}{3\eta^{2}}(\bar{\Lambda}\Gamma_{eg}\bar{\Lambda})(\bar{\Lambda}R)(\Lambda\Gamma^{ejghi}\Lambda)N_{hi}]\times\\
& \{\bar{\Sigma}_{0}^{c} + \frac{2}{\eta^{2}}(\bar{\Lambda}\Gamma_{lm}\bar{\Lambda})(\bar{\Lambda}\Gamma_{np}R)(\Lambda\Gamma^{lmnqc}\Lambda)N^{p}_{\hspace{2mm}q} + \frac{2}{3\eta^{2}}(\bar{\Lambda}\Gamma_{lm}\bar{\Lambda})(\bar{\Lambda}\Gamma_{n}^{\hspace{2mm}c}R)(\Lambda\Gamma^{lmnpq}\Lambda)N_{pq} \notag \\
&  - \frac{2}{3\eta^{2}}(\bar{\Lambda}\Gamma_{ln}\bar{\Lambda})(\bar{\Lambda}R)(\Lambda\Gamma^{lcnpq}\Lambda)N_{pq} + \frac{1}{\eta}(\bar{\Lambda}\Gamma^{c}_{\hspace{2mm}s}\bar{\Lambda})(\Lambda\Gamma^{s}_{\hspace{2mm}k}\Lambda)[\bar{\Sigma}_{0}^{k} + \frac{2}{\eta^{2}}(\bar{\Lambda}\Gamma_{rt}\bar{\Lambda})(\bar{\Lambda}\Gamma_{uw}R)(\Lambda\Gamma^{rtuyk}\Lambda)N^{w}_{\hspace{2mm}y} \notag \\
&  + \frac{2}{3\eta^{2}}(\bar{\Lambda}\Gamma_{rt}\bar{\Lambda})(\bar{\Lambda}\Gamma_{u}^{\hspace{2mm}k}R)(\Lambda\Gamma^{rtuwy}\Lambda)N_{wy} \notag - \frac{2}{3\eta^{2}}(\bar{\Lambda}\Gamma_{ru}\bar{\Lambda})(\bar{\Lambda}R)(\Lambda\Gamma^{rkuwy}\Lambda)N_{wy}]\}\\
\end{align}

\vspace{2mm}
Hence the contributions proportional to $D^{2}$ are:
\begin{eqnarray*}
b^{(2)}_{simp} &=& -\frac{4}{\eta^{2}}(\bar{\Lambda}\Gamma_{ab}\bar{\Lambda})(\bar{\Lambda}\Gamma_{cd}R)(\Lambda\Gamma^{aj}\Lambda)\bar{\Sigma}_{0\,j}[(\Lambda\Gamma^{bd}\Lambda)\bar{\Sigma}_{0}^{c} + \frac{1}{\eta}(\Lambda\Gamma^{bd}\Lambda)(\bar{\Lambda}\Gamma^{cs}\bar{\Lambda})(\Lambda\Gamma_{sk}\Lambda)\bar{\Sigma}_{0}^{k}]\\
&=& -\frac{4}{\eta^{2}}(\bar{\Lambda}\Gamma_{ab}\bar{\Lambda})(\bar{\Lambda}\Gamma_{cd}R)(\Lambda\Gamma^{aj}\Lambda)(\Lambda\Gamma^{bd}\Lambda)\bar{\Sigma}_{0\,j}\bar{\Sigma}_{0}^{c} \\
&& -\frac{4}{\eta^{3}}(\bar{\Lambda}\Gamma_{ab}\bar{\Lambda})(\bar{\Lambda}\Gamma_{cd}R)(\Lambda\Gamma^{aj}\Lambda)(\Lambda\Gamma^{bd}\Lambda)(\bar{\Lambda}\Gamma^{cs}\bar{\Lambda})(\Lambda\Gamma_{sk}\Lambda)\bar{\Sigma}_{0\,j}\bar{\Sigma}_{0}^{k}\\
&=& -\frac{8}{\eta^{3}}(\bar{\Lambda}\Gamma_{ab}\bar{\Lambda})(\bar{\Lambda}\Gamma_{cd}\bar{\Lambda})(\Lambda\Gamma^{bd}\Lambda)(\Lambda\Gamma^{aj}\Lambda)(\bar{\Lambda}\Gamma_{mj}\bar{\Lambda})(\Lambda\Gamma^{m}D)\bar{\Sigma}_{0}^{c}\\
&& - \frac{2}{\eta^{2}}(\bar{\Lambda}\Gamma_{ab}\bar{\Lambda})(\bar{\Lambda}R)(\Lambda\Gamma^{bk}\Lambda)(\Lambda\Gamma^{aj}\Lambda)\bar{\Sigma}_{0\,j}\bar{\Sigma}_{0\,k}\\
&=& -\frac{2}{\eta^{3}}(\bar{\Lambda}\Gamma_{ac}R)(\Lambda\Gamma^{aj}\Lambda)(\bar{\Lambda}\Gamma_{mj}\bar{\Lambda})(\Lambda\Gamma^{m}D)[(\bar{\Lambda}\Gamma_{rs}\bar{\Lambda})(\Lambda\Gamma^{rsc}D) + 2(\bar{\Lambda}\Gamma_{r}^{\hspace{2mm}c}\bar{\Lambda})(\Lambda\Gamma^{r}D)]\\
&& - \frac{1}{\eta}(\bar{\Lambda}R)(\Lambda\Gamma^{jk}\Lambda)\bar{\Sigma}_{0\,j}\bar{\Sigma}_{0\,k}\\
&=& -\frac{2}{\eta^{3}}(\bar{\Lambda}\Gamma_{ac}R)(\Lambda\Gamma^{aj}\Lambda)(\bar{\Lambda}\Gamma_{mj}\bar{\Lambda})(\bar{\Lambda}\Gamma_{rs}\bar{\Lambda})(\Lambda\Gamma^{m}D)(\Lambda\Gamma^{rsc}D)\\
&& - \frac{4}{\eta^{3}}(\bar{\Lambda}R)(\Lambda\Gamma^{aj}\Lambda)(\bar{\Lambda}\Gamma_{mj}\bar{\Lambda})(\bar{\Lambda}\Gamma^{ra}\bar{\Lambda})(\Lambda\Gamma^{m}D)(\Lambda\Gamma^{r}D)\\
&& -\frac{4}{\eta^{3}}(\bar{\Lambda}R)(\Lambda\Gamma^{jk}\Lambda)(\bar{\Lambda}\Gamma_{mj}\bar{\Lambda})(\bar{\Lambda}\Gamma_{nk}\bar{\Lambda})(\Lambda\Gamma^{m}D)(\Lambda\Gamma^{n}D)\\
&=& -\frac{1}{\eta^{2}}(\bar{\Lambda}\Gamma_{rs}\bar{\Lambda})(\bar{\Lambda}\Gamma_{mc}R)(\Lambda\Gamma^{m}D)(\Lambda\Gamma^{rsc}D) + \frac{2}{\eta^{2}}(\bar{\Lambda}R)(\bar{\Lambda}\Gamma^{mr}\bar{\Lambda})(\Lambda\Gamma^{m}D)(\Lambda\Gamma^{r}D)\\
&& - \frac{2}{\eta^{2}}(\bar{\Lambda}R)(\bar{\Lambda}\Gamma^{mr}\bar{\Lambda})(\Lambda\Gamma^{m}D)(\Lambda\Gamma^{r}D)\\
&=& -\frac{1}{\eta^{2}}(\bar{\Lambda}\Gamma_{rs}\bar{\Lambda})(\bar{\Lambda}\Gamma_{mc}R)(\Lambda\Gamma^{m}D)(\Lambda\Gamma^{rsc}D)\\
&=& \frac{1}{\eta^{2}}L^{(1)}_{mc,rs}(\Lambda\Gamma^{m}D)(\Lambda\Gamma^{crs}D)
\end{eqnarray*}
where we have used the equation \eqref{eq13} and the identities \eqref{app3}, \eqref{app4}. 

\vspace{2mm}
The next term to be computed is that proportional to $\eta^{-3}$.  This term can be calculated in two steps. First we focus on the part proportional to $(\Lambda\Gamma^{a}D)$ which will be called $K_{1}$ and then on the part proportional to $(\Lambda\Gamma^{bcd}D)$ which will be called $K_{2}$. Thus
\begin{align}
K_{1} &= - \frac{4}{\eta^{2}}(\bar{\Lambda}\Gamma_{ab}\bar{\Lambda})(\bar{\Lambda}\Gamma_{cd}R)(\Lambda\Gamma^{bd}\Lambda)(\Lambda\Gamma^{a}_{\hspace{2mm}j}\Lambda)\bar{\Sigma}_{0}^{j}\times\{\frac{2}{\eta^{2}}(\bar{\Lambda}\Gamma_{lm}\bar{\Lambda})(\bar{\Lambda}\Gamma_{np}R)(\Lambda\Gamma^{lmnqc}\Lambda)N^{p}_{\hspace{2mm}q} \notag \\
& + \frac{2}{3\eta^{2}}(\bar{\Lambda}\Gamma_{lm}\bar{\Lambda})(\bar{\Lambda}\Gamma_{n}^{\hspace{2mm}c}R)(\Lambda\Gamma^{lmnpq}\Lambda)N_{pq} - \frac{2}{3\eta^{2}}(\bar{\Lambda}\Gamma_{ln}\bar{\Lambda})(\bar{\Lambda}R)(\Lambda\Gamma^{lcnpq}\Lambda)N_{pq}\notag \\
& + \frac{1}{\eta}(\bar{\Lambda}\Gamma^{c}_{\hspace{2mm}s}\bar{\Lambda})(\Lambda\Gamma^{s}_{\hspace{2mm}k}\Lambda)[\frac{2}{\eta^{2}}(\bar{\Lambda}\Gamma_{rt}\bar{\Lambda})(\bar{\Lambda}\Gamma_{uw}R)(\Lambda\Gamma^{rtuyk}\Lambda)N^{w}_{\hspace{2mm}y} \notag \\
&  + \frac{2}{3\eta^{2}}(\bar{\Lambda}\Gamma_{rt}\bar{\Lambda})(\bar{\Lambda}\Gamma_{u}^{\hspace{2mm}k}R)(\Lambda\Gamma^{rtuwy}\Lambda)N_{wy} \notag - \frac{2}{3\eta^{2}}(\bar{\Lambda}\Gamma_{ru}\bar{\Lambda})(\bar{\Lambda}R)(\Lambda\Gamma^{rkuwy}\Lambda)N_{wy}]\}\\ 
\end{align}
We can use the identity \eqref{eq13} to simplify this expression:
\begin{align}
K_{1} &= - \frac{4}{\eta^{2}}(\bar{\Lambda}\Gamma_{ab}\bar{\Lambda})(\bar{\Lambda}\Gamma_{cd}R)(\Lambda\Gamma^{bd}\Lambda)(\Lambda\Gamma^{a}_{\hspace{2mm}j}\Lambda)\bar{\Sigma}_{0}^{j}\times\{\frac{2}{\eta^{2}}(\bar{\Lambda}\Gamma_{lm}\bar{\Lambda})(\bar{\Lambda}\Gamma_{np}R)(\Lambda\Gamma^{lmnqc}\Lambda)N^{p}_{\hspace{2mm}q} \notag \\
& + \frac{2}{3\eta^{2}}(\bar{\Lambda}\Gamma_{lm}\bar{\Lambda})(\bar{\Lambda}\Gamma_{n}^{\hspace{2mm}c}R)(\Lambda\Gamma^{lmnpq}\Lambda)N_{pq} - \frac{2}{3\eta^{2}}(\bar{\Lambda}\Gamma_{ln}\bar{\Lambda})(\bar{\Lambda}R)(\Lambda\Gamma^{lcnpq}\Lambda)N_{pq}\notag \\
& + \frac{1}{\eta}(\bar{\Lambda}\Gamma^{c}_{\hspace{2mm}s}\bar{\Lambda})(\Lambda\Gamma^{s}_{\hspace{2mm}k}\Lambda)[\frac{2}{3\eta^{2}}(\bar{\Lambda}\Gamma_{rt}\bar{\Lambda})(\bar{\Lambda}\Gamma_{u}^{\hspace{2mm}k}R)(\Lambda\Gamma^{rtuwy}\Lambda)N_{wy}]\} \notag \\
 &= - \frac{8}{\eta^{3}}(\bar{\Lambda}\Gamma_{ab}\bar{\Lambda})(\bar{\Lambda}\Gamma_{cd}R)(\Lambda\Gamma^{bd}\Lambda)(\Lambda\Gamma^{aj}\Lambda)(\bar{\Lambda}\Gamma_{xj}\bar{\Lambda})(\Lambda\Gamma^{x}D)\times\{\frac{2}{\eta^{2}}(\bar{\Lambda}\Gamma_{lm}\bar{\Lambda})(\bar{\Lambda}\Gamma_{np}R)(\Lambda\Gamma^{lmnqc}\Lambda)N^{p}_{\hspace{2mm}q} \notag \\
& + \frac{2}{3\eta^{2}}(\bar{\Lambda}\Gamma_{lm}\bar{\Lambda})(\bar{\Lambda}\Gamma_{n}^{\hspace{2mm}c}R)(\Lambda\Gamma^{lmnpq}\Lambda)N_{pq} - \frac{2}{3\eta^{2}}(\bar{\Lambda}\Gamma_{ln}\bar{\Lambda})(\bar{\Lambda}R)(\Lambda\Gamma^{lcnpq}\Lambda)N_{pq}\notag \\
& + \frac{2}{3\eta^{3}}(\bar{\Lambda}\Gamma^{c}_{\hspace{2mm}s}\bar{\Lambda})(\Lambda\Gamma^{s}_{\hspace{2mm}k}\Lambda)(\bar{\Lambda}\Gamma_{rt}\bar{\Lambda})(\bar{\Lambda}\Gamma_{u}^{\hspace{2mm}k}R)(\Lambda\Gamma^{rtuwy}\Lambda)N_{wy}\} \notag \\
&= -\frac{4}{\eta^{2}}(\bar{\Lambda}\Gamma_{xb}\bar{\Lambda})(\bar{\Lambda}\Gamma_{cd}R)(\Lambda\Gamma^{x}D)(\Lambda\Gamma^{bd}\Lambda)\times\{\frac{2}{\eta^{2}}(\bar{\Lambda}\Gamma_{lm}\bar{\Lambda})(\bar{\Lambda}\Gamma_{np}R)(\Lambda\Gamma^{lmnqc}\Lambda)N^{p}_{\hspace{2mm}q} \notag \\
& + \frac{2}{3\eta^{2}}(\bar{\Lambda}\Gamma_{lm}\bar{\Lambda})(\bar{\Lambda}\Gamma_{n}^{\hspace{2mm}c}R)(\Lambda\Gamma^{lmnpq}\Lambda)N_{pq} - \frac{2}{3\eta^{2}}(\bar{\Lambda}\Gamma_{ln}\bar{\Lambda})(\bar{\Lambda}R)(\Lambda\Gamma^{lcnpq}\Lambda)N_{pq}\notag \\
& + \frac{2}{3\eta^{3}}(\bar{\Lambda}\Gamma^{c}_{\hspace{2mm}s}\bar{\Lambda})(\Lambda\Gamma^{s}_{\hspace{2mm}k}\Lambda)(\bar{\Lambda}\Gamma_{rt}\bar{\Lambda})(\bar{\Lambda}\Gamma_{u}^{\hspace{2mm}k}R)(\Lambda\Gamma^{rtuwy}\Lambda)N_{wy}\}
\end{align}
Now it is useful to use the following identity which is followed from \eqref{app4}:
\begin{align}
(\bar{\Lambda}\Gamma_{xb}\bar{\Lambda})(\bar{\Lambda}\Gamma_{cd}R)(\Lambda\Gamma^{bd}\Lambda) & = [\frac{1}{2}(\bar{\Lambda}\Gamma_{xc}\bar{\Lambda})(\bar{\Lambda}\Gamma_{bd}R) + \frac{1}{2}(\bar{\Lambda}\Gamma_{bd}\bar{\Lambda})(\bar{\Lambda}\Gamma_{xc}R) + (\bar{\Lambda}\Gamma_{cd}\bar{\Lambda})(\bar{\Lambda}\Gamma_{bx}R)](\Lambda\Gamma^{bd}\Lambda)
\end{align}
With the additional use of \eqref{app2} we obtain
\begin{align}
K_{1} & = -\frac{4}{\eta^{3}}(\bar{\Lambda}\Gamma_{xc}R)(\Lambda\Gamma^{x}D)(\bar{\Lambda}\Gamma_{lm}\bar{\Lambda})(\bar{\Lambda}\Gamma_{np}R)(\Lambda\Gamma^{lmnqc}\Lambda)N^{p}_{\hspace{2mm}q} \notag  \\
& -\frac{4}{3\eta^{3}}(\bar{\Lambda}\Gamma_{xc}R)(\Lambda\Gamma^{x}D)(\bar{\Lambda}\Gamma_{lm}\bar{\Lambda})(\bar{\Lambda}\Gamma_{n}^{\hspace{2mm}c}R)(\Lambda\Gamma^{lmnpq}\Lambda)N_{pq} \notag \\
& + \frac{4}{3\eta^{3}}(\bar{\Lambda}\Gamma_{xc}R)(\Lambda\Gamma^{x}D)(\bar{\Lambda}\Gamma^{ln}\bar{\Lambda})(\bar{\Lambda}R)(\Lambda\Gamma^{lcnpq}\Lambda)N_{pq} \notag \\
& +\frac{2}{3\eta^{3}}(\bar{\Lambda}\Gamma^{xc}R)(\Lambda\Gamma_{x}D)(\bar{\Lambda}\Gamma_{cu}\bar{\Lambda})(\bar{\Lambda}\Gamma_{rt}R)(\Lambda\Gamma^{rtupq}\Lambda)N_{pq} \notag \\
& = -\frac{4}{\eta^{3}}(\bar{\Lambda}\Gamma_{lm}\bar{\Lambda})(\bar{\Lambda}\Gamma_{np}R)(\bar{\Lambda}\Gamma^{xc}R)(\Lambda\Gamma^{lmnqc}\Lambda)(\Lambda\Gamma^{x}D)N_{pq} \notag  \\
& + \frac{4}{3\eta^{3}}(\bar{\Lambda}\Gamma_{lm}\bar{\Lambda})(\bar{\Lambda}\Gamma_{xc}R)(\bar{\Lambda}\Gamma_{n}^{\hspace{2mm}c}R)(\Lambda\Gamma^{lmnpq}\Lambda)(\Lambda\Gamma^{x}D))N_{pq} \notag \\
& - \frac{4}{3\eta^{3}}(\bar{\Lambda}\Gamma^{ln}\bar{\Lambda})(\bar{\Lambda}\Gamma_{xc}R)(\bar{\Lambda}R)(\Lambda\Gamma^{lcnpq}\Lambda)(\Lambda\Gamma^{x}D)N_{pq} \notag \\
& - \frac{2}{3\eta^{3}}(\bar{\Lambda}\Gamma_{cu}\bar{\Lambda})(\bar{\Lambda}\Gamma^{xc}R)(\bar{\Lambda}\Gamma_{rt}R)(\Lambda\Gamma^{rtupq}\Lambda)(\Lambda\Gamma_{x}D)N_{pq} \notag \\
&= \frac{4}{\eta^{3}}(\bar{\Lambda}\Gamma_{lm}\bar{\Lambda})(\bar{\Lambda}\Gamma_{cx}R)(\bar{\Lambda}\Gamma_{n}^{\hspace{2mm}p}R)(\Lambda\Gamma^{lmcnq}\Lambda)(\Lambda\Gamma^{x}D)N_{qp} \notag\\
& -\frac{2}{\eta^{3}}(\bar{\Lambda}\Gamma_{ln}\bar{\Lambda})(\bar{\Lambda}\Gamma_{xm}R)(\bar{\Lambda}R)(\Lambda\Gamma^{lmnpq}\Lambda)(\Lambda\Gamma^{x}D)N_{pq} \notag \\
&=\frac{4}{\eta^{3}}(\bar{\Lambda}\Gamma_{lm}\bar{\Lambda})(\bar{\Lambda}\Gamma_{cx}R)(\bar{\Lambda}\Gamma_{n}^{\hspace{2mm}p}R)(\Lambda\Gamma^{lmcnq}\Lambda)(\Lambda\Gamma^{x}D)N_{qp} \notag\\
& -\frac{2}{\eta^{3}}(\bar{\Lambda}\Gamma_{lm}\bar{\Lambda})(\bar{\Lambda}\Gamma_{nx}R)(\bar{\Lambda}R)(\Lambda\Gamma^{lmnpq}\Lambda)(\Lambda\Gamma^{x}D)N_{pq}\label{eq403}
\end{align}

\vspace{2mm}
Now let us move on to compute the term proportional to $(\Lambda\Gamma^{abc}D)$, this term comes from the following contribution:
\begin{align}
K_{2} & = -\frac{4}{\eta^{2}}(\bar{\Lambda}\Gamma_{ab}\bar{\Lambda})(\bar{\Lambda}\Gamma_{cd}R)(\Lambda\Gamma^{bd}\Lambda)(\Lambda\Gamma^{a}_{\hspace{2mm}j}\Lambda)[\frac{2}{\eta^{2}}(\bar{\Lambda}\Gamma_{ef}\bar{\Lambda})(\bar{\Lambda}\Gamma_{gh}R)(\Lambda\Gamma^{efgij}\Lambda)N^{h}_{\hspace{2mm}i} \notag \\
& + \frac{2}{3\eta^{2}}(\bar{\Lambda}\Gamma_{ef}\bar{\Lambda})(\bar{\Lambda}\Gamma_{g}^{\hspace{2mm}j}R)(\Lambda\Gamma^{efghi}\Lambda)N_{hi} - \frac{2}{3\eta^{2}}(\bar{\Lambda}\Gamma_{eg}\bar{\Lambda})(\bar{\Lambda}R)(\Lambda\Gamma^{ejghi}\Lambda)N_{hi}]\times \notag \\
& [\bar{\Sigma}_{0}^{c} + \frac{1}{\eta}(\bar{\Lambda}\Gamma^{cs}\bar{\Lambda})(\Lambda\Gamma^{sk}\Lambda)\bar{\Sigma}_{0\,k}] \notag \\
&= -\frac{2}{\eta}(\bar{\Lambda}\Gamma_{ac}R)(\Lambda\Gamma^{a}_{\hspace{2mm}j}\Lambda)[\frac{2}{3\eta^{2}}(\bar{\Lambda}\Gamma_{ef}\bar{\Lambda})(\bar{\Lambda}\Gamma_{g}^{\hspace{2mm}j}R)(\Lambda\Gamma^{efghi}\Lambda)N_{hi}]\times \notag \\
& [\bar{\Sigma}_{0}^{c} + \frac{1}{\eta}(\bar{\Lambda}\Gamma^{cs}\bar{\Lambda})(\Lambda\Gamma^{sk}\Lambda)\bar{\Sigma}_{0\,k}]
\end{align}
where we have just used the identities \eqref{app4} and \eqref{app11}. Therefore
\begin{align}
K_{2} & = -\frac{4}{3\eta^{3}}(\bar{\Lambda}\Gamma_{ac}R)(\Lambda\Gamma^{a}_{\hspace{2mm}j}\Lambda)[(\bar{\Lambda}\Gamma_{ef}\bar{\Lambda})(\bar{\Lambda}\Gamma_{g}^{\hspace{2mm}j}R)(\Lambda\Gamma^{efghi}\Lambda)N_{hi}]\times \notag \\
& [\frac{1}{2\eta}(\bar{\Lambda}\Gamma_{mn}\bar{\Lambda})(\Lambda\Gamma^{mnc}D)] \notag\\
&= -\frac{2}{3\eta^{4}}(\bar{\Lambda}\Gamma_{ac}R)(\Lambda\Gamma^{a}_{\hspace{2mm}j}\Lambda)[(\bar{\Lambda}\Gamma_{ef}\bar{\Lambda})(\bar{\Lambda}\Gamma_{g}^{\hspace{2mm}j}R)(\Lambda\Gamma^{efghi}\Lambda)N_{hi}]\times \notag \\
& (\bar{\Lambda}\Gamma_{mn}\bar{\Lambda})(\Lambda\Gamma^{mnc}D) \notag\\
&= \frac{2}{3\eta^{4}}(\bar{\Lambda}\Gamma_{ac}R)(\Lambda\Gamma^{a}_{\hspace{2mm}j}\Lambda)[(\bar{\Lambda}\Gamma_{g}^{\hspace{2mm}j}\bar{\Lambda})(\bar{\Lambda}\Gamma_{ef}R)(\Lambda\Gamma^{efghi}\Lambda)N_{hi}]\times \notag \\
& (\bar{\Lambda}\Gamma_{mn}\bar{\Lambda})(\Lambda\Gamma^{mnc}D) \notag \\
& = \frac{1}{3\eta^{3}}(\bar{\Lambda}\Gamma_{mn}\bar{\Lambda})(\bar{\Lambda}\Gamma_{gc}R)(\bar{\Lambda}\Gamma_{ef}R)(\Lambda\Gamma^{efghi}\Lambda)N_{hi}(\Lambda\Gamma^{mnc}D) \notag \\
& = -\frac{1}{3\eta^{3}}(\bar{\Lambda}\Gamma_{gc}\bar{\Lambda})(\bar{\Lambda}\Gamma_{mn}R)(\bar{\Lambda}\Gamma_{ef}R)(\Lambda\Gamma^{efghi}\Lambda)N_{hi}(\Lambda\Gamma^{cmn}D) \notag \\
& = \frac{1}{3\eta^{3}}(\bar{\Lambda}\Gamma_{ef}\bar{\Lambda})(\bar{\Lambda}\Gamma_{mn}R)(\bar{\Lambda}\Gamma_{gc}R)(\Lambda\Gamma^{efghi}\Lambda)N_{hi}(\Lambda\Gamma^{cmn}D) \notag \\
& = -\frac{1}{3\eta^{3}}(\bar{\Lambda}\Gamma_{ef}\bar{\Lambda})(\bar{\Lambda}\Gamma_{gc}R)(\bar{\Lambda}\Gamma_{mn}R)(\Lambda\Gamma^{efghi}\Lambda)N_{hi}(\Lambda\Gamma^{cmn}D) \label{eq21}
\end{align}
Thus the term desired is
\begin{align}
b_{simp}^{(3)} &= \frac{4}{\eta^{3}}(\bar{\Lambda}\Gamma_{lm}\bar{\Lambda})(\bar{\Lambda}\Gamma_{cx}R)(\bar{\Lambda}\Gamma_{n}^{\hspace{2mm}p}R)(\Lambda\Gamma^{lmcnq}\Lambda)(\Lambda\Gamma^{x}D)N_{qp} \notag\\
& -\frac{2}{\eta^{3}}(\bar{\Lambda}\Gamma_{lm}\bar{\Lambda})(\bar{\Lambda}\Gamma_{nx}R)(\bar{\Lambda}R)(\Lambda\Gamma^{lmnpq}\Lambda)(\Lambda\Gamma^{x}D)N_{pq} \notag \\
& -\frac{1}{3\eta^{3}}(\bar{\Lambda}\Gamma_{ef}\bar{\Lambda})(\bar{\Lambda}\Gamma_{gc}R)(\bar{\Lambda}\Gamma_{mn}R)(\Lambda\Gamma^{efghi}\Lambda)N_{hi}(\Lambda\Gamma^{cmn}D)\label{eq404}
\end{align}

\vspace{2mm}
The last term to be calculated is that proportional to $\eta^{-4}$. The relevant terms are (after using \eqref{app11}):
\begin{align}
b_{simp}^{(4)} & = -\frac{4}{\eta^{2}}(\bar{\Lambda}\Gamma_{ab}\bar{\Lambda})(\bar{\Lambda}\Gamma_{cd}R)(\Lambda\Gamma^{bd}\Lambda)(\Lambda\Gamma^{aj}\Lambda)(\frac{2}{3\eta^{2}})(\bar{\Lambda}\Gamma_{ef}\bar{\Lambda})(\bar{\Lambda}\Gamma_{gj}R)(\Lambda\Gamma^{efghi}\Lambda)N_{hi}[\frac{2}{\eta^{2}}(\bar{\Lambda}\Gamma_{lm}\bar{\Lambda})\times \notag\\
&(\bar{\Lambda}\Gamma_{np}R)(\Lambda\Gamma^{lmnqc}\Lambda)N_{pq} + \frac{2}{3\eta^{2}}(\bar{\Lambda}\Gamma_{lm}\bar{\Lambda})(\bar{\Lambda}\Gamma_{n}^{\hspace{2mm}c}R)(\Lambda\Gamma^{lmnpq}\Lambda)N_{pq} - \frac{2}{3\eta^{2}}(\bar{\Lambda}\Gamma_{ln}\bar{\Lambda})(\bar{\Lambda}R)(\Lambda\Gamma^{lcnpq}\Lambda)N_{pq} + \notag \\
& + \frac{2}{3\eta{3}}(\bar{\Lambda}\Gamma_{cs}\bar{\Lambda})(\Lambda\Gamma^{s}_{\hspace{2mm}k}\Lambda)(\bar{\Lambda}\Gamma_{rt}\bar{\Lambda})(\bar{\Lambda}\Gamma_{uk}R)(\Lambda\Gamma^{rtupq}\Lambda)N_{pq}] \notag\\
& = -\frac{4}{3\eta^{3}}(\bar{\Lambda}\Gamma_{ac}R)(\Lambda\Gamma^{aj}\Lambda)(\bar{\Lambda}\Gamma_{ef}\bar{\Lambda})(\bar{\Lambda}\Gamma_{gj}R)(\Lambda\Gamma^{efghi}\Lambda)N_{hi}[\frac{2}{\eta^{2}}(\bar{\Lambda}\Gamma_{lm}\bar{\Lambda})(\bar{\Lambda}\Gamma_{np}R)(\Lambda\Gamma^{lmnqc}\Lambda)N_{pq} \notag \\
& + \frac{2}{3\eta^{2}}(\bar{\Lambda}\Gamma_{lm}\bar{\Lambda})(\bar{\Lambda}\Gamma_{n}^{\hspace{2mm}c}R)(\Lambda\Gamma^{lmnpq}\Lambda)N_{pq} - \frac{2}{3\eta^{2}}(\bar{\Lambda}\Gamma_{ln}\bar{\Lambda})(\bar{\Lambda}R)(\Lambda\Gamma^{lcnpq}\Lambda)N_{pq} \notag \\ 
& + \frac{1}{3\eta{2}}(\bar{\Lambda}\Gamma_{rt}\bar{\Lambda})(\bar{\Lambda}\Gamma^{c}_{\hspace{2mm}u}R)(\Lambda\Gamma^{rtupq}\Lambda)N_{pq}] \notag\\
&= \frac{2}{3\eta^{2}}(\bar{\Lambda}\Gamma_{gc}R)(\bar{\Lambda}\Gamma_{ef}R)(\Lambda\Gamma^{efghi}\Lambda)N_{hi}[\frac{2}{\eta^{2}}(\bar{\Lambda}\Gamma_{lm}\bar{\Lambda})(\bar{\Lambda}\Gamma_{np}R)(\Lambda\Gamma^{lmnqc}\Lambda)N_{pq} \notag \\
& + \frac{1}{3\eta^{2}}(\bar{\Lambda}\Gamma_{lm}\bar{\Lambda})(\bar{\Lambda}\Gamma_{n}^{\hspace{2mm}c}R)(\Lambda\Gamma^{lmnpq}\Lambda)N_{pq} - \frac{2}{3\eta^{2}}(\bar{\Lambda}\Gamma_{ln}\bar{\Lambda})(\bar{\Lambda}R)(\Lambda\Gamma^{lcnpq}\Lambda)N_{pq}] \notag \\
&= \frac{2}{3\eta^{2}}(\bar{\Lambda}\Gamma_{gc}R)(\bar{\Lambda}\Gamma_{ef}R)(\Lambda\Gamma^{efghi}\Lambda)N_{hi}[\frac{2}{\eta^{2}}(\bar{\Lambda}\Gamma_{lm}\bar{\Lambda})(\bar{\Lambda}\Gamma_{np}R)(\Lambda\Gamma^{lmnqc}\Lambda)N_{pq} \notag \\
& - \frac{1}{3\eta^{2}}(\bar{\Lambda}\Gamma_{n}^{\hspace{2mm}c}\bar{\Lambda})(\bar{\Lambda}\Gamma_{lm}R)(\Lambda\Gamma^{lmnpq}\Lambda)N_{pq} - \frac{2}{3\eta^{2}}(\bar{\Lambda}\Gamma_{ln}\bar{\Lambda})(\bar{\Lambda}R)(\Lambda\Gamma^{lcnpq}\Lambda)N_{pq}] \notag \\
& = \frac{4}{3\eta^{4}}(\bar{\Lambda}\Gamma_{lm}\bar{\Lambda})(\bar{\Lambda}\Gamma_{np}R)(\bar{\Lambda}\Gamma_{gc}R)(\bar{\Lambda}\Gamma_{ef}R)(\Lambda\Gamma^{efghi}\Lambda)N_{hi}(\Lambda\Gamma^{lmnqc}\Lambda)N_{pq} \notag \\
& -\frac{2}{9\eta^{4}}(\bar{\Lambda}R)(\bar{\Lambda}\Gamma_{ef}R)(\bar{\Lambda}\Gamma_{ng}\bar{\Lambda})(\bar{\Lambda}\Gamma_{lm}R)(\Lambda\Gamma^{efghi}\Lambda)N_{hi}(\Lambda\Gamma^{lmnpq}\Lambda)N_{pq} \notag\\
& - \frac{4}{9\eta^{4}}(\bar{\Lambda}\Gamma_{gm}R)(\bar{\Lambda}\Gamma_{ef}R)(\bar{\Lambda}\Gamma_{ln}\bar{\Lambda})(\bar{\Lambda}R)(\Lambda\Gamma^{efghi}\Lambda)N_{hi}(\Lambda\Gamma^{lmnpq}\Lambda)N_{pq} \notag \\
& = \frac{4}{3\eta^{4}}(\bar{\Lambda}\Gamma_{lm}\bar{\Lambda})(\bar{\Lambda}\Gamma_{np}R)(\bar{\Lambda}\Gamma_{gc}R)(\bar{\Lambda}\Gamma_{ef}R)(\Lambda\Gamma^{efghi}\Lambda)N_{hi}(\Lambda\Gamma^{lmnqc}\Lambda)N_{pq} \notag \\
& +\frac{2}{9\eta^{4}}(\bar{\Lambda}\Gamma_{ef}R)(\bar{\Lambda}\Gamma_{lm}\bar{\Lambda})(\bar{\Lambda}\Gamma_{ng}R)(\bar{\Lambda}R)(\Lambda\Gamma^{efghi}\Lambda)N_{hi}(\Lambda\Gamma^{lmnpq}\Lambda)N_{pq} \notag\\
& - \frac{4}{9\eta^{4}}(\bar{\Lambda}\Gamma_{gm}R)(\bar{\Lambda}\Gamma_{ef}R)(\bar{\Lambda}\Gamma_{ln}\bar{\Lambda})(\bar{\Lambda}R)(\Lambda\Gamma^{efghi}\Lambda)N_{hi}(\Lambda\Gamma^{lmnpq}\Lambda)N_{pq} \notag \\
& = \frac{4}{3\eta^{4}}(\bar{\Lambda}\Gamma_{lm}\bar{\Lambda})(\bar{\Lambda}\Gamma_{np}R)(\bar{\Lambda}\Gamma_{gc}R)(\bar{\Lambda}\Gamma_{ef}R)(\Lambda\Gamma^{efghi}\Lambda)N_{hi}(\Lambda\Gamma^{lmnqc}\Lambda)N_{pq} \notag \\
& - \frac{2}{3\eta^{4}}(\bar{\Lambda}\Gamma_{gm}R)(\bar{\Lambda}\Gamma_{ef}R)(\bar{\Lambda}\Gamma_{ln}\bar{\Lambda})(\bar{\Lambda}R)(\Lambda\Gamma^{efghi}\Lambda)N_{hi}(\Lambda\Gamma^{lmnpq}\Lambda)N_{pq} \notag \\
& = \frac{4}{3\eta^{4}}(\bar{\Lambda}\Gamma_{lm}\bar{\Lambda})(\bar{\Lambda}\Gamma_{np}R)(\bar{\Lambda}\Gamma_{gc}R)(\bar{\Lambda}\Gamma_{ef}R)(\Lambda\Gamma^{efghi}\Lambda)N_{hi}(\Lambda\Gamma^{lmnqc}\Lambda)N_{pq} \notag \\
& - \frac{2}{3\eta^{4}}(\bar{\Lambda}\Gamma_{ln}\bar{\Lambda})(\bar{\Lambda}\Gamma_{mg}R)(\bar{\Lambda}\Gamma_{ef}R)(\bar{\Lambda}R)(\Lambda\Gamma^{efghi}\Lambda)N_{hi}(\Lambda\Gamma^{lnmpq}\Lambda)N_{pq} \notag \\
& = \frac{4}{3\eta^{4}}(\bar{\Lambda}\Gamma_{ef}\bar{\Lambda})(\bar{\Lambda}\Gamma_{gc}R)(\bar{\Lambda}\Gamma_{lm}R)(\bar{\Lambda}\Gamma_{np}R)(\Lambda\Gamma^{efghi}\Lambda)N_{hi}(\Lambda\Gamma^{clmnq}\Lambda)N_{qp} \notag \\
& - \frac{2}{3\eta^{4}}(\bar{\Lambda}\Gamma_{ef}\bar{\Lambda})(\bar{\Lambda}\Gamma_{gm}R)(\bar{\Lambda}\Gamma_{ln}R)(\bar{\Lambda}R)(\Lambda\Gamma^{efghi}\Lambda)N_{hi}(\Lambda\Gamma^{lnmpq}\Lambda)N_{pq} \label{eq20}
\end{align} 
So our simplified $D=11$ $b$-ghost has the following expansion:
\begin{eqnarray}
b_{simpl} &=& P^{i}[\frac{1}{2}\eta^{-1}(\bar{\Lambda}\Gamma_{ab}\bar{\Lambda})(\Lambda\Gamma^{ab}\Gamma_{i}D) + \eta^{-2}L^{(1)}_{ab,cd}[2(\Lambda\Gamma^{abc}_{\hspace{4mm}ki}\Lambda)N^{dk}\nonumber \\
&& 
+ \frac{2}{3}(\eta^{b}_{\hspace{2mm}p}\eta^{d}_{\hspace{2mm}i} - \eta^{bd}\eta_{pi})(\Lambda\Gamma^{apcqj}\Lambda)N_{qj}]] + \frac{1}{\eta^{2}}L^{(1)}_{mc,rs}(\Lambda\Gamma^{m}D)(\Lambda\Gamma^{crs}D) + \nonumber\\
&& + \frac{4}{\eta^{3}}(\bar{\Lambda}\Gamma_{lm}\bar{\Lambda})(\bar{\Lambda}\Gamma_{cx}R)(\bar{\Lambda}\Gamma_{n}^{\hspace{2mm}p}R)(\Lambda\Gamma^{lmcnq}\Lambda)(\Lambda\Gamma^{x}D)N_{qp} \nonumber\\
&& -\frac{2}{\eta^{3}}(\bar{\Lambda}\Gamma_{lm}\bar{\Lambda})(\bar{\Lambda}\Gamma_{nx}R)(\bar{\Lambda}R)(\Lambda\Gamma^{lmnpq}\Lambda)(\Lambda\Gamma^{x}D)N_{pq} \nonumber\\
&& -\frac{1}{3\eta^{3}}(\bar{\Lambda}\Gamma_{ef}\bar{\Lambda})(\bar{\Lambda}\Gamma_{gc}R)(\bar{\Lambda}\Gamma_{mn}R)(\Lambda\Gamma^{efghi}\Lambda)N_{hi}(\Lambda\Gamma^{cmn}D) \nonumber \\
&& + \frac{4}{3\eta^{4}}(\bar{\Lambda}\Gamma_{ef}\bar{\Lambda})(\bar{\Lambda}\Gamma_{gc}R)(\bar{\Lambda}\Gamma_{lm}R)(\bar{\Lambda}\Gamma_{np}R)(\Lambda\Gamma^{efghi}\Lambda)N_{hi}(\Lambda\Gamma^{clmnq}\Lambda)N_{qp} \nonumber \\
&& - \frac{2}{3\eta^{4}}(\bar{\Lambda}\Gamma_{ef}\bar{\Lambda})(\bar{\Lambda}\Gamma_{gm}R)(\bar{\Lambda}\Gamma_{ln}R)(\bar{\Lambda}R)(\Lambda\Gamma^{efghi}\Lambda)N_{hi}(\Lambda\Gamma^{lnmpq}\Lambda)N_{pq} \label{app26}
\end{eqnarray}

\vspace{2mm}
We can compare this result with the expansion of the $b$-ghost in \eqref{eq16}
\begin{align}
b & = \frac{1}{2}\eta^{-1}(\bar{\Lambda}\Gamma_{ab}\bar{\Lambda})(\Lambda\Gamma^{ab}\Gamma^{i}D)P_{i} + \eta^{-2}L^{(1)}_{ab,cd}[(\Lambda\Gamma^{a}D)(\Lambda\Gamma^{bcd}D) + 2(\Lambda\Gamma^{abc}_{\hspace{4mm}ij}\Lambda)N^{di}P^{j} \notag \\
& \frac{2}{3}(\eta^{b}_{\hspace{2mm}p}\eta^{d}_{\hspace{2mm}q} - \eta^{bd}\eta_{pq})(\Lambda\Gamma^{apcij}\Lambda)N_{ij}P^{q}] - \frac{1}{3}\eta^{-3}L^{(2)}_{ab,cd,ef}\{(\Lambda\Gamma^{abcij}\Lambda)(\Lambda\Gamma^{def}D)N_{ij} \notag \\
 & - 12[ (\Lambda\Gamma^{abcei}\Lambda)\eta^{fj} - \frac{2}{3}\eta^{f[a}(\Lambda\Gamma^{bce]ij}\Lambda](\Lambda\Gamma^{d}D)N_{ij}\} \notag \\
& + \frac{4}{3}\eta^{-4}L^{(3)}_{ab,cd,ef,gh}(\Lambda\Gamma^{abcij}\Lambda)[(\Lambda\Gamma^{defgk}\Lambda)\eta^{hl} - \frac{2}{3}\eta^{h[d}(\Lambda\Gamma^{efg]kl}\Lambda)]\{N_{ij},N_{kl}\}
\end{align}
The quadratic term in $D_{\alpha}$ is easy to obtain using the identity \eqref{app5}
\begin{equation}
b^{(2)} = \frac{1}{\eta^{2}}(\bar{\Lambda}\Gamma_{ab}\bar{\Lambda})(\bar{\Lambda}\Gamma_{cd}R)(\Lambda\Gamma^{a}D)(\Lambda\Gamma^{bcd}D)
\end{equation} 

\vspace{2mm}
Now we will find the term proportional to $\eta^{-3}$. Let us do this in two steps: First let us focus on the term proportional to $\Lambda\Gamma^{a}D$ (which will be called $K\ensuremath{'}_{1}$) and then on the term proportional to $\Lambda\Gamma^{abc}D$ (which will be called $K\ensuremath{'}_{2}$):
\begin{eqnarray*}
K\ensuremath{'}_{1} &=& \frac{4}{\eta^{3}}L^{(2)}_{ab,cd,ef}[(\Lambda\Gamma^{abcei}\Lambda)(\Lambda\Gamma^{d}D)\eta^{fj}N_{ij} - \frac{2}{3}\eta^{f[a}(\Lambda\Gamma^{bce]ij}\Lambda)(\Lambda\Gamma^{d}D)N_{ij}]\\
&=& \frac{4}{\eta^{3}}(\bar{\Lambda}\Gamma_{ab}\bar{\Lambda})(\bar{\Lambda}\Gamma_{cd}R)(\bar{\Lambda}\Gamma_{e}^{\hspace{2mm}j}R)(\Lambda\Gamma^{abcei}\Lambda)(\Lambda\Gamma^{d}D)N_{ij}\\
&& - \frac{8}{3\eta^{3}}L^{(2)}_{ab,cd,ef}(\frac{1}{4})[\eta^{fa}(\Lambda\Gamma^{bceij}\Lambda) - \eta^{fb}(\Lambda\Gamma^{aceij}\Lambda) + \eta^{fc}(\Lambda\Gamma^{abeij}\Lambda)](\Lambda\Gamma^{d}D)N_{ij}\\
&=& \frac{4}{\eta^{3}}(\bar{\Lambda}\Gamma_{ab}\bar{\Lambda})(\bar{\Lambda}\Gamma_{cd}R)(\bar{\Lambda}\Gamma_{e}^{\hspace{2mm}j}R)(\Lambda\Gamma^{abcei}\Lambda)(\Lambda\Gamma^{d}D)N_{ij}\\
&& - \frac{2}{3\eta^{3}}L^{(2)}_{ab,cd,ef}[2\eta^{fa}(\Lambda\Gamma^{bceij}\Lambda) + \eta^{fc}(\Lambda\Gamma^{abeij}\Lambda)](\Lambda\Gamma^{d}D)N_{ij}\\
&=& \frac{4}{\eta^{3}}(\bar{\Lambda}\Gamma_{ab}\bar{\Lambda})(\bar{\Lambda}\Gamma_{cd}R)(\bar{\Lambda}\Gamma_{e}^{\hspace{2mm}j}R)(\Lambda\Gamma^{abcei}\Lambda)(\Lambda\Gamma^{d}D)N_{ij}\\
&& - \frac{2}{3\eta^{3}}[2L^{(2)\hspace{5mm}a}_{ab,cd,e}(\Lambda\Gamma^{bceij}\Lambda) + L^{(2)\hspace{5mm}c}_{ab,cd,e}(\Lambda\Gamma^{abeij}\Lambda)](\Lambda\Gamma^{d}D)N_{ij}
\end{eqnarray*}
where the identity \eqref{app7} was used from the first to the second line. Now we make use of the identities \eqref{app8} and \eqref{app9}:
\begin{eqnarray*}
K\ensuremath{'}_{1} &=& \frac{4}{\eta^{3}}(\bar{\Lambda}\Gamma_{ab}\bar{\Lambda})(\bar{\Lambda}\Gamma_{cd}R)(\bar{\Lambda}\Gamma_{e}^{\hspace{2mm}j}R)(\Lambda\Gamma^{abcei}\Lambda)(\Lambda\Gamma^{d}D)N_{ij}\\
&& - \frac{2}{9\eta^{3}}\{[4(\bar{\Lambda}\Gamma_{eb}\bar{\Lambda})(\bar{\Lambda}\Gamma_{cd}R)(\bar{\Lambda}R) - 2(\bar{\Lambda}\Gamma_{cd}\bar{\Lambda})(\bar{\Lambda}\Gamma_{ab}R)(\bar{\Lambda}\Gamma_{e}^{\hspace{2mm}a}R)](\Lambda\Gamma^{bceij}\Lambda)\\
&& + [(\bar{\Lambda}\Gamma_{ab}\bar{\Lambda})(\bar{\Lambda}\Gamma_{cd}R)(\bar{\Lambda}\Gamma_{e}^{\hspace{2mm}c}R) - 2(\bar{\Lambda}\Gamma_{ed}\bar{\Lambda})(\bar{\Lambda}\Gamma_{ab}R)(\bar{\Lambda}R)](\Lambda\Gamma^{abeij}\Lambda)\}(\Lambda\Gamma^{d}D)N_{ij}\\
&=&\frac{4}{\eta^{3}}(\bar{\Lambda}\Gamma_{ab}\bar{\Lambda})(\bar{\Lambda}\Gamma_{cd}R)(\bar{\Lambda}\Gamma_{e}^{\hspace{2mm}j}R)(\Lambda\Gamma^{abcei}\Lambda)(\Lambda\Gamma^{d}D)N_{ij}\\
&& - \frac{2}{9\eta^{3}}\{[6(\bar{\Lambda}\Gamma_{eb}\bar{\Lambda})(\bar{\Lambda}\Gamma_{cd}R)(\bar{\Lambda}R) - 2(\bar{\Lambda}\Gamma_{cd}\bar{\Lambda})(\bar{\Lambda}\Gamma_{ab}R)(\bar{\Lambda}\Gamma_{e}^{\hspace{2mm}a}R)](\Lambda\Gamma^{bceij}\Lambda)\\
&& + (\bar{\Lambda}\Gamma_{ab}\bar{\Lambda})(\bar{\Lambda}\Gamma_{cd}R)(\bar{\Lambda}\Gamma_{e}^{\hspace{2mm}c}R)(\Lambda\Gamma^{abeij}\Lambda)\}(\Lambda\Gamma^{d}D)N_{ij}\\
&=& \frac{4}{\eta^{3}}(\bar{\Lambda}\Gamma_{ab}\bar{\Lambda})(\bar{\Lambda}\Gamma_{cd}R)(\bar{\Lambda}\Gamma_{e}^{\hspace{2mm}j}R)(\Lambda\Gamma^{abcei}\Lambda)(\Lambda\Gamma^{d}D)N_{ij}\\
&& -\frac{2}{9\eta^{3}}\{[6(\bar{\Lambda}\Gamma_{eb}\bar{\Lambda})(\bar{\Lambda}\Gamma_{cd}R)(\bar{\Lambda}R) - 2(\bar{\Lambda}\Gamma_{ce}\bar{\Lambda})(\bar{\Lambda}\Gamma_{db}R)(\bar{\Lambda}R) - (\bar{\Lambda}\Gamma_{cb}\bar{\Lambda})(\bar{\Lambda}\Gamma_{ad}R)(\bar{\Lambda}\Gamma_{e}^{\hspace{2mm}a}R)\\
&& - (\bar{\Lambda}\Gamma_{ed}\bar{\Lambda})(\bar{\Lambda}\Gamma_{cb}R)(\bar{\Lambda}R)](\Lambda\Gamma^{bceij}\Lambda) - (\bar{\Lambda}\Gamma_{cb}\bar{\Lambda})(\bar{\Lambda}\Gamma_{ad}R)(\bar{\Lambda}\Gamma_{e}^{\hspace{2mm}a}R)(\Lambda\Gamma^{bceij}\Lambda)\}(\Lambda\Gamma^{d}D)N_{ij}\\
&=& \frac{4}{\eta^{3}}(\bar{\Lambda}\Gamma_{ab}\bar{\Lambda})(\bar{\Lambda}\Gamma_{cd}R)(\bar{\Lambda}\Gamma_{e}^{\hspace{2mm}j}R)(\Lambda\Gamma^{abcei}\Lambda)(\Lambda\Gamma^{d}D)N_{ij}\\
&& -\frac{2}{9\eta^{3}}\{[6(\bar{\Lambda}\Gamma_{eb}\bar{\Lambda})(\bar{\Lambda}\Gamma_{cd}R)(\bar{\Lambda}R) - 2(\bar{\Lambda}\Gamma_{ce}\bar{\Lambda})(\bar{\Lambda}\Gamma_{db}R)(\bar{\Lambda}R) - 2(\bar{\Lambda}\Gamma_{cb}\bar{\Lambda})(\bar{\Lambda}\Gamma_{ad}R)(\bar{\Lambda}\Gamma_{e}^{\hspace{2mm}a}R)\\
&& - (\bar{\Lambda}\Gamma_{ed}\bar{\Lambda})(\bar{\Lambda}\Gamma_{cb}R)(\bar{\Lambda}R)](\Lambda\Gamma^{bceij}\Lambda)\}(\Lambda\Gamma^{d}D)N_{ij}\\
&=& \frac{4}{\eta^{3}}(\bar{\Lambda}\Gamma_{ab}\bar{\Lambda})(\bar{\Lambda}\Gamma_{cd}R)(\bar{\Lambda}\Gamma_{e}^{\hspace{2mm}j}R)(\Lambda\Gamma^{abcei}\Lambda)(\Lambda\Gamma^{d}D)N_{ij}\\
&& -\frac{2}{9\eta^{3}}\{[6(\bar{\Lambda}\Gamma_{eb}\bar{\Lambda})(\bar{\Lambda}\Gamma_{cd}R)(\bar{\Lambda}R) - 2(\bar{\Lambda}\Gamma_{ce}\bar{\Lambda})(\bar{\Lambda}\Gamma_{db}R)(\bar{\Lambda}R) - 2(\bar{\Lambda}\Gamma_{cb}\bar{\Lambda})(\bar{\Lambda}\Gamma_{ed}R)(\bar{\Lambda}R)\\
&& + (\bar{\Lambda}\Gamma_{cb}\bar{\Lambda})(\bar{\Lambda}\Gamma_{ed}R)(\bar{\Lambda}R)](\Lambda\Gamma^{bceij}\Lambda)\}(\Lambda\Gamma^{d}D)N_{ij}
\end{eqnarray*}
where we have made use of the identity \eqref{app4}. By using the antisymmetry in $(b,c,e)$ we show that:
\begin{eqnarray}\label{eq15}
K\ensuremath{'}_{1} &=& \frac{4}{\eta^{3}}(\bar{\Lambda}\Gamma_{ab}\bar{\Lambda})(\bar{\Lambda}\Gamma_{cd}R)(\bar{\Lambda}\Gamma_{e}^{\hspace{2mm}j}R)(\Lambda\Gamma^{abcei}\Lambda)(\Lambda\Gamma^{d}D)N_{ij}\nonumber \\
&& - \frac{2}{\eta^{3}}(\bar{\Lambda}\Gamma_{bc}\bar{\Lambda})(\bar{\Lambda}\Gamma_{ed}R)(\bar{\Lambda}R)(\Lambda\Gamma^{bceij}\Lambda)(\Lambda\Gamma^{d}D)N_{ij}
\end{eqnarray}

\vspace{2mm}
Now let us focus on the term proportional to $(\Lambda\Gamma^{bcd}D)$. This term appears in the expression \eqref{eq16} in the form (after using \eqref{app7}):
\begin{equation}\label{eq17}
K\ensuremath{'}_{2} = -\frac{1}{3\eta^{2}}(\bar{\Lambda}\Gamma_{ab}\bar{\Lambda})(\bar{\Lambda}\Gamma_{cd}R)(\bar{\Lambda}\Gamma_{ef}R)(\Lambda\Gamma^{abcpq}\Lambda)(\Lambda\Gamma^{def}D)N_{pq}
\end{equation}
Putting these results together we obtain
\begin{eqnarray}
b^{(3)} &=& \frac{4}{\eta^{3}}(\bar{\Lambda}\Gamma_{ab}\bar{\Lambda})(\bar{\Lambda}\Gamma_{cd}R)(\bar{\Lambda}\Gamma_{e}^{\hspace{2mm}j}R)(\Lambda\Gamma^{abcei}\Lambda)(\Lambda\Gamma^{d}D)N_{ij}\nonumber \\
&& - \frac{2}{\eta^{3}}(\bar{\Lambda}\Gamma_{bc}\bar{\Lambda})(\bar{\Lambda}\Gamma_{ed}R)(\bar{\Lambda}R)(\Lambda\Gamma^{bceij}\Lambda)(\Lambda\Gamma^{d}D)N_{ij}\nonumber \\
&& -\frac{1}{3\eta^{2}}(\bar{\Lambda}\Gamma_{ab}\bar{\Lambda})(\bar{\Lambda}\Gamma_{cd}R)(\bar{\Lambda}\Gamma_{ef}R)(\Lambda\Gamma^{abcpq}\Lambda)(\Lambda\Gamma^{def}D)N_{pq}
\end{eqnarray}
which it should be compared with the analog expression corresponding to $b_{simpl}$, equation \eqref{eq404}.

\vspace{2mm}
The last term to be computed is that proportional to $\eta^{-4}$ in \eqref{eq16}:
\begin{equation}
b^{(4)} = \frac{4}{3}\eta^{-4}L^{(3)}_{ab,cd,ef,gh}(\Lambda\Gamma^{abcij}\Lambda)[(\Lambda\Gamma^{defgk}\Lambda)\eta^{hl} - \frac{2}{3}\eta^{h[d}(\Lambda\Gamma^{efg]kl}\Lambda)]\{N_{ij}, N_{kl}\}
\end{equation}
Let us use some identities in order to write this expression in a simpler way. It is more convenient to do this in two steps: First we will focus on the first term ($P_{1}$) and then on the second term ($P_{2}$):
\begin{align}\label{eq18}
P_{1} &= \frac{4}{3}\eta^{-4}L^{(3)}_{ab,cd,ef,gh}(\Lambda\Gamma^{abcij}\Lambda)(\Lambda\Gamma^{defgk}\Lambda)\eta^{hl}\{N_{ij}, N_{kl}\} \notag \\
& = \frac{4}{3\eta^{4}}(\frac{1}{4})[(\bar{\Lambda}\Gamma_{ab}\bar\Lambda)(\bar{\Lambda}\Gamma_{cd}R)(\bar{\Lambda}\Gamma_{ef}R)(\bar{\Lambda}\Gamma_{gh}R) - (\bar{\Lambda}\Gamma_{cd}\bar\Lambda)(\bar{\Lambda}\Gamma_{ab}R)(\bar{\Lambda}\Gamma_{ef}R)(\bar{\Lambda}\Gamma_{gh}R) + \notag \\
& + (\bar{\Lambda}\Gamma_{ef}\bar\Lambda)(\bar{\Lambda}\Gamma_{ab}R)(\bar{\Lambda}\Gamma_{cd}R)(\bar{\Lambda}\Gamma_{gh}R) - (\bar{\Lambda}\Gamma_{gh}\bar\Lambda)(\bar{\Lambda}\Gamma_{ab}R)(\bar{\Lambda}\Gamma_{cd}R)(\bar{\Lambda}\Gamma_{ef}R)](\Lambda\Gamma^{abcij}\Lambda)\times \notag \\
& (\Lambda\Gamma^{defgk}\Lambda)\eta^{hl}\{N_{ij}, N_{kl}\} \notag \\
& = \frac{2}{3\eta^{4}}[(\bar{\Lambda}\Gamma_{ab}\bar\Lambda)(\bar{\Lambda}\Gamma_{cd}R)(\bar{\Lambda}\Gamma_{ef}R)(\bar{\Lambda}\Gamma_{gh}R) + (\bar{\Lambda}\Gamma_{ef}\bar\Lambda)(\bar{\Lambda}\Gamma_{ab}R)(\bar{\Lambda}\Gamma_{cd}R)(\bar{\Lambda}\Gamma_{gh}R)](\Lambda\Gamma^{abcij}\Lambda)\times \notag \\
& (\Lambda\Gamma^{defgk}\Lambda)\eta^{hl}\{N_{ij}, N_{kl}\} \notag \\
& = \frac{2}{3\eta^{4}}[(\bar{\Lambda}\Gamma_{ab}\bar\Lambda)(\bar{\Lambda}\Gamma_{cd}R)(\bar{\Lambda}\Gamma_{ef}R)(\bar{\Lambda}\Gamma_{gh}R) - (\bar{\Lambda}\Gamma_{cd}\bar\Lambda)(\bar{\Lambda}\Gamma_{ab}R)(\bar{\Lambda}\Gamma_{ef}R)(\bar{\Lambda}\Gamma_{gh}R)](\Lambda\Gamma^{abcij}\Lambda)\times \notag \\
& (\Lambda\Gamma^{defgk}\Lambda)\eta^{hl}\{N_{ij}, N_{kl}\} \notag \\
& = \frac{4}{3\eta^{4}}(\bar{\Lambda}\Gamma_{ab}\bar\Lambda)(\bar{\Lambda}\Gamma_{cd}R)(\bar{\Lambda}\Gamma_{ef}R)(\bar{\Lambda}\Gamma_{gh}R)(\Lambda\Gamma^{abcij}\Lambda)(\Lambda\Gamma^{defgk}\Lambda)\eta^{hl}\{N_{ij}, N_{kl}\}
\end{align}
The simplifications made here are result of repeated uses of the identities \eqref{app5} and \eqref{app6}. Now let us focus on $P_{2}$:
\begin{align}
P_{2} & = -\frac{8}{9\eta^{4}}L^{(3)}_{ab,cd,ef,gh}(\Lambda\Gamma^{abcij}\Lambda)\eta^{h[d}(\Lambda\Gamma^{efg]kl}\Lambda)\{N_{ij}, N_{kl}\} \notag \\
& = -\frac{2}{9\eta^{4}}L^{(3)}_{ab,cd,ef,gh}(\Lambda\Gamma^{abcij}\Lambda)[\eta^{hd}(\Lambda\Gamma^{efgkl}\Lambda) - 2\eta^{he}(\Lambda\Gamma^{dfgkl}\Lambda)]\{N_{ij}, N_{kl}\} \notag \\
& = -\frac{2}{9\eta^{4}}[L^{(3)}_{ab,cd,ef,gh}\eta^{hd}(\Lambda\Gamma^{abcij}\Lambda)(\Lambda\Gamma^{efgkl}\Lambda) - 2L^{(3)}_{ab,cd,ef,gh}\eta^{he}(\Lambda\Gamma^{abcij}\Lambda)(\Lambda\Gamma^{dfgkl}\Lambda)]\{N_{ij}, N_{kl}\}
\end{align}
We can simplify each term separately:
\begin{align}
(P\ensuremath{'}_{2})^{ijkl} & = L^{(3)}_{ab,cd,ef,gh}\eta^{hd}(\Lambda\Gamma^{abcij}\Lambda)(\Lambda\Gamma^{efgkl}\Lambda) \notag \\
& = \frac{1}{4}[(\bar{\Lambda}\Gamma_{ab}\bar{\Lambda})(\bar{\Lambda}\Gamma_{cd}R)(\bar{\Lambda}\Gamma_{ef}R)(\bar{\Lambda}\Gamma_{g}^{\hspace{2mm}d}R) - (\bar{\Lambda}\Gamma_{cd}\bar{\Lambda})(\bar{\Lambda}\Gamma_{ab}R)(\bar{\Lambda}\Gamma_{ef}R)(\bar{\Lambda}\Gamma_{g}^{\hspace{2mm}d}R) \notag \\
& + (\bar{\Lambda}\Gamma_{ef}\bar{\Lambda})(\bar{\Lambda}\Gamma_{ab}R)(\bar{\Lambda}\Gamma_{cd}R)(\bar{\Lambda}\Gamma_{g}^{\hspace{2mm}d}R) - (\bar{\Lambda}\Gamma_{g}^{\hspace{2mm}d}\Lambda)(\bar{\Lambda}\Gamma_{ab}R)(\bar{\Lambda}\Gamma_{cd}R)(\bar{\Lambda}\Gamma_{ef}R)](\Lambda\Gamma^{abcij}\Lambda)(\Lambda\Gamma^{efgkl}\Lambda) \notag \\
& = \frac{1}{4}[ - 2(\bar{\Lambda}\Gamma_{cg}\bar{\Lambda})(\bar{\Lambda}\Gamma_{ab}R)(\bar{\Lambda}\Gamma_{ef}R)(\bar{\Lambda}R) - 2(\bar{\Lambda}\Gamma_{g}^{\hspace{2mm}d}\Lambda)(\bar{\Lambda}\Gamma_{ab}R)(\bar{\Lambda}\Gamma_{cd}R)(\bar{\Lambda}\Gamma_{ef}R)](\Lambda\Gamma^{abcij}\Lambda)(\Lambda\Gamma^{efgkl}\Lambda) \notag \\
& = \frac{1}{4}[ - 2(\bar{\Lambda}\Gamma_{cg}\bar{\Lambda})(\bar{\Lambda}\Gamma_{ab}R)(\bar{\Lambda}\Gamma_{ef}R)(\bar{\Lambda}R) - 2(\bar{\Lambda}\Gamma_{gc}\Lambda)(\bar{\Lambda}\Gamma_{ab}R)(\bar{\Lambda}R)(\bar{\Lambda}\Gamma_{ef}R)](\Lambda\Gamma^{abcij}\Lambda)(\Lambda\Gamma^{efgkl}\Lambda) \notag \\
& = (\bar{\Lambda}\Gamma_{ab}\bar{\Lambda})(\bar{\Lambda}\Gamma_{cg}R)(\bar{\Lambda}\Gamma_{ef}\bar{\Lambda})(\bar{\Lambda}R)(\Lambda\Gamma^{abcij}\Lambda)(\Lambda\Gamma^{efgkl}\Lambda)
\end{align} 

\begin{align}
(P\ensuremath{''}_{2})^{ijkl} & = -2L^{(3)}_{ab,cd,ef,gh}(\Lambda\Gamma^{dfgkl}\Lambda)(\Lambda\Gamma^{abcij}\Lambda) \notag \\
& = -\frac{1}{2}[(\bar{\Lambda}\Gamma_{ab}\bar{\Lambda})(\bar{\Lambda}\Gamma_{cd}R)(\bar{\Lambda}\Gamma_{ef}R)(\bar{\Lambda}\Gamma_{g}^{\hspace{2mm}e}R) - (\bar{\Lambda}\Gamma_{cd}\bar{\Lambda})(\bar{\Lambda}\Gamma_{ab}R)(\bar{\Lambda}\Gamma_{ef}R)(\bar{\Lambda}\Gamma_{g}^{\hspace{2mm}e}R) \notag \\
& + (\bar{\Lambda}\Gamma_{ef}\bar{\Lambda})(\bar{\Lambda}\Gamma_{ab}R)(\bar{\Lambda}\Gamma_{cd}R)(\bar{\Lambda}\Gamma_{g}^{\hspace{2mm}e}R) - (\bar{\Lambda}\Gamma_{g}^{\hspace{2mm}e}\bar{\Lambda})(\bar{\Lambda}\Gamma_{ab}R)(\bar{\Lambda}\Gamma_{cd}R)(\bar{\Lambda}\Gamma_{ef}R)](\Lambda\Gamma^{dfgkl}\Lambda)(\Lambda\Gamma^{abcij}\Lambda) \notag \\
&=[(\bar{\Lambda}\Gamma_{cd}\bar{\Lambda})(\bar{\Lambda}\Gamma_{ab}R)(\bar{\Lambda}\Gamma_{ef}R)(\bar{\Lambda}\Gamma_{g}^{\hspace{2mm}e}R) + (\bar{\Lambda}\Gamma_{fg}\bar{\Lambda})(\bar{\Lambda}\Gamma_{ab}R)(\bar{\Lambda}\Gamma_{cd}R)(\bar{\Lambda}R)](\Lambda\Gamma^{dfgkl}\Lambda)(\Lambda\Gamma^{abcij}\Lambda) \notag \\
&=[-(\bar{\Lambda}\Gamma_{cd}\bar{\Lambda})(\bar{\Lambda}\Gamma_{ab}R)(\bar{\Lambda}\Gamma_{fg}R)(\bar{\Lambda}R) + (\bar{\Lambda}\Gamma_{fg}\bar{\Lambda})(\bar{\Lambda}\Gamma_{ab}R)(\bar{\Lambda}\Gamma_{cd}R)(\bar{\Lambda}R)](\Lambda\Gamma^{dfgkl}\Lambda)(\Lambda\Gamma^{abcij}\Lambda) \notag \\
& = 2(\bar{\Lambda}\Gamma_{fg}\bar{\Lambda})(\bar{\Lambda}\Gamma_{ab}R)(\bar{\Lambda}\Gamma_{cd}R)(\bar{\Lambda}R)(\Lambda\Gamma^{dfgkl}\Lambda)(\Lambda\Gamma^{abcij}\Lambda) \notag \\
& = 2(\bar{\Lambda}\Gamma_{fg}\bar{\Lambda})(\bar{\Lambda}\Gamma_{dc}R)(\bar{\Lambda}\Gamma_{ab}R)(\bar{\Lambda}R)(\Lambda\Gamma^{fgdkl}\Lambda)(\Lambda\Gamma^{cabij}\Lambda)
\end{align}
Therefore
\begin{align}
P_{2} & = -\frac{2}{3\eta^{4}}(\bar{\Lambda}\Gamma_{ab}\bar{\Lambda})(\bar{\Lambda}\Gamma_{cg}R)(\bar{\Lambda}\Gamma_{ef}\bar{\Lambda})(\bar{\Lambda}R)(\Lambda\Gamma^{abcij}\Lambda)(\Lambda\Gamma^{efgkl}\Lambda)\{N_{ij}, N_{kl}\}
\end{align}
This result together with \eqref{eq18} gives us the following result for $b^{(4)}$:
\begin{align}
b^{(4)} & = \frac{4}{3\eta^{4}}(\bar{\Lambda}\Gamma_{ab}\bar\Lambda)(\bar{\Lambda}\Gamma_{cd}R)(\bar{\Lambda}\Gamma_{ef}R)(\bar{\Lambda}\Gamma_{gh}R)(\Lambda\Gamma^{abcij}\Lambda)(\Lambda\Gamma^{defgk}\Lambda)\eta^{hl}\{N_{ij}, N_{kl}\} \notag \\
& -\frac{2}{3\eta^{4}}(\bar{\Lambda}\Gamma_{ab}\bar{\Lambda})(\bar{\Lambda}\Gamma_{cg}R)(\bar{\Lambda}\Gamma_{ef}\bar{\Lambda})(\bar{\Lambda}R)(\Lambda\Gamma^{abcij}\Lambda)(\Lambda\Gamma^{efgkl}\Lambda)\{N_{ij}, N_{kl}\}
\end{align}
which it should be compared with the analog expression corresponding to $b_{simpl}$, equation \eqref{eq20}.

\vspace{2mm}
Hence we can write the full expansion for the $b$-ghost given in \eqref{eq16}
\begin{eqnarray}
b &=& P^{i}[\frac{1}{2}\eta^{-1}(\bar{\Lambda}\Gamma_{ab}\bar{\Lambda})(\Lambda\Gamma^{ab}\Gamma_{i}D) + \eta^{-2}L^{(1)}_{ab,cd}[2(\Lambda\Gamma^{abc}_{\hspace{4mm}ki}\Lambda)N^{dk}\nonumber \\
&& 
+ \frac{2}{3}(\eta^{b}_{\hspace{2mm}p}\eta^{d}_{\hspace{2mm}i} - \eta^{bd}\eta_{pi})(\Lambda\Gamma^{apcqj}\Lambda)N_{qj}]] + \frac{1}{\eta^{2}}L^{(1)}_{ab,cd}(\Lambda\Gamma^{a}D)(\Lambda\Gamma^{bcd}D) \nonumber \\
&& +\frac{4}{\eta^{3}}(\bar{\Lambda}\Gamma_{ab}\bar{\Lambda})(\bar{\Lambda}\Gamma_{cd}R)(\bar{\Lambda}\Gamma_{e}^{\hspace{2mm}j}R)(\Lambda\Gamma^{abcei}\Lambda)(\Lambda\Gamma^{d}D)N_{ij}\nonumber \\
&& - \frac{2}{\eta^{3}}(\bar{\Lambda}\Gamma_{bc}\bar{\Lambda})(\bar{\Lambda}\Gamma_{ed}R)(\bar{\Lambda}R)(\Lambda\Gamma^{bceij}\Lambda)(\Lambda\Gamma^{d}D)N_{ij}\nonumber \\
&& -\frac{1}{3\eta^{2}}(\bar{\Lambda}\Gamma_{ab}\bar{\Lambda})(\bar{\Lambda}\Gamma_{cd}R)(\bar{\Lambda}\Gamma_{ef}R)(\Lambda\Gamma^{abcpq}\Lambda)(\Lambda\Gamma^{def}D)N_{pq} \nonumber \\
&& + \frac{4}{3\eta^{4}}(\bar{\Lambda}\Gamma_{ab}\bar\Lambda)(\bar{\Lambda}\Gamma_{cd}R)(\bar{\Lambda}\Gamma_{ef}R)(\bar{\Lambda}\Gamma_{gh}R)(\Lambda\Gamma^{abcij}\Lambda)(\Lambda\Gamma^{defgk}\Lambda)\eta^{hl}\{N_{ij}, N_{kl}\} \notag \\
&& -\frac{2}{3\eta^{4}}(\bar{\Lambda}\Gamma_{ab}\bar{\Lambda})(\bar{\Lambda}\Gamma_{cg}R)(\bar{\Lambda}\Gamma_{ef}\bar{\Lambda})(\bar{\Lambda}R)(\Lambda\Gamma^{abcij}\Lambda)(\Lambda\Gamma^{efgkl}\Lambda)\{N_{ij}, N_{kl}\}\label{app27}
\end{eqnarray}
This expression differs from \eqref{app26} in two points. First, the position of $N_{hi}$ in the last term proportional to $\eta^{-3}$ is not at the end of the expression as it is in \eqref{app27}. Second, in the terms proportional to $\eta^{-4}$ we do not have the anticommutator of $N_{ab}$'s in \eqref{app26} as we do in \eqref{app27}, and once again the position of $N_{hi}$ is not at the end of the expressions in \eqref{app26} as it is in \eqref{app27}.

\vspace{2mm}
In order to have a clearer idea on what is happening, we will move all of the $N_{ab}$'s at the end of the expressions mentioned above in \eqref{app26}. Let us start with the term proportional to $\eta^{-3}$. We should put the ghost current $N^{hi}$ to the right hand side of $(\Lambda\Gamma^{cmn}D)$. For this purpose we compute the commutator between $N^{hi}$ and $(\Lambda\Gamma^{cmn}D)$ with the symmetry properties written in \eqref{eq21}:
\begin{eqnarray}
[N^{hi}, (\Lambda\Gamma^{cmn}D)] &=& -2\eta^{hi}_{mn}(\Lambda\Gamma^{c}D) - 4\eta^{cm}_{hi}(\Lambda\Gamma^{n}D) - 4\eta^{hm}(\Lambda\Gamma^{cin}D) - 2\eta^{ch}(\Lambda\Gamma^{imn}D)\nonumber \\
&& + (\Lambda\Gamma^{chimn}D)
\end{eqnarray}
The use of the identities \eqref{app4}, \eqref{app15} allows us to cancel out all of the terms except the last one, so
\begin{align}
K_{2} & = -\frac{1}{3\eta^{3}}(\bar{\Lambda}\Gamma_{ef}\bar{\Lambda})(\bar{\Lambda}\Gamma_{gc}R)(\bar{\Lambda}\Gamma_{mn}R)(\Lambda\Gamma^{efghi}\Lambda)(\Lambda\Gamma^{cmn}D)N_{hi} \notag\\
& - \frac{1}{3\eta^{3}}(\bar{\Lambda}\Gamma_{ef}\bar{\Lambda})(\bar{\Lambda}\Gamma_{gc}R)(\bar{\Lambda}\Gamma_{mn}R)(\Lambda\Gamma^{efghi}\Lambda)(\Lambda\Gamma^{chimn}D)  
\end{align}
Therefore $b^{(3)}_{simpl}$ changes by the factor $- \frac{1}{3\eta^{3}}(\bar{\Lambda}\Gamma_{ef}\bar{\Lambda})(\bar{\Lambda}\Gamma_{gc}R)(\bar{\Lambda}\Gamma_{mn}R)(\Lambda\Gamma^{efghi}\Lambda)(\Lambda\Gamma^{chimn}D)$ when $N_{mn}$ is placed at the end of the full expression. 

\vspace{2mm}
Now we move on to the terms proportional to $\eta^{-4}$. We should move $N_{hi}$ to the right hand side of $(\Lambda\Gamma^{clmnq}\Lambda)$:
\begin{align}
[N^{hi}, (\Lambda\Gamma^{clmnq}\Lambda)] & = 4\eta^{hq}(\Lambda\Gamma^{cilmn}\Lambda) - 4\eta^{hn}(\Lambda\Gamma^{cilmq}\Lambda) - 8\eta^{hl}(\Lambda\Gamma^{cimnq}\Lambda) - 4\eta^{ch}(\Lambda\Gamma^{ilmnq}\Lambda)
\end{align}
These are the relevant terms in \eqref{eq20} and it is a direct consequence of the equation \eqref{app14} and the symmetry properties of the expression \eqref{eq20}. The last two terms do not contribute as it can be seen after using \eqref{app4}, \eqref{app15}. So we are left with:
\begin{align}
Z_{1} & = \frac{4}{3\eta^{4}}(\bar{\Lambda}\Gamma_{ef}\bar{\Lambda})(\bar{\Lambda}\Gamma_{gc}R)(\bar{\Lambda}\Gamma_{lm}R)(\bar{\Lambda}\Gamma_{np}R)(\Lambda\Gamma^{efghi}\Lambda)(\Lambda\Gamma^{clmnq}\Lambda)N_{hi}N_{qp} \notag \\
& + \frac{16}{3\eta^{4}}(\bar{\Lambda}\Gamma_{ef}\bar{\Lambda})(\bar{\Lambda}\Gamma_{gc}R)(\bar{\Lambda}\Gamma_{lm}R)(\bar{\Lambda}\Gamma_{np}R)[(\Lambda\Gamma^{efgqi}\Lambda)(\Lambda\Gamma^{cilmn}\Lambda) - (\Lambda\Gamma^{efgni}\Lambda)(\Lambda\Gamma^{cilmq}\Lambda)]N_{qp} \notag \\
& = \frac{4}{3\eta^{4}}(\bar{\Lambda}\Gamma_{ef}\bar{\Lambda})(\bar{\Lambda}\Gamma_{gc}R)(\bar{\Lambda}\Gamma_{lm}R)(\bar{\Lambda}\Gamma_{np}R)(\Lambda\Gamma^{efghi}\Lambda)(\Lambda\Gamma^{clmnq}\Lambda)N_{hi}N_{qp} \notag \\
& + \frac{16}{3\eta^{4}}[(\bar{\Lambda}\Gamma_{ef}\bar{\Lambda})(\bar{\Lambda}\Gamma_{gc}R)(\bar{\Lambda}\Gamma_{lm}R)(\bar{\Lambda}\Gamma_{np}R)(\Lambda\Gamma^{efgqi}\Lambda)(\Lambda\Gamma^{cilmn}\Lambda) \notag \\
& - \frac{16}{3\eta^{4}}(\bar{\Lambda}\Gamma_{lm}\bar{\Lambda})(\bar{\Lambda}\Gamma_{cg}R)(\bar{\Lambda}\Gamma_{ef}R)(\bar{\Lambda}\Gamma_{np}R)(\Lambda\Gamma^{lmcni}\Lambda)(\Lambda\Gamma^{giefq}\Lambda)]N_{qp} \notag \\
& = \frac{4}{3\eta^{4}}(\bar{\Lambda}\Gamma_{ef}\bar{\Lambda})(\bar{\Lambda}\Gamma_{gc}R)(\bar{\Lambda}\Gamma_{lm}R)(\bar{\Lambda}\Gamma_{np}R)(\Lambda\Gamma^{efghi}\Lambda)(\Lambda\Gamma^{clmnq}\Lambda)N_{hi}N_{qp} \notag \\
& + \frac{16}{3\eta^{4}}[(\bar{\Lambda}\Gamma_{ef}\bar{\Lambda})(\bar{\Lambda}\Gamma_{gc}R)(\bar{\Lambda}\Gamma_{lm}R)(\bar{\Lambda}\Gamma_{np}R)(\Lambda\Gamma^{efgqi}\Lambda)(\Lambda\Gamma^{cilmn}\Lambda) \notag \\
& - \frac{16}{3\eta^{4}}(\bar{\Lambda}\Gamma_{gc}\bar{\Lambda})(\bar{\Lambda}\Gamma_{lm}R)(\bar{\Lambda}\Gamma_{ef}R)(\bar{\Lambda}\Gamma_{np}R)(\Lambda\Gamma^{clmni}\Lambda)(\Lambda\Gamma^{efgiq}\Lambda)]N_{qp} \notag \\
& = \frac{4}{3\eta^{4}}(\bar{\Lambda}\Gamma_{ef}\bar{\Lambda})(\bar{\Lambda}\Gamma_{gc}R)(\bar{\Lambda}\Gamma_{lm}R)(\bar{\Lambda}\Gamma_{np}R)(\Lambda\Gamma^{efghi}\Lambda)(\Lambda\Gamma^{clmnq}\Lambda)N_{hi}N_{qp} \notag \\
& + \frac{16}{3\eta^{4}}[(\bar{\Lambda}\Gamma_{ef}\bar{\Lambda})(\bar{\Lambda}\Gamma_{gc}R)(\bar{\Lambda}\Gamma_{lm}R)(\bar{\Lambda}\Gamma_{np}R)(\Lambda\Gamma^{efgqi}\Lambda)(\Lambda\Gamma^{cilmn}\Lambda) \notag \\
& - \frac{16}{3\eta^{4}}(\bar{\Lambda}\Gamma_{ef}\bar{\Lambda})(\bar{\Lambda}\Gamma_{gc}R)(\bar{\Lambda}\Gamma_{lm}R)(\bar{\Lambda}\Gamma_{np}R)(\Lambda\Gamma^{climn}\Lambda)(\Lambda\Gamma^{efgqi}\Lambda)]N_{qp} \notag \\
& = \frac{4}{3\eta^{4}}(\bar{\Lambda}\Gamma_{ef}\bar{\Lambda})(\bar{\Lambda}\Gamma_{gc}R)(\bar{\Lambda}\Gamma_{lm}R)(\bar{\Lambda}\Gamma_{np}R)(\Lambda\Gamma^{efghi}\Lambda)(\Lambda\Gamma^{clmnq}\Lambda)N_{hi}N_{qp}
\end{align}

Now let us make the same procedure with the remaining term. The relevant commutation relation is:
\begin{align}
[N^{hi}, (\Lambda\Gamma^{lmnpq}\Lambda)] & = -8\eta^{hp}(\Lambda\Gamma^{ilmnq}\Lambda) - 4\eta^{hm}(\Lambda\Gamma^{ilnpq}\Lambda) + 8\eta^{hl}(\Lambda\Gamma^{imnpq}\Lambda)
\end{align}
Once again, we have obtained this result by using the identity \eqref{app14} and the symmetry properties of the corresponding expression in \eqref{eq20}. When applying the equations \eqref{app4}, \eqref{app15}, the last two terms vanish and we obtain
\begin{align}
Z_{2} & = - \frac{2}{3\eta^{4}}(\bar{\Lambda}\Gamma_{ef}\bar{\Lambda})(\bar{\Lambda}\Gamma_{gm}R)(\bar{\Lambda}\Gamma_{ln}R)(\bar{\Lambda}R)(\Lambda\Gamma^{efghi}\Lambda)(\Lambda\Gamma^{lnmpq}\Lambda)N_{hi}N_{pq} \notag \\
& + \frac{16}{3\eta^{4}}(\bar{\Lambda}\Gamma_{ef}\bar{\Lambda})(\bar{\Lambda}\Gamma_{gm}R)(\bar{\Lambda}\Gamma_{ln}R)(\bar{\Lambda}R)(\Lambda\Gamma^{efgpi}\Lambda)(\Lambda\Gamma^{ilmnq}\Lambda)N_{pq}
\end{align}
Now we will show that the last term is zero. Let us see how this happens:
\begin{align}
M & = \frac{16}{3\eta^{4}}(\bar{\Lambda}\Gamma_{ef}\bar{\Lambda})(\bar{\Lambda}\Gamma_{gm}R)(\bar{\Lambda}\Gamma_{ln}R)(\bar{\Lambda}R)(\Lambda\Gamma^{efgpi}\Lambda)(\Lambda\Gamma^{ilmnq}\Lambda)N_{pq} \notag \\
& =  -\frac{16}{3\eta^{4}}(\bar{\Lambda}\Gamma_{ln}\bar{\Lambda})(\bar{\Lambda}\Gamma_{mg}R)(\bar{\Lambda}\Gamma_{ef}R)(\bar{\Lambda}R)(\Lambda\Gamma^{lnmqi}\Lambda)(\Lambda\Gamma^{iegfp}\Lambda)N_{pq} \notag \\
& = \frac{16}{3\eta^{4}}(\bar{\Lambda}\Gamma_{ef}\bar{\Lambda})(\bar{\Lambda}\Gamma_{mg}R)(\bar{\Lambda}\Gamma_{ln}R)(\bar{\Lambda}R)(\Lambda\Gamma^{lmnqi}\Lambda)(\Lambda\Gamma^{efgpi}\Lambda)N_{pq} \notag \\
& = -\frac{16}{3\eta^{4}}(\bar{\Lambda}\Gamma_{ef}\bar{\Lambda})(\bar{\Lambda}\Gamma_{gm}R)(\bar{\Lambda}\Gamma_{ln}R)(\bar{\Lambda}R)(\Lambda\Gamma^{lmnqi}\Lambda)(\Lambda\Gamma^{efgpi}\Lambda)N_{pq} \notag \\
& = -M \notag \\
\rightarrow M & = 0
\end{align}
Therefore,
\begin{equation}
Z_{2} = - \frac{2}{3\eta^{4}}(\bar{\Lambda}\Gamma_{ef}\bar{\Lambda})(\bar{\Lambda}\Gamma_{gm}R)(\bar{\Lambda}\Gamma_{ln}R)(\bar{\Lambda}R)(\Lambda\Gamma^{efghi}\Lambda)(\Lambda\Gamma^{lnmpq}\Lambda)N_{hi}N_{pq}
\end{equation}
This means that $b^{(4)}_{simpl}$ does not change when we move $N_{mn}$ to the end of the full expression. We can summarize the result in the following expression for the simplified $b$-ghost:
\begin{eqnarray}
b_{simpl} &=& P^{i}[\frac{1}{2}\eta^{-1}(\bar{\Lambda}\Gamma_{ab}\bar{\Lambda})(\Lambda\Gamma^{ab}\Gamma_{i}D) + \eta^{-2}L^{(1)}_{ab,cd}[2(\Lambda\Gamma^{abc}_{\hspace{4mm}ki}\Lambda)N^{dk}\nonumber \\
&& 
+ \frac{2}{3}(\eta^{b}_{\hspace{2mm}p}\eta^{d}_{\hspace{2mm}i} - \eta^{bd}\eta_{pi})(\Lambda\Gamma^{apcqj}\Lambda)N_{qj}]] + \frac{1}{\eta^{2}}L^{(1)}_{mc,rs}(\Lambda\Gamma^{m}D)(\Lambda\Gamma^{crs}D) + \nonumber\\
&& + \frac{4}{\eta^{3}}(\bar{\Lambda}\Gamma_{lm}\bar{\Lambda})(\bar{\Lambda}\Gamma_{cx}R)(\bar{\Lambda}\Gamma_{n}^{\hspace{2mm}p}R)(\Lambda\Gamma^{lmcnq}\Lambda)(\Lambda\Gamma^{x}D)N_{qp} \nonumber\\
&& -\frac{2}{\eta^{3}}(\bar{\Lambda}\Gamma_{lm}\bar{\Lambda})(\bar{\Lambda}\Gamma_{nx}R)(\bar{\Lambda}R)(\Lambda\Gamma^{lmnpq}\Lambda)(\Lambda\Gamma^{x}D)N_{pq} \nonumber\\
&& -\frac{1}{3\eta^{3}}(\bar{\Lambda}\Gamma_{ef}\bar{\Lambda})(\bar{\Lambda}\Gamma_{gc}R)(\bar{\Lambda}\Gamma_{mn}R)(\Lambda\Gamma^{efghi}\Lambda)(\Lambda\Gamma^{cmn}D)N_{hi} \nonumber \\
&& + \frac{4}{3\eta^{4}}(\bar{\Lambda}\Gamma_{ef}\bar{\Lambda})(\bar{\Lambda}\Gamma_{gc}R)(\bar{\Lambda}\Gamma_{lm}R)(\bar{\Lambda}\Gamma_{np}R)(\Lambda\Gamma^{efghi}\Lambda)(\Lambda\Gamma^{clmnq}\Lambda)N_{hi}N_{qp} \nonumber \\
&& - \frac{2}{3\eta^{4}}(\bar{\Lambda}\Gamma_{ef}\bar{\Lambda})(\bar{\Lambda}\Gamma_{gm}R)(\bar{\Lambda}\Gamma_{ln}R)(\bar{\Lambda}R)(\Lambda\Gamma^{efghi}\Lambda)(\Lambda\Gamma^{lnmpq}\Lambda)N_{hi}N_{pq}\nonumber \\
&& - \frac{1}{3\eta^{3}}(\bar{\Lambda}\Gamma_{ef}\bar{\Lambda})(\bar{\Lambda}\Gamma_{gc}R)(\bar{\Lambda}\Gamma_{mn}R)(\Lambda\Gamma^{efghi}\Lambda)(\Lambda\Gamma^{chimn}D) \label{app29}
\end{eqnarray}
We can write the anticommutator instead of the ordinary product of $N_{ab}$'s recalling the relation $2N_{hi}N_{pq} = [N_{hi}, N_{pq}] + \{N_{hi}, N_{pq}\}$. The commutator contribution vanishes because of the identity \eqref{app11}. Therefore the final form for the expansion of the simplified $b$-ghost is
\begin{eqnarray}
b_{simpl} &=& P^{i}[\frac{1}{2}\eta^{-1}(\bar{\Lambda}\Gamma_{ab}\bar{\Lambda})(\Lambda\Gamma^{ab}\Gamma_{i}D) + \eta^{-2}L^{(1)}_{ab,cd}[2(\Lambda\Gamma^{abc}_{\hspace{4mm}ki}\Lambda)N^{dk}\nonumber \\
&& 
+ \frac{2}{3}(\eta^{b}_{\hspace{2mm}p}\eta^{d}_{\hspace{2mm}i} - \eta^{bd}\eta_{pi})(\Lambda\Gamma^{apcqj}\Lambda)N_{qj}]] + \frac{1}{\eta^{2}}L^{(1)}_{mc,rs}(\Lambda\Gamma^{m}D)(\Lambda\Gamma^{crs}D) + \nonumber\\
&& + \frac{4}{\eta^{3}}(\bar{\Lambda}\Gamma_{lm}\bar{\Lambda})(\bar{\Lambda}\Gamma_{cx}R)(\bar{\Lambda}\Gamma_{n}^{\hspace{2mm}p}R)(\Lambda\Gamma^{lmcnq}\Lambda)(\Lambda\Gamma^{x}D)N_{qp} \nonumber\\
&& -\frac{2}{\eta^{3}}(\bar{\Lambda}\Gamma_{lm}\bar{\Lambda})(\bar{\Lambda}\Gamma_{nx}R)(\bar{\Lambda}R)(\Lambda\Gamma^{lmnpq}\Lambda)(\Lambda\Gamma^{x}D)N_{pq} \nonumber\\
&& -\frac{1}{3\eta^{3}}(\bar{\Lambda}\Gamma_{ef}\bar{\Lambda})(\bar{\Lambda}\Gamma_{gc}R)(\bar{\Lambda}\Gamma_{mn}R)(\Lambda\Gamma^{efghi}\Lambda)(\Lambda\Gamma^{cmn}D)N_{hi} \nonumber \\
&& + \frac{2}{3\eta^{4}}(\bar{\Lambda}\Gamma_{ef}\bar{\Lambda})(\bar{\Lambda}\Gamma_{gc}R)(\bar{\Lambda}\Gamma_{lm}R)(\bar{\Lambda}\Gamma_{np}R)(\Lambda\Gamma^{efghi}\Lambda)(\Lambda\Gamma^{clmnq}\Lambda)\{N_{hi}, N_{qp}\} \nonumber \\
&& - \frac{1}{3\eta^{4}}(\bar{\Lambda}\Gamma_{ef}\bar{\Lambda})(\bar{\Lambda}\Gamma_{gm}R)(\bar{\Lambda}\Gamma_{ln}R)(\bar{\Lambda}R)(\Lambda\Gamma^{efghi}\Lambda)(\Lambda\Gamma^{lnmpq}\Lambda)\{N_{hi}, N_{pq}\}\nonumber \\
&& - \frac{1}{3\eta^{3}}(\bar{\Lambda}\Gamma_{ef}\bar{\Lambda})(\bar{\Lambda}\Gamma_{gc}R)(\bar{\Lambda}\Gamma_{mn}R)(\Lambda\Gamma^{efghi}\Lambda)(\Lambda\Gamma^{chimn}D) \label{app30}
\end{eqnarray}
The differences between this equation and \eqref{app27} is a factor of 2 in the coefficients proportional to $\eta^{-4}$ and the non-zero extra term proportional to $(\Lambda\Gamma^{chimn}D)$. However this mismatch could be fixed in a possible (normal-ordered) quantum version of both expressions.

\end{appendices}

\providecommand{\href}[2]{#2}\begingroup\raggedright\endgroup


\begin{thebibliography}{10}

\bibitem{Berkovits:2002uc}
N.~Berkovits, ``{Towards covariant quantization of the supermembrane},''
  \href{http://dx.doi.org/10.1088/1126-6708/2002/09/051}{{\em JHEP} {\bf 09}
  (2002)  051},
\href{http://arxiv.org/abs/hep-th/0201151}{{\tt arXiv:hep-th/0201151
  [hep-th]}}.

\bibitem{Howe:1991bx}
P.~S. Howe, ``{Pure spinors, function superspaces and supergravity theories in
  ten-dimensions and eleven-dimensions},''
\href{http://dx.doi.org/10.1016/0370-2693(91)90558-8}{{\em Phys. Lett.} {\bf
  B273} (1991)  90--94}.

\bibitem{Green:1999by}
M.~B. Green, M.~Gutperle, and H.~H. Kwon, ``{Light cone quantum mechanics of
  the eleven-dimensional superparticle},''
  \href{http://dx.doi.org/10.1088/1126-6708/1999/08/012}{{\em JHEP} {\bf 08}
  (1999)  012},
\href{http://arxiv.org/abs/hep-th/9907155}{{\tt arXiv:hep-th/9907155
  [hep-th]}}.

\bibitem{Cederwall:2001dx}
M.~Cederwall, B.~E.~W. Nilsson, and D.~Tsimpis, ``{Spinorial cohomology and
  maximally supersymmetric theories},''
  \href{http://dx.doi.org/10.1088/1126-6708/2002/02/009}{{\em JHEP} {\bf 02}
  (2002)  009},
\href{http://arxiv.org/abs/hep-th/0110069}{{\tt arXiv:hep-th/0110069
  [hep-th]}}.

\bibitem{Berkovits:2005bt}
N.~Berkovits, ``{Pure spinor formalism as an N=2 topological string},''
  \href{http://dx.doi.org/10.1088/1126-6708/2005/10/089}{{\em JHEP} {\bf 10}
  (2005)  089},
\href{http://arxiv.org/abs/hep-th/0509120}{{\tt arXiv:hep-th/0509120
  [hep-th]}}.

\bibitem{Cederwall:2009ez} 
  M.~Cederwall, ``{Towards a manifestly supersymmetric action for 11-dimensional supergravity},''
  \href{http://dx.doi.org/10.1007/JHEP01(2010)117}{{\em JHEP} {\bf 1001} (2010) 117},
  \href{http://arxiv.org/abs/0912.1814}{{\tt arXiv:0912.1814 [hep-th]}}.
   
\bibitem{Cederwall:2010tn} 
  M.~Cederwall, ``{D=11 supergravity with manifest supersymmetry},''
  \href{http://dx.doi.org/10.1142/S0217732310034407}{{\em Phys. Lett.}  {\bf A25}, (2010) 3201},
  \href{http://arxiv.org/abs/1001.0112}{{\tt arXiv:1001.0112 [hep-th]}}.

\bibitem{Kugo:1979gm}
T.~Kugo and I.~Ojima, ``{Local Covariant Operator Formalism of Nonabelian Gauge
  Theories and Quark Confinement Problem},''
\href{http://dx.doi.org/10.1143/PTPS.66.1}{{\em Prog. Theor. Phys. Suppl.} {\bf
  66} (1979)  1--130}.

\bibitem{Cederwall:2012es}
M.~Cederwall and A.~Karlsson, ``{Loop amplitudes in maximal supergravity with
  manifest supersymmetry},''
  \href{http://dx.doi.org/10.1007/JHEP03(2013)114}{{\em JHEP} {\bf 03} (2013)
  114},
\href{http://arxiv.org/abs/1212.5175}{{\tt arXiv:1212.5175 [hep-th]}}.

\bibitem{Karlsson:2014xva}
A.~Karlsson, ``{Ultraviolet divergences in maximal supergravity from a pure
  spinor point of view},''
  \href{http://dx.doi.org/10.1007/JHEP04(2015)165}{{\em JHEP} {\bf 04} (2015)
  165},
\href{http://arxiv.org/abs/1412.5983}{{\tt arXiv:1412.5983 [hep-th]}}.

\bibitem{Berkovits:2013pla}
N.~Berkovits, ``{Dynamical twisting and the b ghost in the pure spinor
  formalism},'' \href{http://dx.doi.org/10.1007/JHEP06(2013)091}{{\em JHEP}
  {\bf 06} (2013)  091},
\href{http://arxiv.org/abs/1305.0693}{{\tt arXiv:1305.0693 [hep-th]}}.

\bibitem{Berkovits:2001rb}
N.~Berkovits, ``{Covariant quantization of the superparticle using pure
  spinors},'' \href{http://dx.doi.org/10.1088/1126-6708/2001/09/016}{{\em JHEP}
  {\bf 09} (2001)  016},
\href{http://arxiv.org/abs/hep-th/0105050}{{\tt arXiv:hep-th/0105050
  [hep-th]}}.

\bibitem{Brink:1981nb}
L.~Brink and J.~H. Schwarz, ``{Quantum Superspace},''
\href{http://dx.doi.org/10.1016/0370-2693(81)90093-9}{{\em Phys. Lett.} {\bf
  B100} (1981)  310--312}.

\bibitem{Berkovits:2002zk}
N.~Berkovits, ``{ICTP lectures on covariant quantization of the superstring},''
  in {\em {Superstrings and related matters. Proceedings, Spring School,
  Trieste, Italy, March 18-26, 2002}}, pp.~57--107.
\newblock 2002.
\newblock \href{http://arxiv.org/abs/hep-th/0209059}{{\tt arXiv:hep-th/0209059
  [hep-th]}}.
\newblock
\url{http://www.ictp.trieste.it/~pub_off/lectures/lns013/Berkovits/Berkovits.pdf}.
\newblock

\bibitem{Bedoya:2009np}
O.~A. Bedoya and N.~Berkovits, ``{GGI Lectures on the Pure Spinor Formalism of
  the Superstring},'' in {\em {New Perspectives in String Theory Workshop
  Arcetri, Florence, Italy, April 6-June 19, 2009}}.
\newblock 2009.
\newblock \href{http://arxiv.org/abs/0910.2254}{{\tt arXiv:0910.2254
  [hep-th]}}.
\newblock
\url{https://inspirehep.net/record/833767/files/arXiv:0910.2254.pdf}.
\newblock

\bibitem{Bjornsson:2010wm}
J.~Bjornsson and M.~B. Green, ``{5 loops in 24/5 dimensions},''
  \href{http://dx.doi.org/10.1007/JHEP08(2010)132}{{\em JHEP} {\bf 08} (2010)
  132},
\href{http://arxiv.org/abs/1004.2692}{{\tt arXiv:1004.2692 [hep-th]}}.

\bibitem{Bjornsson:2010wu}
J.~Bjornsson, ``{Multi-loop amplitudes in maximally supersymmetric pure spinor
  field theory},'' \href{http://dx.doi.org/10.1007/JHEP01(2011)002}{{\em JHEP}
  {\bf 01} (2011)  002},
\href{http://arxiv.org/abs/1009.5906}{{\tt arXiv:1009.5906 [hep-th]}}.

\bibitem{Jusinskas:2013yca}
R.~Lipinski~Jusinskas, ``{Nilpotency of the b ghost in the non-minimal pure
  spinor formalism},'' \href{http://dx.doi.org/10.1007/JHEP05(2013)048}{{\em
  JHEP} {\bf 05} (2013)  048},
\href{http://arxiv.org/abs/1303.3966}{{\tt arXiv:1303.3966 [hep-th]}}.

\bibitem{Berkovits:2014ama}
N.~Berkovits and O.~Chandia, ``{Simplified Pure Spinor b Ghost in a Curved
  Heterotic Superstring Background},''
  \href{http://dx.doi.org/10.1007/JHEP06(2014)001}{{\em JHEP} {\bf 06} (2014)
  001},
\href{http://arxiv.org/abs/1403.2429}{{\tt arXiv:1403.2429 [hep-th]}}.

\bibitem{Karlsson:2015qda}
A.~Karlsson, ``{Pure spinor indications of ultraviolet finiteness in D=4
  maximal supergravity},''
\href{http://arxiv.org/abs/1506.07505}{{\tt arXiv:1506.07505 [hep-th]}}.

\bibitem{Gran:2001yh}
U.~Gran, ``{GAMMA: A Mathematica package for performing gamma matrix algebra
  and Fierz transformations in arbitrary dimensions},''
\href{http://arxiv.org/abs/hep-th/0105086}{{\tt arXiv:hep-th/0105086
  [hep-th]}}.


\end{thebibliography}
\end{document}